\documentclass[prd,nofootinbib,showpacs,superscriptaddress, preprint]{revtex4-1}
\pdfoutput=1
\usepackage{amsmath,amsfonts,amssymb,graphicx,natbib,bm}
\usepackage{color}
\usepackage{slashed}    
\usepackage[colorlinks,citecolor=blue]{hyperref}
\usepackage{color}
\usepackage{url}
\usepackage{cancel}
\usepackage{slashed}
\usepackage{multirow}
\usepackage{rotating}
\usepackage{braket}
\usepackage{float}
\usepackage{subfigure}

\newcommand{\be}{\begin{eqnarray*}}
\newcommand{\ee}{\end{eqnarray*}}

\newcommand{\bee}{\begin{eqnarray}}
\newcommand{\eee}{\end{eqnarray}}
\newcommand{\beeq}{\begin{equation}}
\newcommand{\eeq}{\end{equation}}

\usepackage{xspace}

\newcommand{\ba}{\begin{array}}
\newcommand{\ea}{\end{array}}
\newcommand{\bd}{\begin{displaymath}}
\newcommand{\ed}{\end{displaymath}}
\newcommand{\besub}{\begin{subequations}}
\newcommand{\eesub}{\end{subequations}}
\newcommand{\bea}{\begin{eqnarray}}
\newcommand{\eea}{\end{eqnarray}}

\def\a{\alpha}

\def\b{\beta}
\def\g{\gamma}

\def\l{\lambda}
\def\m{\mu}

\def\G{\Gamma}
\def\D{\Delta}

\def\q2 {q^2}

\def\bt{\begin{table}}
\def\et{\end{table}}

\begin{document}
\title{Neutrino mass and asymmetric dark matter: study with inert Higgs doublet and high scale validity}

\author{Amit Dutta Banik}
\email{amitdbanik@mail.ccnu.edu.cn}
\affiliation{Key Laboratory of Quark and Lepton Physics (MoE) and Institute of Particle Physics, Central China
Normal University, Wuhan 430079, China }
\author{Rishav Roshan}
\email{rishav.roshan@iitg.ac.in}
\author{Arunansu Sil}
\email{asil@iitg.ac.in}
\affiliation{Department of Physics, Indian Institute of Technology Guwahati, Assam-781039, India }
\vspace{1.5 cm}
\begin{abstract}  
We consider an inert Higgs doublet (IHD) extension of the Standard Model 
accompanied with three right handed neutrinos and a dark sector, consisting of 
a singlet fermion and a scalar, in order to provide a common framework for dark 
matter, leptognesis and neutrino mass. While the Yukawa coupling of the right 
handed neutrinos with IHD (having mass in the intermediate regime: 80-500 GeV) 
is responsible for explaining the observed baryon 
asymmetry through leptogenesis, its coupling with the dark sector explains the 
dark matter relic density. The presence of IHD also explains the neutrino mass 
through radiative correction.  We find that study of the high scale validity 
of the model in this context becomes crucial as it restricts the parameter space significantly. It turns out 
that there exists a small, but non-zero contribution to the relic density of DM from IHD 
too. Considering all the constraints from dark matter, leptogenesis, neutrino mass 
and high scale validity of the model, we perform a study to find out the viable parameter space. 
\end{abstract}

\maketitle
\section{Introduction}  
\label{int}

Despite its overwhelming success, the Standard Model (SM) of particle physics is yet unable to provide 
answers to several questions involving modern day particle physics and cosmology originated from the 
results of many terrestrial experiments and cosmological observations. Among them, some of the 
most astounding ones involve explanation of tiny neutrino masses and mixing \cite{Fukuda:1998mi,Ahmad:2002jz,Hsu:2006gt,Ahn:2002up}, 
existence and nature of dark matter \cite{Julian:1967zz}-\cite{Tegmark:2003ud} and origin of the 
baryon asymmetry of the Universe \cite{Riotto:1999yt}-\cite{Dine:2003ax}. All these 
indicate that the SM needs to be extended in order to address resolutions of these problems. 
It would be interesting to further investigate whether these problems can have any common framework or 
origin. 

Among various extensions of the SM leading to such a common framework, one interesting possibility 
is to have an inert Higgs doublet (IHD) 
along with three right handed (RH) neutrinos\footnote{Another very economical scenario is the $\nu$MSM model \cite{Asaka:2005an, Asaka:2005pn}  which is based on the extension of SM with three RH neutrinos only.}. 
Provided 
the dark matter (DM) is identified with the lightest neutral component of the IHD, whose 
stability is guaranteed by an imposed $Z_2$ symmetry, it can also generate 
light neutrino mass 
through radiative correction \cite{Ma:2006km}. Simultaneously, the out-of-equilibrium 
decay of the lightest RH neutrino can also be responsible for explaining the baryon asymmetry 
through leptogenesis \cite{Fukugita:1986hr,Buchmuller:2004nz,Anisimov:2007mw,Davidson:2008bu,Baek:2013qwa,Buchmuller:2005eh,Davoudiasl:2015jja,Guo:2016ixx,Hernandez:2016kel,Narendra:2017uxl,Dolan:2018qpy,Ipek:2018sai,Das:2019ntw,Domcke:2020ety,Das:2020vca,Chen:2019etb}. The same Yukawa couplings involved in this asymmetry generation 
are also present in the radiative mass of light neutrinos. It was shown in \cite{Kashiwase:2012xd} that in such a scenario, sufficient lepton asymmetry 
can be generated with heavy RH neutrinos $\gtrsim 10^8$ GeV. For a lighter RH neutrinos, the 
asymmetry generation requires the use of resonant  leptogenesis \cite{Pilaftsis:1997jf,Pilaftsis:2003gt,Dev:2017wwc} in this framework. 
Now turning back to DM status, separate studies of IHD model alone \cite{Barbieri:2006dq,Cirelli:2005uq,LopezHonorez:2006gr,Cao:2007rm,Majumdar:2006nt,Lundstrom:2008ai,Dolle:2009fn,Honorez:2010re,LopezHonorez:2010tb,Chowdhury:2011ga,Arhrib:2013ela,Plascencia:2015xwa,Borah:2017dfn} suggest that there mainly 
exists two mass ($m_{\rm DM}$) ranges of dark matter where the relic density and direct detection 
(DD) limits are satisfied: one is below 80 GeV and other is above 500 GeV. Presence 
of heavy RH neutrinos and neutrino Yukawa coupling (between RH neutrino and IHD) would not 
have significant effect on this conclusion \cite{Borah:2017dfn}.

In view of the above discussion, we pose a question: the lightest neutral component of IHD being in 
the intermediate mass range, 
{\it{i.e.}} 80 - 500 GeV 
\cite{Borah:2017dfn}-\cite{Biswas:2017dxt} what could be an extension or modification of the IHD assisted with radiative neutrino mass 
generation that can account for DM, neutrino mass and baryon asymmetry of the Universe? Note that this intermediate region is 
otherwise interesting from collider search point of view, although ruled out the possibility of being a DM\footnote{The neutral component of the inert doublet being heavier than the 
$W$ boson, annihilation cross-section of DM increases and hence relic density of DM becomes under-abundant in this region.} that satisfies the required 
relic density. 

It is to be noted that the baryon asymmetry can perhaps be best explained by Leptogenesis 
scenario in which out-of-equilibrium decay of 
RH neutrinos take place. Moreover, the ratio of the baryon density to dark matter density $\sim$1/5 
indicates that they may have a common origin. In fact in ref \cite{Falkowski:2011xh}, it was shown a RH neutrino decay 
can simultaneously produce lepton asymmetries in two different sectors: in the SM sector and in 
a hidden DM sector. Finally asymmetric components of the SM and dark sector lepton 
asymmetries would be converted into baryon asymmetry and DM number density respectively. This 
scenario is different from the standard asymmetric dark matter (ADM) \cite{Kaplan:2009ag,Zurek:2013wia,Hamze:2014wca,Kitabayashi:2015oda,Frandsen:2016bke,Murase:2016nwx,Agrawal:2016uwf,Nagata:2016knk,Baldes:2017gzw,Gresham:2017cvl,HajiSadeghi:2017zrl,Tsao:2017vtn,Gresham:2018anj,Narendra:2018vfw,Ibe:2018juk,Dong:2018aak,Narendra:2019cyt} 
scenario where the asymmetry is generated in one sector and then it is transferred to the other sector.
Different models of ADM simultaneously generating visible sector and dark sector asymmetry have been explored
extensively \cite{An:2009vq,Arina:2011cu,JosseMichaux:2011ba,Arina:2012jp,Gu:2016xno,Fornal:2017owa,Yang:2018zrj,Biswas:2018sib,Narendra:2019pag}.

Now returning to the question we have raised above, we consider here the existence of a dark hidden 
sector (secluded from the SM one), the RH neutrinos couple to this hidden dark sector as well as the 
IHD extended SM sector. Such a construction can be fulfilled in an economic way if we consider the dark sector 
comprised of a SM singlet fermion and scalar as in \cite{Falkowski:2011xh}. A different $Z_2$ charge prevails in 
the dark sector under which the dark fermion and scalar are charged and all other particles remain even. This 
way, the lightest among them can be a stable dark matter candidate. Since RH neutrinos carry a lepton number, 
a lepton number conserving interaction of it with dark sector fermion indicates that we also need to assign a lepton 
number to the dark fermion field. This initiates the possibility that a CP violating decay of the RH neutrinos 
generate lepton number asymmetries in both the sectors similar to the two-sector leptogenesis by 
\cite{Falkowski:2011xh}. With the remaining asymmetry being different in two sectors, this will finally lead to 
an asymmetric dark matter and baryon asymmetry of the Universe.

Note that our construction has some interesting differences from the one in \cite{Falkowski:2011xh}. For example, 
the usual Yukawa coupling involving SM lepton doublet, the Higgs doublet and RH neutrino in \cite{Falkowski:2011xh} 
is absent in this construction.Instead the lepton asymmetry progresses through the decay of RH neutrinos into SM 
lepton doublet and IHD. This IHD is also involved in achieving radiative neutrino mass. The absence of neutrino Yukawa interaction involving SM lepton doublet, the Higgs and the RH neutrinos
is also beneficial from electroweak (EW) vacuum stability point of view as such interaction may pose a threat \cite{Ghosh:2017fmr,Bhattacharya:2019fgs,Bandyopadhyay:2020vfc} to it . 
On the other hand, the presence of such interaction involving IHD (as in the present setup) does no harm 
to the stability of the electroweak vacuum. In fact, the presence of IHD turns out to be useful in keeping the 
Higgs quartic coupling positive till a large scale such that electroweak  Vacuum stability can be achieved. Although IHD 
can be a candidate for dark matter by itself, being in the intermediate mass region in this work, its contribution to the 
relic density is expected to be sub-dominant. One could make this contribution as negligible one by considering 
sufficiently large mass splitting among the components of the IHD so as to obtain 
the asymmetric dark matter component as the sole contribution to the relic. However with such large splitting, it turns 
out that the high scale validity of the framework ($i.e.$ the stability of the electroweak vacuum\footnote{Although within 
the present mass limits (3$\sigma$) of Higgs and the top quark mass, the EW vacuum seems to be metastable 
\cite{Isidori:2001bm,Greenwood:2008qp,Ellis:2009tp,EliasMiro:2011aa,Alekhin:2012py,Degrassi:2012ry,Buttazzo:2013uya, Anchordoqui:2012fq,Tang:2013bz,Salvio:2015cja,Salvio:2018rv}, the presence of additional scalars and fermions such as IHD, RH neutrinos, dark sector fields in our model may 
affect this conclusion \cite{Ghosh:2017fmr,DuttaBanik:2018emv,Bhattacharya:2019fgs,Bhattacharya:2019tqq,Borah:2020nsz,Bandyopadhyay:2020vfc,Jangid:2020qgo,Bandyopadhyay:2020djh}.} and the perturbativity of the couplings involved) experiences a challenge as some of the 
parameters may become non-perturbative way before the Planck scale. Hence in this work, we plan to find out the relevant 
parameter space which would not only validate the dark matter, leptogenesis and neutrino mass but also be consistent 
with the high scale validity. In doing so, it is found that the IHD contributes to dark matter abundance to a negligible but 
non-zero extent in addition to the ADM resulting a multi-particle dark matter scenario \cite{Borah:2019aeq,Bhattacharya:2019fgs,Bhattacharya:2019tqq,DuttaBanik:2020jrj} with symmetric and asymmetric 
components.

The paper is organized as follows. We introduce our  dark matter model in 
Sec.~\ref{model} where the particle spectrum of our model and charges under different 
symmetry groups have been discussed. 
Various theoretical and experimental constraints in our model are presented in 
Sec.~\ref{constraints}. In Sec.~\ref{asymmetry} we discuss how asymmetries 
(lepton and dark sector) in two sectors and radiative neutrino masses are  generated in the model. 
The Boltzmann equations that produces the final lepton and dark matter asymmetry is presented in 
Sec.~\ref{Boltzmann}.
In Sec.~\ref{strategy}, we briefly explain the strategy to evaluate asymmetries using neutrino parameters consistent with the vacuum stability constraints and relic abundance of symmetric dark matter 
is reported. In Sec.~\ref{result} we discuss our results by evaluating 
asymmetry in visible and dark sector for different set of parameters considered. 
Bounds from dark matter direct detection and flavour violating decay is also discussed in 
Sec.~\ref{result}. Finally in Sec.~\ref{summary} we conclude.

\section{The Model}
\label{model}

We consider an extension of the SM by an IHD ($\Phi$), and three RH neutrinos ($N_{i=1,2,3}$), 
which can accommodate radiatively generated light neutrinos mass \cite{Ma:2006km}. A $Z_2$ 
symmetry is imposed under which both $\Phi$ and $N_{i}$ are odd while all SM fields are even. 
This prohibits the Yukawa coupling involving lepton doublets ($l_{L}$), $N_{i}$ and 
the SM Higgs doublet $H$ and hence the neutrino Dirac mass 
term is absent. We also consider the existence of a dark sector which is composed of a SM singlet 
Dirac fermion $\Psi$ and a  real singlet scalar $S$. This sector is secluded by an additional $Z'_2$ 
symmetry under which only these dark sector fields remain odd.  Charge assignments of the various 
fields involved are shown in Table~\ref{t1}.

\begin{table}[H]
\begin{center}
\vskip 0.5 cm
\begin{tabular}{|c|c|c|c|c|c|c|c|}
\hline
              & $l_{L}$ & $e_{R}$ & $H$ &   $\Phi$  & $S$ & 
$N_i$      & $\Psi$             \\
\hline        
$SU(2)$       &       2       &  1     &  2     &  2          &  1   &  1       
    & 1             \\
\hline
$U(1)_Y$      & $-\frac{1}{2}$&  -1     &  $\frac{1}{2}$     & $\frac{1}{2}$&   
0&0 & 0            \\
\hline
$Z_2$         &       +       &  +     &   +    &  -          &   -  & -        
    & +            
    \\

\hline
${Z_{2}^{'}}$      &        +      &  +     &  +     &  +          &   - & +            & -       

 \\
\hline
\end{tabular}
\end{center}  
\caption{Particles and their charges under different symmetries.}
\label{t1}
\end{table}


The Lagrangian describing the Yukawa interaction between the additional fields and the SM ones is then given by
\bea
\mathcal{L}_{\rm{Int}}=m_{\Psi}\bar{\Psi}\Psi+ \frac{1}{2}M_{i}\bar{N}^{c}_{i}N_{i}
+Y_{i \alpha }\bar{N}_{i} \tilde{\Phi}^{\dagger}l_{L\alpha}+\lambda_{D_i}\bar{N}_{i}S\Psi+ h.c.\,\,,
\label{e5}
\eea
with indices $i,\alpha$ run as 1,2,3 (generation indices) and $\lambda_D$ denotes the coupling among 
the dark sector and the RH neutrinos.
In the above Lagrangian, we also include the masses for heavy RH  Majorana 
neutrinos and dark sector Dirac fermion as $M_{i}$ and 
$m_{\Psi}$ respectively. For simplicity, we consider the RH neutrino mass matrix 
to be diagonal. With the above construction, these heavy RH neutrinos 
couple not only with the IHD extended SM but also with the dark sector (consisting of  
$S$ and $\Psi$). Therefore, RHNs in the present model serve as the mediator 
between these two sectors. Being heavy, it can decay into both the sectors and can in principle be 
responsible for lepton asymmetry in case  
the associated Yukawa couplings are complex. We assume that the charged 
lepton mass matrix is diagonal. Note that due to the presence of other $Z'_2$, direct 
interaction of $\Psi$ with IHD extended SM is forbidden. The construction has a global 
$B-L$ symmetry under which the RHNs are charged as -1 and the dark sector field $\Psi$ (${\bar{\Psi}}$) is also charged 
as -1 (1). However introduction of Majorana mass for the RH neutrinos break this 
symmetry explicitly.


The scalar sector of the present model consists of the SM Higgs doublet $H$, inert Higgs doublet $\Phi$ 
and the real singlet scalar $S$. Therefore the most general potential, invariant under the chosen symmetry, 
can be written as
\bea
V &=& -\mu_{H}^2 {H}^\dagger {H} + \mu_{\Phi}^2  {\Phi}^{\dagger}\Phi + \frac{1}{2}
\mu_s^2 S^2 +\lambda_H (H^\dagger {H})^2
+ \lambda_{\Phi} ({\Phi}^\dagger {\Phi})^2 + \lambda_{1}
(H^\dagger H)(\Phi^\dagger \Phi) \nonumber \\ &+&
\lambda_2 (\Phi^\dagger H)(H^\dagger \Phi) + \frac{1}{2} \lambda_3
[(\Phi^\dagger H)^2 + (H^\dagger \Phi)^2]+  \frac{\l_{H S}}{2} S^2 (H^\dagger H) +  \frac{\l_{\Phi S}}{2}
S^2 (\Phi^\dagger \Phi)  \nonumber
\\  &+& \frac{1}{4!} \l_S S^4.
\label{e1}
\eea 
After electroweak symmetry breaking (EWSB) of the SM, the SM Higgs doublet and the IHD can be written as 
\bea                 
H = \left( \begin{array}{c}
                         0  \\
        \frac{1}{\sqrt{2}}(v+h)  
                 \end{array}  \right) \, ,                     
&& \Phi =\left( \begin{array}{c}
                           \Phi^+   \\
        \frac{1}{\sqrt{2}}(\Phi_0+iA_0)  
                 \end{array}  \right) \,\, ,
\label{e2a}
\eea  
where $v=246$ GeV. 
Masses of different physical scalars are given as
\bea
m_h^2&=&2 \lambda_H v^2, \nonumber \\
m^{2}_{\Phi^{\pm}}&=&\mu_{\Phi}^{2}+\lambda_{1}\frac{v^{2}}{2}, \nonumber \\
m_{\Phi_0}^{2}&=&\mu_{\Phi}^{2}+(\lambda_{1}+\lambda_{2}+\lambda_{3})\frac{v^{2}}{2}, \nonumber \\
m_{A_0}^{2}&=&\mu_{\Phi}^{2}+(\lambda_{1}+\lambda_{2}-\lambda_{3})\frac{v^{2}}{2}, \nonumber\\
m_{S}^{2}&=&\mu_{s}^{2}+\frac{\l_{H S}}{2} v^{2} 
\,\, .
\label{e3}   
\eea
We consider all the couplings in the expression of potential in Eq.~(\ref{e1}) are real with $\mu_{H}^2,\mu_{\Phi}^{2},\mu_{S}^{2}>0$ and $\lambda_3<0$ such that $\Phi_0$ is the lightest among inert particles. We define $\l_{L} \equiv \frac{\l_{1} + \l_{2} + \l_{3}}{2}$, which denote the  Higgs portal couplings of $\Phi_0$. For our analysis purpose, we choose the following sets of independent parameters: 
$$(m_{\Phi_0}, m_{A_0},m_{\Phi^+},m_{S},M_i,m_{\Psi} ,\l_L,\l_{HS},\l_{\Phi S},\l_{D_i}).$$  

\section{Constraints}
\label{constraints}

\subsection{Theoretical constraints}
\begin{itemize}
\item[(i)]{\bf{Stability:}} In view of stability of the scalar potential as mentioned in Eq. (\ref{e1}), the quartic couplings involved there 
are required to satisfy the co-positivity conditions \cite{Kannike:2012pe, Chakrabortty:2013mha} as below: 
\bea
\rm{cpc}(1,2,3): ~~~~~~~~~~~~~~~\lambda_{\it{H}},\,\lambda_{\Phi},\,\l_{\it{S}} \geq 0,\nonumber\\
\rm{cpc}(4): ~~~~~~~~~~~~~~~~~\lambda_{1} + 2\sqrt{\lambda_{\it{H}}\lambda_{\Phi}}  &\geq&  0,\nonumber\\
\rm{cpc}(5,6): ~~~\lambda_{1} +\lambda_2 -|\lambda_3| + 2\sqrt{\lambda_{\it{H}}\lambda_{\Phi}} & \geq & 0, \nonumber\\
\rm{cpc}(7): ~~~~~~~~~~~~~~~\l_{\it{HS}} +\sqrt{\frac{2}{3}\lambda_{\it{H}}\l_{\it{S}}}  &\geq & 0,\nonumber\\
\rm{cpc}(8): ~~~~~~~~~~~~~~~\l_{\Phi \it{S}} +\sqrt{\frac{2}{3}\lambda_{\Phi}\l_{\it{S}}}  &\geq &  0\
\label{copo}
\eea
where cpc(i) denotes $i^{th}$ copositivity condition.  

\item[(ii)]{\bf{Perturbativity:}} In order to keep the model parameter perturbative one expects: 
\bea
|\l_{i}|< 4\pi~ {\rm{and}}~ |g_i|,|Y_{i \b}|,|\l_{D_i}|<\sqrt{4\pi}  .
\label{pert} 
\eea
where $\l_i$  represents the scalar quartic couplings involved in the present setup whereas $g_i$ denotes the SM gauge couplings and finally, $Y_{i \b}$ and $\l_{D_i}$ denote the Yukawa couplings respectively. We will investigate the perturbativity of the couplings present in the model by employing the renormalisation group equations (RGE).

\end{itemize}

\subsection{Experimental constraints}

\begin{itemize}
\item[(i)]{\bf{Electroweak precision parameters:}} For a multi-Higgs scenario, the strongest constraint is imposed by the $T$ \cite{Peskin:1991sw,Grimus:2008nb,Arhrib:2012ia}. More precisely, this restricts the mass splitting
between the scalars belonging to an $SU(2)_L$ multiplet. The contribution coming from the IHD is expressed as given by \cite{Grimus:2008nb,Arhrib:2012ia}:
\bea
\Delta T&=& \frac{g^2}{64\pi^2m_W^2\a}\bigg{[}F(m_{\Phi^+}^2,m_{\phi_0}^2)+F(m_{\Phi^+}^2,m_{A_0}^2)-F(m_{A_0}^2,m_{\phi_0}^2)\bigg{]}\nonumber\\.  
\label{STU} 
\eea 

where $F(x,y) = \frac{1}{2}(x + y)-\frac{xy}{x-y}
\rm{ln}( \frac{x}{y} )$ for $x \neq y$ and $F(x,y) =0$ for $x=y$. We use the latest bound \cite{Tanabashi:2018oca} as $\Delta T = 0.07\pm0.12.$

\item[(ii)]{\bf{LHC diphoton signal strength:}} Due to the presence of the interactions among the SM Higgs and the 
IHD (see Eq. (\ref{e1})), the charged component $\Phi^{\pm}$ of the IHD provides a significant contribution to the 
$h\rightarrow \gamma\g$ at one loop in addition to the SM contribution. The analytic expression of the entire contribution 
can be expressed as \cite{Arhrib:2012ia,Swiezewska:2012eh}
\bea
\Gamma(h\rightarrow \g\g)_{\rm{IHD+SM}}&=&\frac{G_f \a^2 m_h^3}{128\sqrt{2}\pi^3}\bigg{|}\frac{4}{3}\mathcal{A}_{1/2}(x_i)+\mathcal{A}_{1}(x_i)+\frac{\l_{1} v^2}{2m_{\Phi^{\pm}}^2}\mathcal{A}_{0}(x_i)\bigg{|}^2,
\label{hgg4}   
\eea 
where $G_f$, is the Fermi constant. The form factors $\mathcal{A}_{1/2}(x_i),\mathcal{A}_{1}(x_i)~{\rm and}~\mathcal{A}_{0}(x_i),$ are induced by top quark, $W$ gauge boson and $\Phi^{\pm}$ loop respectively. The formula for the form factors are listed below:
\besub
\bea
\mathcal{A}_{1/2}(x_i)&=& 2[x_i+(x_i-1)f(x_i)]x_i^{-2}, \\ 
\mathcal{A}_{1}(x_i)&=& -[3x_i+2x_i^2+3(2x_i-1)f(x_i)]x_i^{-2}, \\ 
\mathcal{A}_{0}(x_i)&=& -[x_i-f(x_i)]x_i^{-2}, 
\eea 
\eesub
where $x_i=\frac{m_h^2}{4m_i^2}$ and $f(x)= (\sin^{-1}\sqrt{x})^2$.

The Higgs to diphoton signal strength is conventionally parametrized as 
\bea
\mu_{\g\g}&=&\frac{\Gamma(h\rightarrow \g\g)_{\rm{IHD+SM}}}{\Gamma(h\rightarrow \g\g)_{\rm{SM}}} .
\label{hgg3}   
\eea 
In order to ensure that $\mu_{\g\g}$ lies within the experimental uncertainties, the analysis should respect the latest signal 
strength from LHC as given by $\mu_{\g\g}=0.99\pm 0.14$ from ATLAS \cite{Aaboud:2018xdt} 
and $\mu_{\g\g}=1.17\pm 0.10$ from CMS \cite{Sirunyan:2018ouh}.
 
 \item[(iii)]{\bf{Baryon asymmetry of the Universe:}} The baryon asymmetry of the Universe is usually expressed in terms $Y_{\Delta B}$ which is the  ratio of the baryon density $n_{\Delta B}$ to the entropy density $s$ of the Universe measured today. The present bound on this ratio is \cite{Tanabashi:2018oca}:
 \bea 
 Y_{\Delta B}=\frac{n_{\Delta B}}{s}\simeq(8.24-9.38)\times10^{-11}\, ,
 \label{BAU}
 \eea
 where $n_{\Delta B} =n_B-n_{\bar{B}}$.

\item[(iv)]{\bf{Relic density and Direct detection of DM:}} The relic density bound obtained from the Planck 
experiment \cite{Aghanim:2018eyx} is given by
\begin{gather}
   \Omega_{\rm DM}  h^2 = 0.120{\pm 0.001} \, ,
 \label{planck}
 \end{gather}
which is used to restrict the parameter space of the current setup. 
In addition, the parameter space can be  further restricted by applying bounds on the DM direct 
detection cross-section coming from various experiments like LUX \cite{Akerib:2016vxi}, XENON-1T 
\cite{Aprile:2018dbl}, PandaX-II \cite{Tan:2016zwf,Cui:2017nnn}. 

\item[(v)]{\bf{Neutrino mass and mixing}} Global fits to neutrino 
oscillation parameters (in terms of light neutrino masses and mixing) are summarized in Table~\ref{nuparameters} \cite{Tanabashi:2018oca}.

\begin{table}[htb]
\centering
\begin{tabular}{|c|c|c|}
\hline
Parameters & Normal Hierarchy (NH) & Inverted Hierarchy (IH) \\
\hline
$ \frac{\Delta m_{21}^2}{10^{-5} \text{eV}^2}$ & $6.79-8.01$ & $6.79-8.01 $ \\
$ \frac{|\Delta m_{31}^2|}{10^{-3} \text{eV}^2}$ & $2.427-2.625$ & $2.412-2.611 $ \\
$ \sin^2\theta_{12} $ &  $0.275-0.350 $ & $0.275-0.350 $ \\
$ \sin^2\theta_{23} $ & $0.418-0.627$ &  $0.423-0.629 $ \\
$\sin^2\theta_{13} $ & $0.02045-0.02439$ & $0.02068-0.02463 $ \\
$ \delta (^\circ) $ & $125-392$ & $196-360$ \\
\hline
\end{tabular}
\caption{Global fit $3\sigma$ values of neutrino oscillation parameters \cite{Tanabashi:2018oca}.}
\label{nuparameters}
\end{table}

\item[(vi)]{\bf{Lepton flavour violation (LFV):}} It is to be noted that the present setup includes 
right handed neutrinos and inert Higgs doublet which may enhance flavour violating 
decays \cite{Ma:2001mr,Toma:2013zsa,Baek:2014awa,Das:2017ski}. Flavour violating decays are highly 
suppressed in Standard Model of particle physics. Therefore it is necessary
to ensure that such processes do not get enhanced significantly. The primary 
contribution to such flavour violating decays in the present set-up originates from the exchange of right 
handed neutrino $N_k$ and charged inert scalar $\Phi^{\pm}$ at one loop. 
The branching ratio of such decays involving the lepton sector Yukawa interactions are given by \cite{Baek:2014awa}:
\bea
Br(\mu\rightarrow e\gamma) = \frac{3\alpha_{em}}{64\pi G_{F}^{2} 
m_{\Phi^{\pm}}^{4}}|\sum_{k}Y_{k\mu}Y^{*}_{ke}F(M_{k}^2/m_{\Phi^{\pm}}^2)|^2\, ,
\label{lfv} 
\eea
with $\alpha_{em}$ = $e^2/4\pi$ the electromagnetic fine structure constant, $G_F$ is the Fermi constant and 
$F(x)$ = $(1-6x+3x^2+2x^3-6x^2\rm{ln}\textit{x})/6(1-\textit{x})^4$\cite{Ma:2001mr}. As mentioned 
earlier, non-observation of these flavour violating decay imposes a strong 
upper bound on the branching ratio of these decay modes. The present upper 
bound on the $Br(\mu\rightarrow e\gamma)$ reported by the MEG collaboration 
\cite{TheMEG:2016wtm} is $Br(\mu\rightarrow e\gamma) < 4.2\times10^{-13}$ at 
$90 \%$ C.L. 
\end{itemize}

\section {Asymmetric dark matter, leptogenesis and neutrino mass}
\label{asymmetry}
The model is constructed primarily with an aim to realize dark matter and lepton asymmetry 
through the so-called `two-sector leptogenesis' scenario as proposed in \cite{Falkowski:2011xh} 
and to include radiative generation of neutrino masses. Now in view of the 
detailed construction based on the imposed symmetries, we find that the decay of 
the lightest RH neutrino can be responsible for such a realization.  The 
lightest RH neutrino $N_1$  decays into $S$ and $\Psi$ through the Yukawa 
interaction proportional to $\lambda_{D_1}$. The $Z'_2$ odd fermion field $\Psi$ is 
the dark matter candidate and its stability is ensured by assuming the scalar 
$S$ to be heavier than $\Psi$. Simultaneously, $N_1$ decays into the IHD and the SM 
lepton doublet through the other Yukawa interaction 
(proportional to $Y$). Assuming the presence of complex Yukawa couplings (in $Y$ and/or 
$\lambda_{D}$), such decays can produce 
(a) lepton asymmetry in IHD extended SM sector and (b) a dark matter number asymmetry. Assuming 
the symmetric component of the DM being washed out, it is the asymmetry in number densities 
of DM particles ($n_{\Psi} - n_{{\bar{\Psi}}}$) which would determine the relic density of DM. A 
typical characteristic of such an asymmetric DM model is found to be: as the DM particle $\Psi$ carries a lepton 
number, lepton asymmetries are generated in both the sectors. The relative difference in 
these two asymmetries will however depend on the branching ratio of the $N_1$ decay and the 
wash-out effects discussed later in Sec.~\ref{Boltzmann} while solving the Boltzmann equations. 
Below we provide a brief discussion on the expressions for asymmetries as well as light neutrino 
mass.

As already stated, we consider the RH neutrino mass matrix as diagonal with $M_{1}<<M_{{2,3}}$ 
and $Y$ is the Yukawa coupling defined in basis where the charged lepton mass matrix 
is also diagonal.
\begin{figure}[h]
\centering
\subfigure[]{
\includegraphics[scale=0.35]{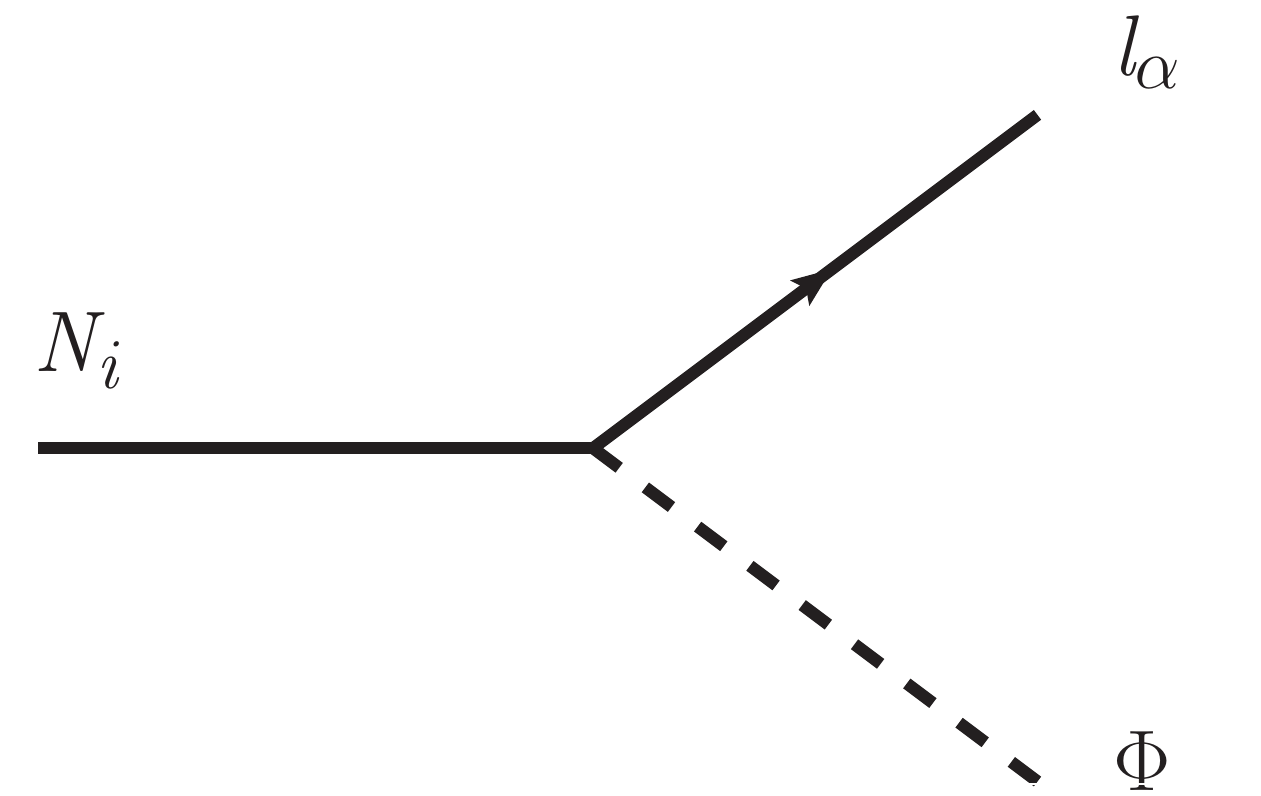}}
\subfigure []{
\includegraphics[scale=0.35]{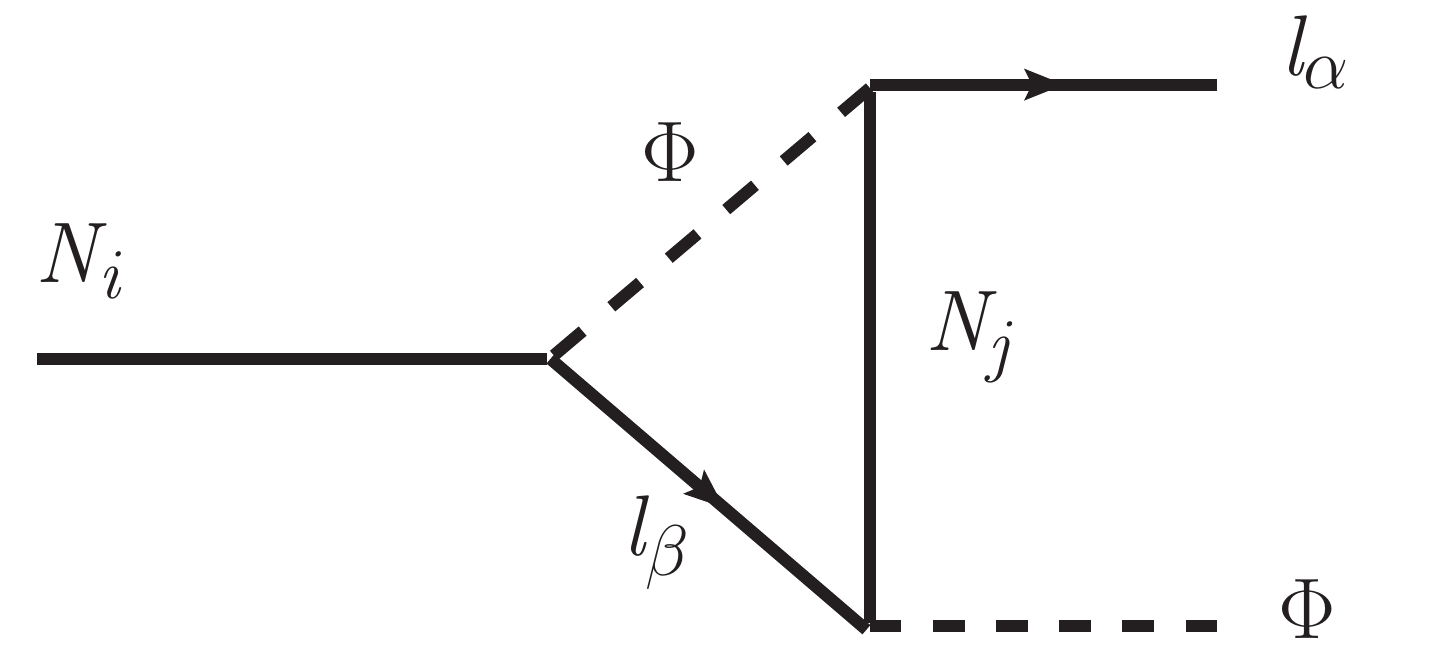}}
\subfigure []{
\includegraphics[scale=0.35]{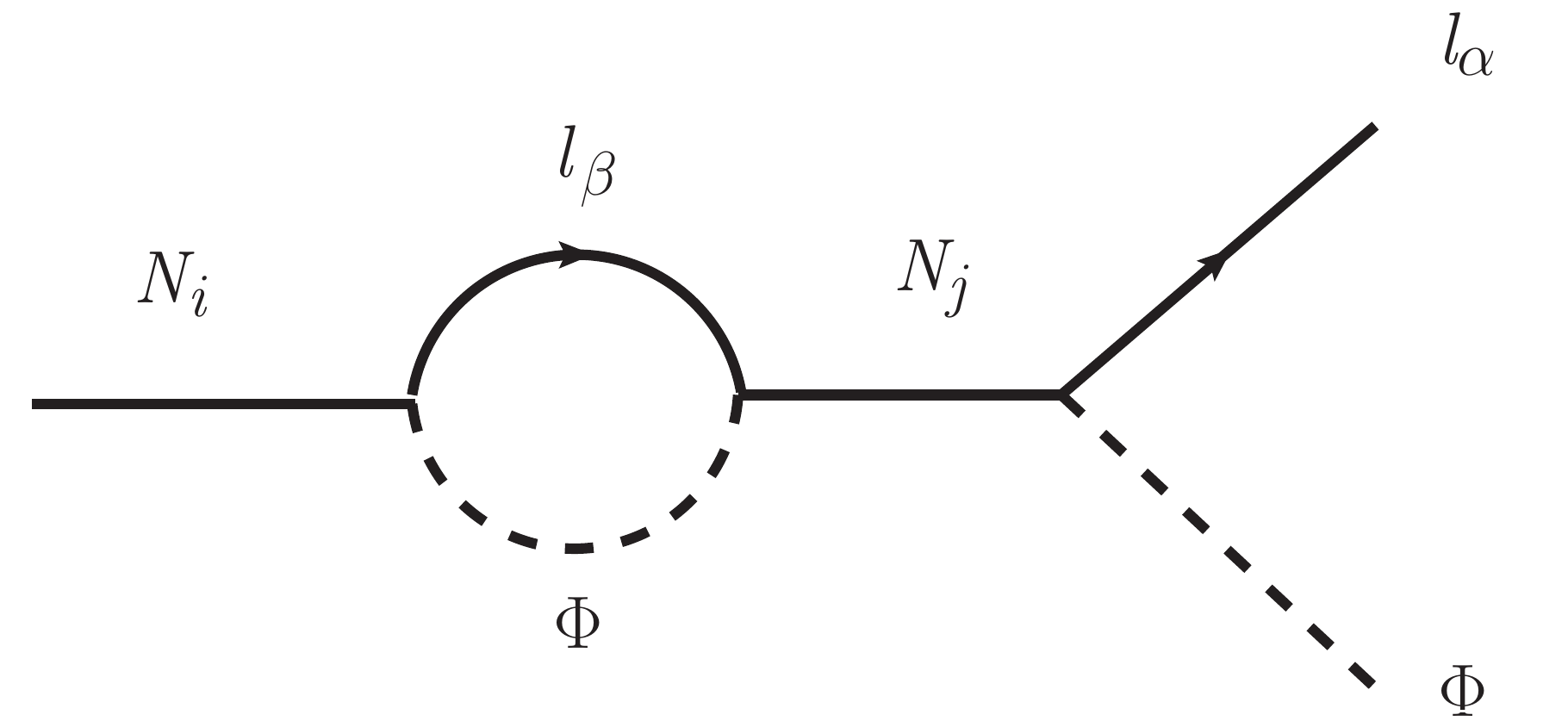}}
\subfigure []{
\includegraphics[scale=0.35]{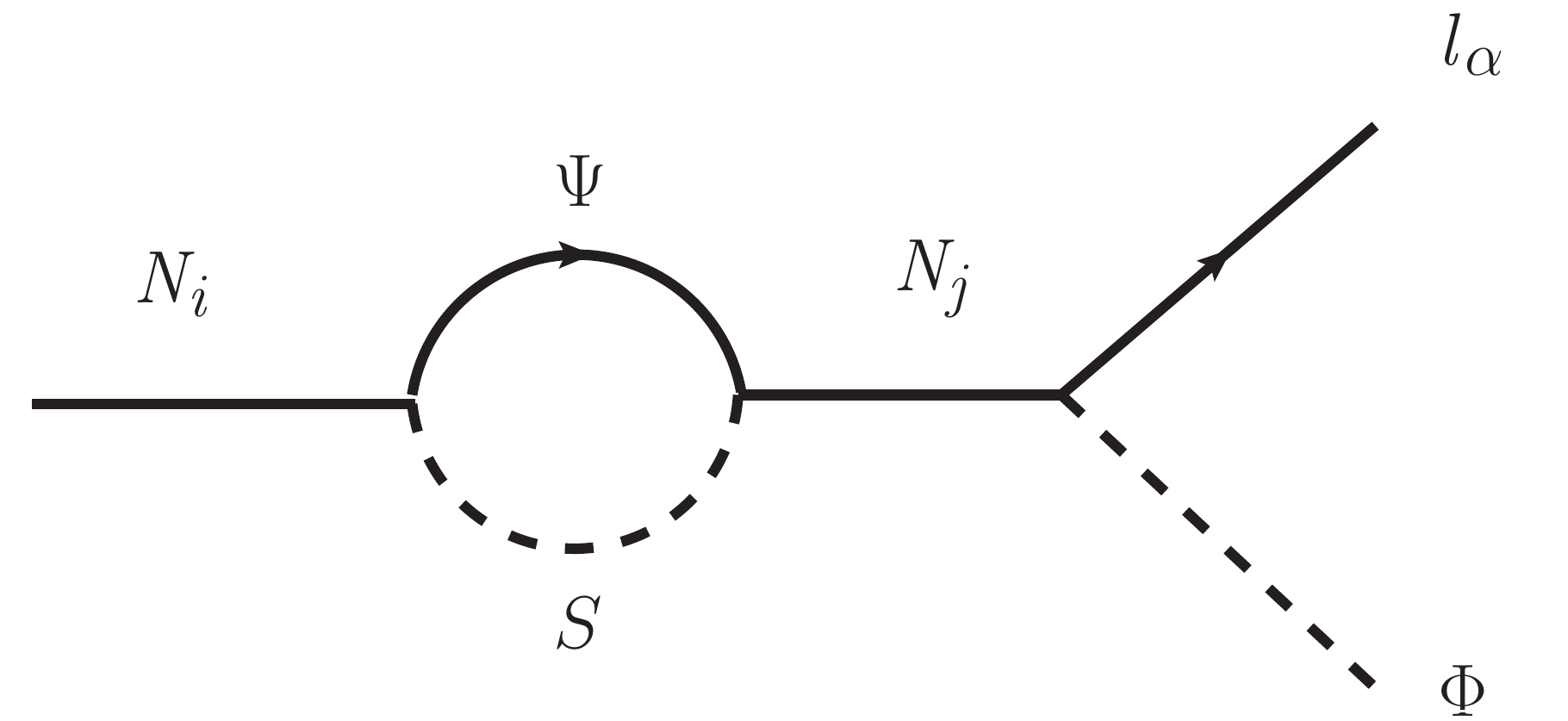}}
\caption{Tree level, vertex and the self energy diagrams required for the generation of the asymmetry in the lepton sector}
\label{figa}
\end{figure}
\begin{figure}[]
\centering
\centering
\subfigure[]{
\includegraphics[scale=0.35]{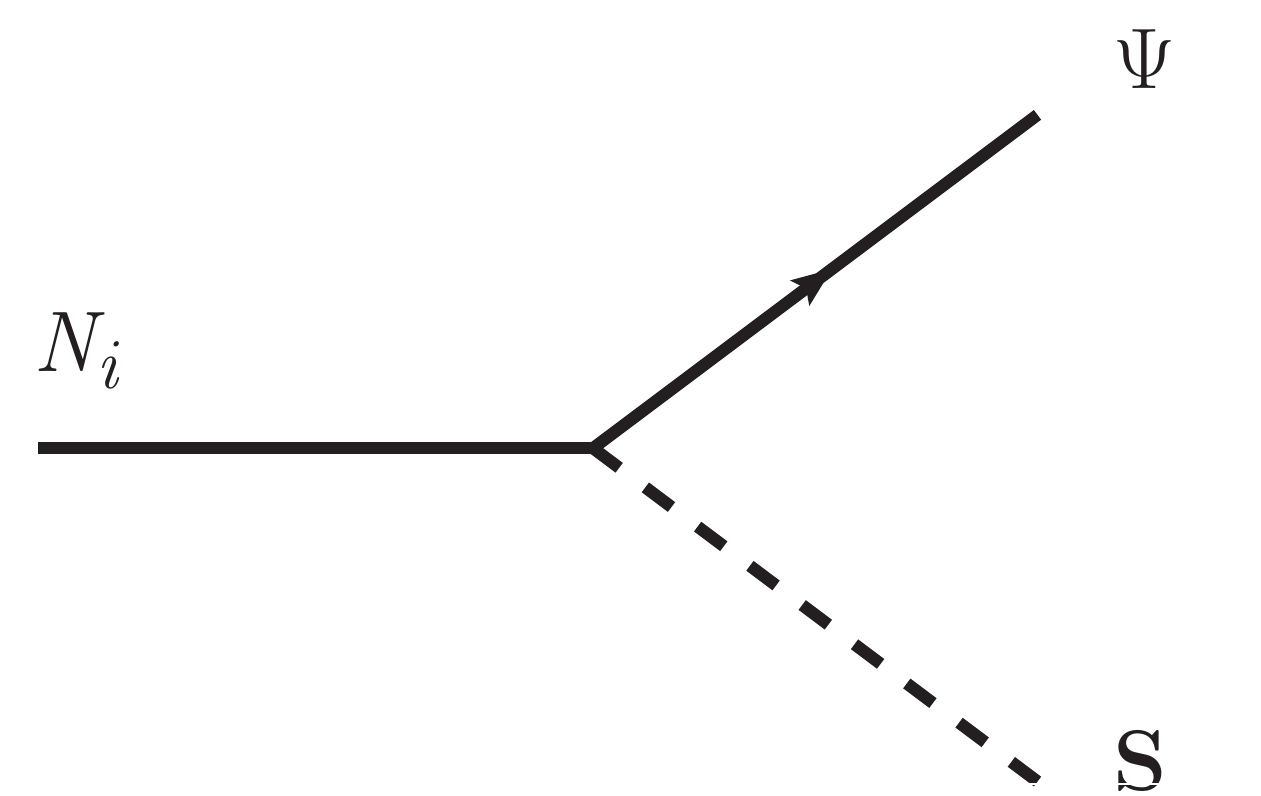}}
\subfigure []{
\includegraphics[scale=0.35]{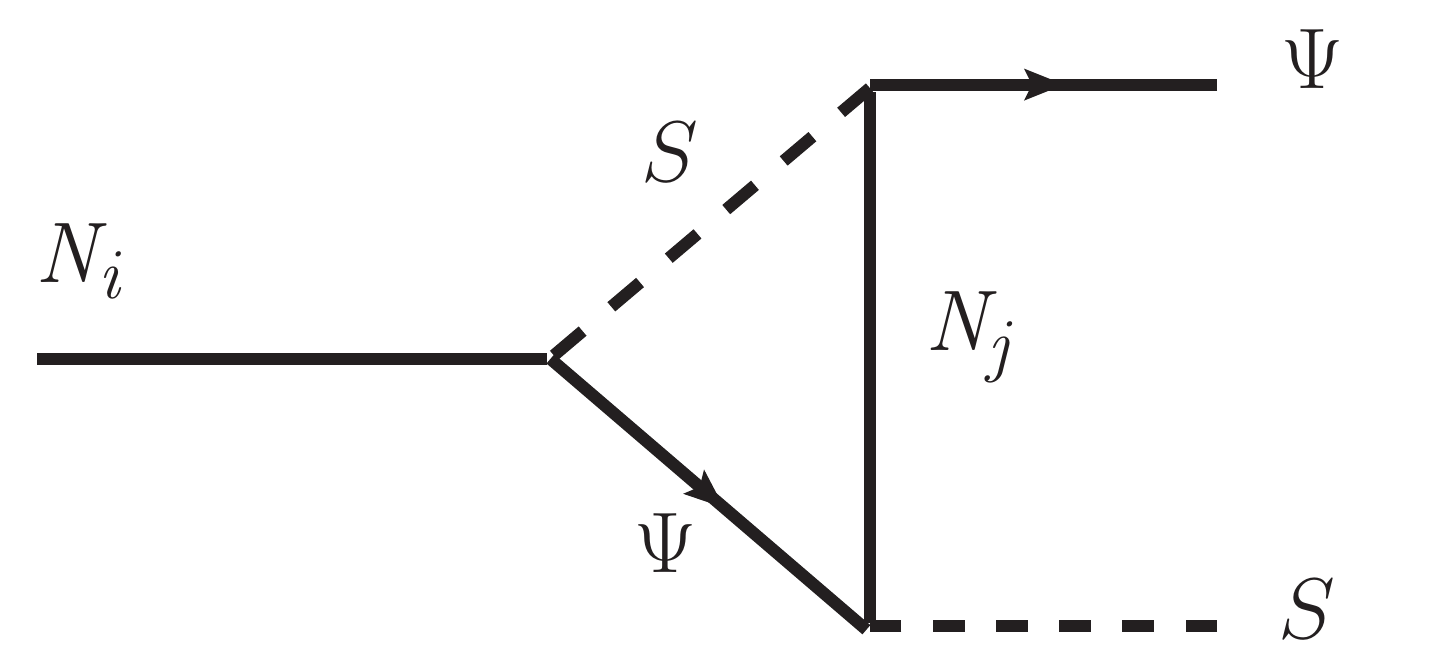}}
\subfigure []{
\includegraphics[scale=0.35]{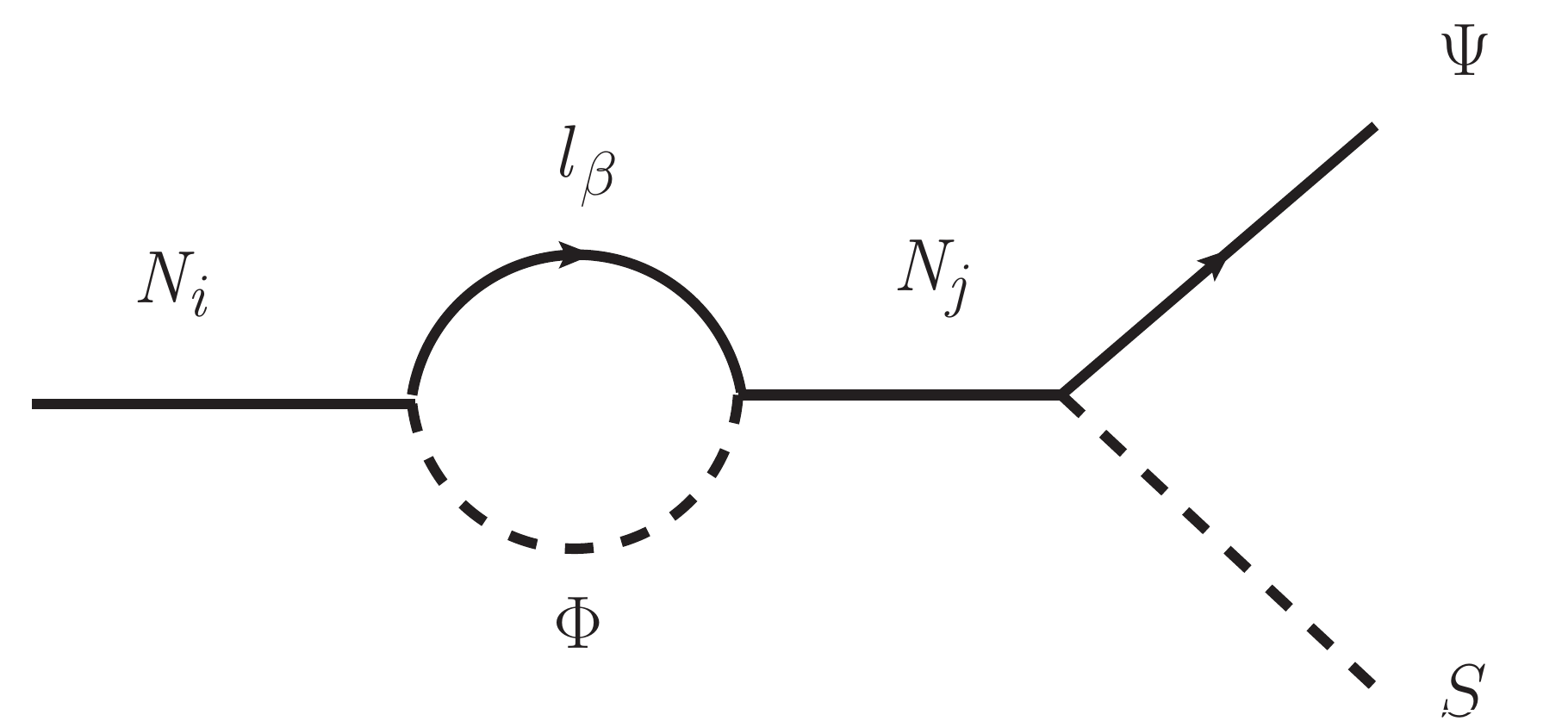}}
\subfigure []{
\includegraphics[scale=0.35]{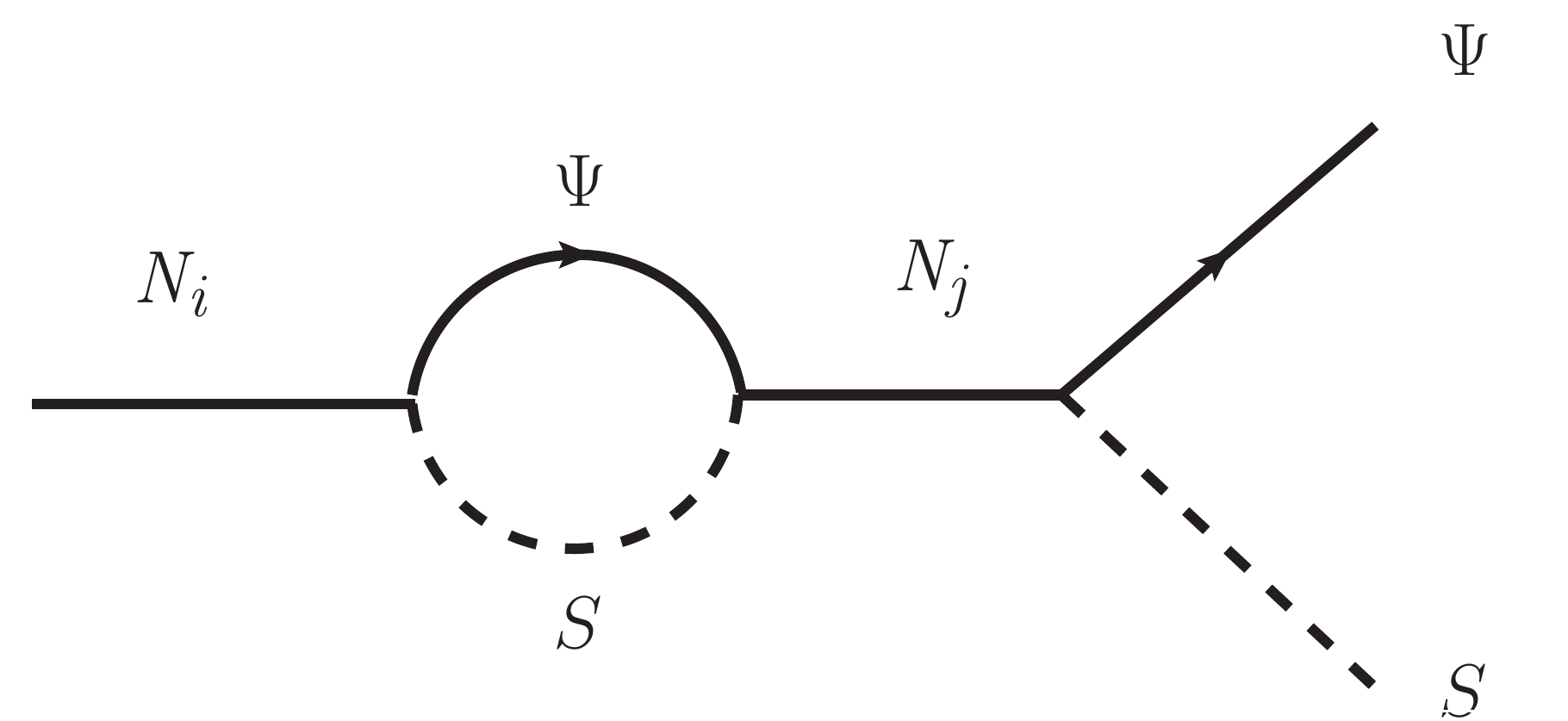}}
\caption{Tree level, vertex and the self energy diagrams required for the generation of the asymmetry in the dark sector}
\label{figb}
\end{figure}
In Fig.~\ref{figa} and Fig.~\ref{figb} we present 
the corresponding diagrams that produce these asymmetries from RH neutrino decay. Below we express the 
asymmetries 
produced in the two sectors  from the decay of the lightest RH neutrino $N_1$ given by, 
\bea
\label{eq:epsilonSM1}
\epsilon_L  &= & \frac{\sum_{\alpha}[\Gamma(N_1 \rightarrow l_{\alpha}+\Phi)-\Gamma(N_1 \rightarrow \bar{l_{\alpha}}+\Phi^{*})]}{\Gamma_1}
\\
&\simeq& \frac{M_1}{8\pi} \frac{{\rm Im}[(3Y^*Y^T + \lambda_{D}^*\lambda_{D}^T) 
M^{-1} YY^\dagger]_{11}}{[2YY^\dagger + \lambda_{D}\lambda_{D}^{\dagger}]_{11}}\,,
\label{eq:epsilonSM1a}
\eea
and
\bea
\label{eq:epsilonDM1}
\epsilon_{\Psi} &=&
\frac{\Gamma(N_1 \rightarrow \Psi+S)-\Gamma(N_1 \rightarrow \bar{\Psi}+S^{*})}{\Gamma_1} 
\\
& \simeq &  \frac{M_1}{8\pi} \frac{{\rm Im}[(Y^*Y^T + \lambda_{D}^*\lambda_{D}^T) 
M^{-1} \lambda_{D}\lambda_{D}^\dagger]_{11}}{[2YY^\dagger + \lambda_{D}\lambda_{D}^{\dagger}]_{11}}\, ,
\label{eq:epsilonDM1a}
\eea
where {$\Gamma_1=\frac{M_{1}}{16\pi}(2YY^{\dagger}+\lambda_{D}\lambda_{D}^{\dagger})_{11}$} is the 
total decay width of $N_1$ and we have employed Eq.(\ref{e5}). Since we have a single generation of 
$\Psi$, a $3 \times 3$ structure of $\lambda_{D}$ matrix can be taken to be of the form,
\bea
{\lambda_{D}} = \left(
\begin{array}{ccc}
 \lambda_{D1} & 0\,~ 0 \\
\lambda_{D2} & 0\, ~0 \\
\lambda_{D3} & 0\, ~0 \\
\end{array}
\right). \,\,
\label{lam}
\eea

\begin{figure}[h]
\begin{center}
\includegraphics[scale=0.55]{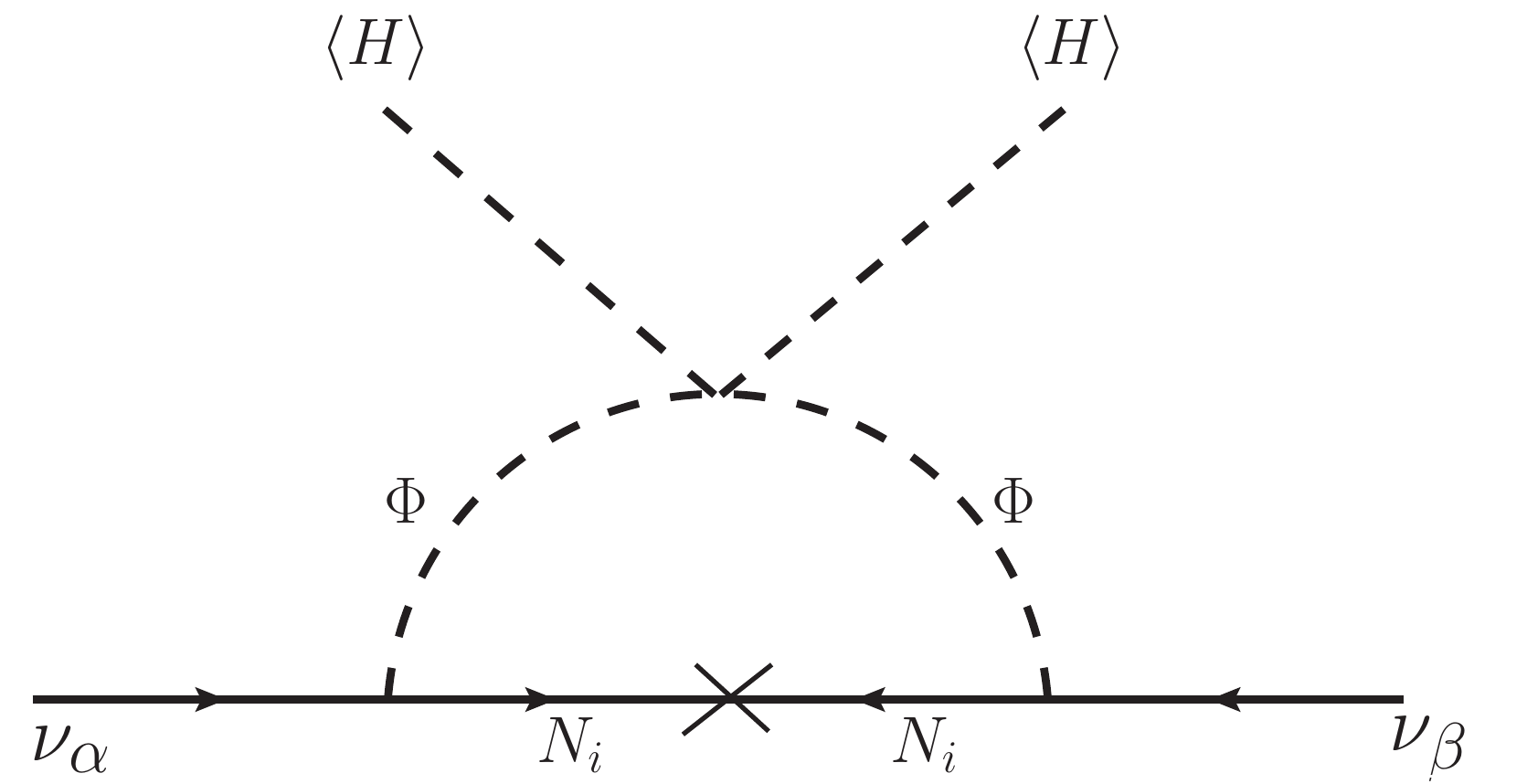}
\caption{One-loop generation of neutrino mass.}
\label{rad-mass}
\end{center}
\end{figure}

\noindent Note that in the expression for $\epsilon_L$ and $\epsilon_{\Psi}$, the Yukawa couplings involved ($Y$) are also part of light 
neutrino mass matrix which is generated by one loop radiative correction\cite{Ma:2006km} as shown in Fig. \ref{rad-mass}. 
The light neutrino mass in our set-up is given by \cite{Ahriche:2017iar}
\bea
(m_{\nu})_{\alpha\beta}=\sum_{i}\frac{Y_{i\alpha}Y_{i\beta}M_{i}}{
{32}\pi^2}\bigg{[}\frac{m_{\Phi_{0}}^2}{m_{\Phi_{0}}^2-M_{i}^2} {\rm{ln}}\frac{m_{\Phi_{0}}^{2}}{
M_{i}^{2}}-\frac{m_{A_0}^2}{m_{A_0}^2-M_{i}^2} {\rm{ln}}\frac{m_{A_0}^{2}}{M_{i}^{2}} 
\bigg{]},
\label{numass1}
\eea
where expressions of $m_{\Phi_0}$ and $m_{A_0}$ can be obtained from Eq. (\ref{e3}). 
The mass eigenvalues and mixing are then obtained by diagonalizing the light neutrino mass matrix as:
\be
m_{\nu} =U^{*}(m_{\nu}^{d})U^{\dagger}
\label{numass2}
\ee
with $m_{\nu}^{d}={\rm{diag}}(m_1,m_2,m_3)$ consisting of mass eigenvalues and 
$U$ is the  Pontecorvo-Maki-Nakagawa-Sakata (PMNS) matrix 
\cite{Maki:1962mu}  (the charged lepton mass matrix is considered 
to be diagonal) which can be written as:
\be
U=\left(
\begin{array}{ccc}
 c_{12} c_{13} & c_{13} s_{12} & e^{-i \delta_{CP} } s_{13} \\
 -c_{23} s_{12}-e^{i \delta_{CP} } c_{12} s_{13} s_{23} & c_{12} c_{23}-e^{i \delta_{CP} } s_{12} s_{13} s_{23} & c_{13} s_{23} \\
 s_{12} s_{23}-e^{i \delta_{CP} } c_{12} c_{23} s_{13} & -e^{i \delta_{CP} } c_{23} s_{12} s_{13}-c_{12} s_{23} & c_{13} c_{23} \\
\end{array}
\right)\times {\rm{diag}}(e^{i\alpha_{1}/2},e^{i\alpha_{2}/2},1),
\label{pmns}
\ee
where $\theta_{12},\theta_{23},\theta_{13}$ are the mixing angles ($c_{ij}=\cos\theta_{ij},s_{ij}=\sin\theta_{ij}$).
Dirac CP phase is $\delta_{CP}$ and Majorana CP phases are denoted by $\alpha_{1},\alpha_{2}$. For simplicity we 
will consider the Majorana phases to be zero. The best fit values of these parameters along with their 3$\sigma$ ranges
are mentioned in Table~\ref{nuparameters}.

\section{Boltzmann Equations}
\label{Boltzmann}
In this section we present the Boltzmann equations that lead to the final lepton asymmetry and dark matter 
abundance. As we have mentioned earlier, asymmetries in both the sectors are generated from the decay of 
the lightest right handed 
Majorana neutrinos $N_1$ (in this work the right handed Majorana neutrinos follow a hierarchical structure 
$M_1 < M_2 < M_3$). Once  the lepton and dark matter asymmetries ($\epsilon_L$ and $\epsilon_{\Psi}$) 
are obtained via Eqs.~(\ref{eq:epsilonSM1a}) and (\ref{eq:epsilonDM1a}), their evolution can be studied by a 
set of Boltzmann equations involving all the relevant interactions. This 
is represented by the abundance yields $Y_{j}=\frac{n_j}{s}$ ($j=N_1, x$ with $x = l,\Psi$), $n_j$ being 
the respective number density of particle $j$ and $s$ is the entropy density at certain temperature $T$. Then denoting $Y_{\Delta L}=Y_l-Y_{\bar l}$ and $Y_{\Delta\Psi}= 
Y_{\Psi}-Y_{\bar {\Psi}}$, the Boltzmann equations  involving the corresponding quantities 
 become \cite{Falkowski:2011xh} 
\begin{eqnarray}
\label{eq:BE}
\frac{d Y_{N_1}}{dz}&=& -z\frac{\Gamma_1}{H_1} 
\frac{K_1(z)}{K_2(z)}\left(Y_{N_1}-Y_{N_1}^{\rm eq}\right) \, \,,
\\ 
\label{eq:BE1}
\frac{d Y_{\Delta x}}{dz} &=&   \frac{\Gamma_1}{H_1} 
\left (\epsilon_x  z \frac{K_1(z)}{K_2(z)}(Y_{N_1}- Y_{N_1}^{eq})  -  Br_x  
\frac{z^3 K_1(z)}{4} Y_{\Delta x}  \right)\,, (x=L,\Psi).
\end{eqnarray}
Here $Y_{N_1}^{\rm eq}$ represents the equilibrium abundance of $N_1$ and $z$ is given by $z=\frac{M_1}{T}$. 
$H_1$ is the Hubble parameter at $T=M_1$. $K_i(z),~i=1,2$ are the modified Bessel functions of first and second 
kinds respectively  and $Br_{x}$ denotes the branching ratios of $N_1$ decay into the SM and dark sectors. 

The evolution of the lightest RHN $N_1$ abundance due to its decay and inverse decay is described by Eq. (\ref{eq:BE}). Here, $\frac{\G_1}{H_1}$ illustrates the strength of these interactions and controls the departure of $N_1$ from thermal equilibrium. 
Eq. (\ref{eq:BE1}) describes the evolution of the asymmetries generated in both visible as well as the dark sectors. The term 
proportional to $\epsilon_{L (\Psi)}$ is responsible for the production of the asymmetry $Y_{\Delta L ({\Delta \Psi})}$ 
once $N_1$ drops out of thermal equilibrium whereas the second term (in the first bracketed term of r.h.s of  Eq. (\ref{eq:BE1})) 
proportional to $Br_{L(\Psi)}$ is responsible for the washout of the asymmetries due to the inverse decay of $N_1$.

In principle, the final yields of lepton (matter) as well as dark matter should be obtained by solving the coupled 
Boltzmann equations. However here they turn out to be independent (see Eq. (\ref{eq:BE1})) due to our consideration 
of narrow width approximation\footnote{We discuss the validity of the narrow width approximation in our set-up in section 
\ref{result}, after we get to know the $Y$ matrix.}. 
Using the relevant neutrino Yukawa coupling (turns out to be below $\mathcal{O}(1)$) 
and so with $\lambda_{D1}$ (see Sec.~\ref{strategy} for detail), we find that $\frac{\Gamma_1}{M_1}<<1$  with RH 
neutrino mass $M_1$ being $10^9$ GeV or above. We further find that with sufficiently large RH 
neutrino mass $M_1\sim 10^{12}~ \rm{GeV}$ (as in this work), the other requirement for realizing narrow width approximation, $i.e.$, $\frac{\Gamma_1^2}{M_1H_1} <1$ 
is also satisfied. Such a choice of heavy RH neutrino mass is also consistent with Davidson Ibarra (DI) bound \cite{Davidson:2002qv} 
and allows us not to consider the flavor effects in the analysis.
Within this approximation, the $2 \leftrightarrow 2$ transfer ($l \Phi \leftrightarrow \bar{\Psi} S,~l \Phi \leftrightarrow \Psi S,~\bar{l} \Phi^{\dagger} \leftrightarrow \bar{\Psi} S,~\bar{l} \Phi^{\dagger} \leftrightarrow \Psi S$ and washout terms ($l 
\Phi \leftrightarrow \bar{l} \Phi^{\dagger},~\bar{\Psi} S \leftrightarrow \Psi S$, where $S$ is a real scalar) mediated by RHNs are 
neglected. Therefore, transfer of asymmetries between the SM and dark sectors are absent. However the presence of $2 \rightarrow 
1$ inverse decay processes ( $l \Phi, \bar{l} \Phi^{\dagger} \rightarrow N_1$ and  $\Psi S, ~\bar{\Psi} S\rightarrow N_1$) 
are included through the term proportional to $Br_x$, which corresponds to the washout of asymmetries.

Finally a part of the lepton asymmetry (yield $Y_{\Delta L}$)  is further converted into the baryon asymmetry 
($i.e.$ to the yield of $Y_{\Delta B}$) via Sphaleron transition (see \cite{Davidson:2008bu} and references therein). On the other hand, since Sphaleron does 
not interact with dark sector, dark sector asymmetry will not be converted. Considering the Spahleron is in equilibrium above 
the electroweak phase transition (EWPT) temperature \cite{Davidson:2008bu}-\cite{Buchmuller:2005eh}, 
the net baryon asymmetry can be expressed as \cite{Davidson:2008bu}
\bea
Y_{\Delta B}=\frac{8}{23}Y_{\Delta L}\,.
\eea
We now denote the final yields of the lepton and the dark sector (obtained by solving the Boltzmann equations) at present temperature $T_0$ by $Y_{\Delta L}^0$ and $Y_{\Delta\Psi}^0$ respectively. In order to satisfy the observed baryon 
asymmetry in the Universe $Y_{\Delta B}^0=(8.24-9.38)\times10^{-11}$\cite{Tanabashi:2018oca}, $Y_{\Delta L}^0$ must 
be within the range $(2.37-2.70)\times10^{-10}$. %
The relic abundance of dark matter follows the relation \cite{Edsjo:1997bg}
\bea
\Omega_{\Psi}h^2= 2.755 \times 10^8 \left(\frac{m_{\Psi}}{{\rm{GeV}}}\right)Y_{\Delta \Psi}^0\, .
\label{relic}
\eea

The asymmetric dark matter candidate $\Psi$ has no other interaction except the one in Eq.~(\ref{e5}) 
and therefore it decouples from thermal bath once the decay of right handed neutrino completed. However, the 
other decay products of $N_1$ such as the scalar $S$ and inert Higgs doublet remain in thermal equilibrium. It 
is to be noted that any asymmetry in the inert Higgs doublet (as a decay product) is expected to be restored fast 
due to its copious interactions and hence its number density reaches equilibrium satisfying the condition 
$n^{eq}_{\Phi}=n^{eq}_{\Phi^{\dagger}}$. Also, $S$ being a real scalar, any asymmetry 
in $S$ is redundant. 
However as the temperature of the Universe decreases, it may leave us a symmetric dark matter component 
($\Phi_0$) too which decouples at some lower temperature from the thermal bath when thermal annihilation 
freezes out. Such a contribution would provide an additional contribution to the relic in the set-up.

The relic abundance of the symmetric dark matter can be obtained by solving Boltzmann equation
\bea
\frac{dY_{\Phi_0}}{dz'} &=& -\frac{1}{z'^2} \langle \sigma v_{\Phi_0\Phi_0\rightarrow XX} \rangle 
\left (Y_{\Phi_0}^{2}-(Y_{\Phi_0}^{eq})^2\right )\, ,
\eea
\label{BEidm}
where $z'=m_{\Phi_0}/T$ and $Y_{\Phi_0}^{eq}$ denotes equilibrium number density of $\Phi_0$. 
The relic density of inert scalar $\Phi_0$ is then expressed as  
\bea
\Omega_{\Phi_0}h^2= 2.755 \times 10^8 \left(\frac{m_{\Phi_0}}{{\rm{GeV}}}\right)Y_{\Phi_0}^0\, .
\label{relicidm}
\eea
Therefore in the present framework, one must satisfy the condition for total DM relic abundance
\bea
\Omega_{\rm DM}h^2=\Omega_{\Psi}h^2 + \Omega_{\Phi_0}h^2\, .
\label{relictot}
\eea

\section{Strategy for evaluating $\epsilon_L$ and $\epsilon_{\Psi}$}
\label{strategy}

In order to obtain initial values of $\epsilon_{L}$ and $\epsilon_{\Psi}$ via Eqs.~(\ref{eq:epsilonSM1a}) and (\ref{eq:epsilonDM1a}), 
we note that evaluations of $Y$ matrix, $\lambda_{D_i}$ and $M_1$ are required. The same $Y$ being involved 
in neutrino mass matrix $m_{\nu}$ via  Eq. (\ref{numass1}) must be evaluated so as to obey the constraints on neutrino parameters 
(see section \ref{constraints}). On the other hand, $\lambda_{D_i}$ turn out to be free parameters. For simplicity, we 
consider the source of CP violation to follow only from the complex neutrino Yukawa coupling ($Y$). Hence $\lambda_{D_i}$ 
are considered to be real. We further assume all three $\lambda_{D_i}$ are same, denoted by $\lambda_D$. As discussed 
before, to generate lepton asymmetry, dark matter and neutrino mass, we rely here on heavy RH neutrinos having mass $ \sim 
10^9$ GeV and above\footnote{with some exceptions \cite{Hugle:2018qbw,Mahanta:2019gfe,Mahanta:2019sfo}} so as to satisfy the narrow width approximation and the DI bound. 

The remaining ingredient is to find out whether the asymmetric component is the sole contribution to the dark matter relic or there 
could be a subdominant, but non-negligible, symmetric contribution to follow from IHD. In determining this, we note that the relic abundance $\Phi_0$ depends on mass splitting between the particles of the inert multiplet\footnote{For simplification purpose, we 
choose $\lambda_2 = \lambda_3$ so that $m_{\Phi^{\pm}} = m_{A_0}$.}: $\Delta M = m_{(\Phi^\pm, A^0)} - 
m_{\Phi_0}$. Keeping in mind that we mostly focus on the intermediate mass range for the IHD, it results under-abundance of 
relic density with $\Delta M \sim \mathcal{O}(1)$ GeV. As a limiting case, with sufficiently large $\Delta M$, the contribution 
seems to be negligible and we may end up having the dark matter abundance constituted only by the asymmetric component 
from $\Psi$. However as we will see below that such a large $\Delta M$ poses a threat to the high scale validity. Hence a 
balance is required in choosing $\Delta M$ in our scenario so as to keep the IHD's contribution to relic small and simultaneously 
the stability of the set-up till a large scale can be achieved. 

\subsection {Determination of $Y$}

We understand that the neutrino Yukawa coupling $Y$ plays vital role in determining the neutrino mass 
as well as asymmetries $\epsilon_{L}$ and $\epsilon_{\Psi}$. 
The neutrino mass expressed in Eq. (\ref{numass1}), can be redefined as
\bea 
({m_{\nu}})_{\alpha\beta}=(Y\Lambda Y^T)_{\alpha\beta}\,\,
\label{mnu1}
\eea
where $\Lambda =\rm{diag}(\Lambda_1,\Lambda_2,\Lambda_3)$  and $\Lambda_i $ (for i=1,2,3) is expressed by 
\bea
\Lambda_{i}&=&\bigg{[}\frac{M_i}{
{32}\pi^2}\bigg{(}\frac{m_{\Phi_{0}}^2}{m_{\Phi_0}^2-M_{i}^2} {\rm{ln}}\frac{m_{\Phi_0}^{2}}{
M_i^{2}}-\frac{m_{A_0}^2}{m_{A_0}^2-M_i^2} {\rm{ln}}\frac{m_{A_0}^{2}}{M_i^{2}} 
\bigg{)}\bigg{]}.
\label{numass2}
\eea

Note that within $\Lambda$, parameters $m_{\Phi_0}, m_{A_0}$ are part of the DM phenomenology. Therefore 
with their fixed values, $\Lambda$ would only be a function of $M_i$ which can be evaluated with a choice of $M_1$ and 
a fixed mass ratio of RHNs (we consider hierarchical RHNs). Afterward, we use the Casas-Ibarra (CI) parametrization 
\cite{Casas:2001sr} to evaluate the Yukawa couplings via 
\bea
Y&=&\sqrt{\Lambda^{-1}} R \sqrt{m^d_{\nu}} U^{\dagger},\, 
\label{yuk}
\eea
where $R$ is a complex orthogonal matrix $R^TR=1$, taken as
\begin{align}
\label{eq:R-matrix}
R=\begin{pmatrix}
    0 & \cos{z} & \sin{z}\\
    0 & -\sin{z} & \cos{z}\\
    1 & 0 & 0\\
\end{pmatrix}\hspace{0.5 cm} &\text{for NH}\,,&\hspace{1.0 cm}
R=\begin{pmatrix}
     \cos{z} & -\sin{z} & 0\\
     \sin{z} & \cos{z} & 0\\
     0 & 0 & 1\\ 
\end{pmatrix}\hspace{0.5 cm} &\text{for IH}&,
\end{align}

with $z=z_R+i z_I$, a complex number. In order to obtain $m_{\nu}^d = diag(m_1, m_2, m_3)$, 
we consider the lightest neutrino mass eigenvalue to be zero with $m_{1}=0$ ($m_1 < m_2 < m_3$) 
for NH and $m_3=0$ ($m_3< m_1 < m_2$) for IH. Then $m_{\nu}^d$ is calculable using the best fit values of solar 
and atmospheric mass splittings of Table \ref{nuparameters}. Therefore, elements of Yukawa coupling matrix, 
$Y_{i \alpha}$, for  a specific $z$ value can be easily obtained for different choices of model 
parameters $m_{\Phi_0},~\Delta M,~M_1$, ratio of RHN masses etc.

\subsection{Fixing $\Delta M$} 

By knowing the elements of neutrino Yukawa coupling, $Y_{i \alpha}$, we are now able to calculate initial asymmetries, 
$\epsilon_{\Psi}$ and $\epsilon_{L}$. Now following the discussion in the beginning of this section, we plan to discuss 
the suitable choice of $\Delta M$ as it controls the high scale validity, $i.e.$ perturbativity and vacuum stability, of the 
present construction. In doing that, we recall that the top Yukawa coupling $y_t\sim\mathcal{O}(1)$, drags the Higgs 
quartic coupling towards the negative value at a scale around $10^{10}$ GeV in the SM \cite{Buttazzo:2013uya,Degrassi:2012ry,Tang:2013bz,Ellis:2009tp,EliasMiro:2011aa} which is suggestive of 
a metastable EW vacuum within the present $3\sigma$ limits of $m_t$ \cite{Tanabashi:2018oca}. Involvement of the new scalar degree of 
freedoms like IHD and scalar singlet in the present scenario can modify the fate of EW vacuum as the running of the 
Higgs quartic coupling $\lambda_H$ would be affected by their presence. On the other hand, due to the presence of 
heavy RHNs, we must ensure that the couplings involved should obey the stability and pertubativity conditions (see 
Eq. (\ref{copo}), $\l_i<4\pi$) at least up to the highest RHN mass scale. As we will see, choice of $\Delta M$ would be crucial in this regard. 

\subsubsection{Vacuum stability}
For this purpose, we first study the running of different couplings of the model. Beta functions of all such coupling are provided\footnote{These are generated using the model implementation in SARAH \cite{Staub:2013tta}.} 
in Appendix \ref{RGE}. The modification of the beta function of $\l_H$ (through their Higgs portal coupling) 
can be expressed (at one loop) as:
\bea
\beta_{\l_H} & = \b_{\l_H}^{\rm{SM}}+\b_{\l_H}^{\rm{IHD}} +\b_{\l_H}^{\rm{S}}~ = ~\b_{\l_H}^{\rm{SM}}+2 \lambda_{1}^{2} +2 \lambda_{1} \lambda_{2} +\lambda_{2}^{2}+\lambda_{3}^{2}+\frac{1}{2}\l^2_{HS}
\label{lamH}
\eea
The condition, $\l_H > 0$ till the Planck scale ($M_{Pl}$) ensures the absolute stability of the EW vacuum, if 
violated at any scale below $M_{Pl}$ the EW vacuum can become metastable or unstable. Now, if the Higgs 
quartic coupling $\l_{H}(\m)$ turns negative at any scale (as happens for SM at $\mu \sim 10^{9-10}$ GeV), there 
may exist another deeper minimum other than the EW one. In such a scenario, one requires to confirm 
the metastability of the Higgs vacuum by estimating the tunneling probability $\mathcal{P}_T$ of the EW 
vacuum to the second minimum such that the associated decay time is longer than the age of the Universe. 
The tunneling probability is given by \cite{Isidori:2001bm,Buttazzo:2013uya},
\bea
\mathcal{P}_T=T^4_U\m_B^4 e^{-\frac{8\pi^2}{3|\l_{H}(\m_B)|}},
\eea 
where $T_U$ is the age of the Universe, $\m_B$ is the scale at which the tunneling probability is maximized, 
determined from $\b_{\l_{H}}(\m_B)=0$. Solving the above equation, the metastability requires:
$\l_{H}(\m_B) > -0.065/[1-\rm{ln}\big{(}\frac{v}{\m_B}\big{)}]$. 

At high energies, one can write the RG improved effective potential as \cite{Degrassi:2012ry}
\bea
V^{\rm{eff}}_H&=&\frac{\l^{\rm{eff}}_H(\m)}{4}h^4, 
\label{effpot}   
\eea        
\noindent where $\l^{\rm{eff}}_H(\m)=\l^{\rm{SM,eff}}_H(\m)+\l^{\rm{IHD},\rm{eff}}_H(\m)+\l^{\rm{S,eff}}_H(\m)$. Here, $\l^{\rm{SM,eff}}_H(\m)$ is the contribution coming from the SM fields to $\l_H$ whereas $\l^{\rm{IHD,eff}}_H(\m)$ and $\l^{\rm{S,eff}}_H(\m)$ are contribution to the $\l_H$ coming from the IHD ($\Phi$) and the scalar singlet ($S$) in the current setup. These new contributions can be expressed as:
\besub
\bea
\l^{\rm{IHD,eff}}_H(\m)&=& e^{4\G(h=\m)}\frac{1}{16\pi^2} \bigg[2\frac{\l_{1}^2}{4}\bigg(\rm{ln}\frac{\l_{1}}{2}-\frac{3}{2}\bigg)+\frac{(\l_{1}+\l_2+\l_3)^2}{4}\bigg(\rm{ln}\frac{\l_{1}+\l_2+\l_3}{2}-\frac{3}{2}\bigg)\nonumber \\
&+&\frac{(\l_{1}+\l_2-\l_3)^2}{4}\bigg(\rm{ln}\frac{\l_{1}+\l_2-\l_3}{2}-\frac{3}{2}\bigg)\bigg],\\ 
\l^{\rm{S,eff}}_H(\m)&=& e^{4\G(h=\m)}\frac{1}{16\pi^2} \bigg[\frac{\l^2_{HS}}{4}\bigg(\rm{ln}\frac{\l_{HS}}{2}-\frac{3}{2}\bigg)\bigg].  
\eea 
\label{eff_lamH} 
\eesub
Here, $\G(h)=\int_{m_t}^{h}\g(\m)~\rm{d~ln(\m)}$ and $\g(\m)$ is the anomalous dimension of the Higgs
 field \cite{Buttazzo:2013uya}.
 
\begin{table}[]
\centering
\begin{tabular}{|c|c| c| c | c|c| c| c|c|}
\hline
Scale & $\l_{H} $  & ~$y_{t}$ &  ~$g_{1}$& ~$g_{2}$& ~$g_3$\\  
\hline
$\mu=m_t$ &$0.125932$ & $0.93610$ & $0.357606$ & $0.648216$ & $1.16655$  \\
 \hline
\end{tabular}
\caption{Values of the relevant SM couplings (top-quark Yukawa $y_t$ , gauge couplings $g_i (i = 1, 2, 3)$ and Higgs quartic
coupling $\l_H$ ) at energy scale $\mu= m_t= 173.2$ GeV with $m_h =125.09$ GeV and $\a_S(m_Z)= 0.1184$.}
\label{initial_conditions}
\end{table} 

For the stability analysis purpose, we choose $\mu = m_t$ as an initial scale and the running of the scalar couplings (inclusive 
of $\lambda_H$ and other relevant scalar couplings in the set-up) is done at two-loops\footnote{We only provide the one-loop $\b$ functions in Appendix \ref{RGE}.}  till $M_{Pl}$. Since RHNs do not couple to the SM Higgs, it does not affect the running of $\lambda_H$ 
and other scalar couplings significantly and hence for this part of analysis, we do not consider the presence of RHNs.  
In table \ref{initial_conditions}, we provide the initial conditions of relevant SM couplings at $\mu=m_t$ \cite{Buttazzo:2013uya}. 
Here, we consider $m_h=125.09$ GeV, $m_t=173.2$ GeV, and $\a_S(m_Z)= 0.1184$. For scalar couplings 
such as $\lambda_{1,2,3}$ we use the following relations of them with the mass parameters (see Eq. (\ref{e3})) given by,
\besub
\bea
\l_{1} &=& 2\l_{L} + \frac{2(m_{\Phi^+}^2-m_{\Phi_0}^2 )}{v^2}, \\
\l_{2} &=& \frac{m_{\Phi_0}^2 + m_{A_0}^2 - 2 m_{\Phi^+}^2}{v^2}, \\
\l_{3} &=& \frac{(m^2_{\Phi_0} - m^2_{A_0})}{v^2}. 
\eea
\eesub
\label{IHD_couplings}
Note that the mass splitting $\Delta M$ controls the values of these scalar couplings (we consider $\lambda_2 = \lambda_3$). 
Since we are interested in keeping the IHD's mass in the intermediate range, we choose $m_{\Phi_0}$ to be 300 GeV and $\l_L=0.01$ as a 
benchmark values and consider $\Delta M$ as the parameter to be varied. We consider portal 
couplings $\l_{HS}= \lambda_{\Phi S} = 0.001$ so that effect of $\Delta M$ on the running of $\l_{H}$ and the other quartic 
couplings becomes prominent.

\begin{figure}[]
\centering
\subfigure[]{
\includegraphics[height=7 cm, width=8 cm,angle=0]{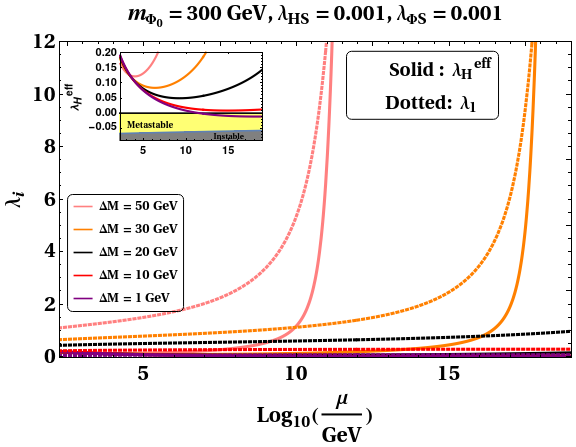}}
\subfigure []{
\includegraphics[height=6.85 cm, width=8 cm,angle=0]{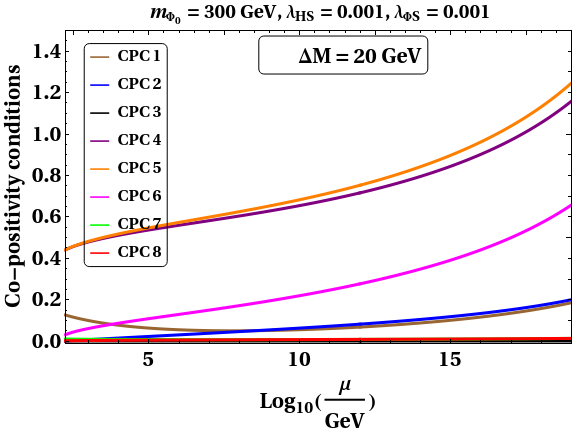}}
\caption{Left panel: Evolution of effective Higgs quartic coupling ($\l_{H}^{\rm eff}$) and $\l_1$ against the scale $\mu$, Right panel: Evolution of all the copositivity conditions (Eq.\ref{copo}) against the scale $\mu$ . In both the plots we have kept $m_{\Phi_0}=300$ GeV, $\l_L=0.01$, $\l_{HS}=0.001$ and $\l_{\Phi S}=0.001$ .}
\label{hsv_quartic}
\end{figure}

In Fig.\ref{hsv_quartic} (a), we show the evolution of the couplings $\lambda_H^{\rm eff}$ (solid line) and $\l_1$ (dotted line) 
with the energy scale $\m$ for the different choices of mass splittings: $\Delta M$ = 50 GeV (pink), 30 GeV (orange), 20 GeV (black),
10 GeV (red) and 1 GeV (purple). One finds that for $\Delta M=30$ GeV and above\footnote{For $\Delta M = 50$ GeV, the couplings ($\l_1$ and $\l_{H}^{\rm eff}$) become non-perturbative at around $\m \sim 10^{10} $ GeV which is even below the lightest RHN mass $M_1$. Hence such a $\Delta M$ can be disregarded.}, the coupling $\l_1$ 
becomes non-perturbative ($\l_1 (\m)>4\pi$) well before the $M_{pl}$. This is because for $\Delta M=30$ GeV itself, 
$\l_1$ turns out to be quite large, $\sim \mathcal{O}(0.6)$ at $\m=m_t$. However, the values of the coupling 
$|\l_2|$ (or $|\l_3|$) is found to be $\sim\mathcal{O}(0.3)$ at $m_t$ and it remains perturbative till $M_{pl}$. In this plot, 
we also provide evolution of the effective Higgs quartic coupling $\l_{H}^{\rm eff}$.  Due to the involvement of term 
proportional to $\l_1^2$ (also significant contribution follows from $\lambda_{2,3}^2$) as seen in Eq. (\ref{lamH}), 
the rapid increase of $\l_H^{\rm eff}$ is observed for $\Delta M=30$ GeV (and above). This observation suggests that we need 
to keep our choice of $\Delta M$ below 30 GeV. Note that with $\Delta M$ below 10 GeV, the EW vacuum remains 
metastable (yellow shade region) as shown in the inset figure of the left panel. In the right panel, Fig.\ref{hsv_quartic} (b), 
we show that all the co-positivity conditions (discussed in Eq. (\ref{copo})) are maintained till the Planck scale with a 
choice $\Delta M=20$ GeV. 

\subsubsection{Relic contribution from IHD}

As we have already discussed, we aim for asymmetric DM in this work to satisfy the entire relic. However, 
involvement of the IHD in the set-up naturally puts the question whether there can be any symmetric DM 
contribution to the relic. In this regard we know that the relic of the IHD doublet remains under abundant 
in the mass regime we are interested in, $i.e. ~80-500$ GeV. This is due to the fact that IHD being a $SU(2)$ 
doublet it can annihilate into the SM gauge bosons ($W^+,Z$) with a large cross-section. Beyond this range, 
the relic satisfaction by the lightest component of IHD happens with small $\Delta M \sim \mathcal{O}(1)$ GeV.

\begin{figure}[h]
\centering
\includegraphics[height=6 cm, width=8.5 cm,angle=0]{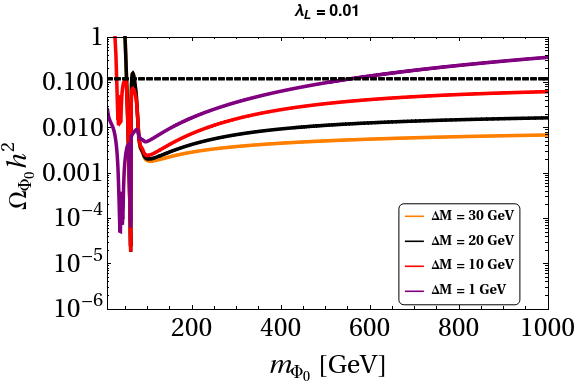}
\caption{Variation of the relic density $\Omega_{\Phi_0}h^2$ of the IHD with its mass $m_{\Phi_0}$  keeping $\l_L=0.01$ for different values of $\Delta M$.}
\label{pure_IHD}
\end{figure}

In Fig \ref{pure_IHD}, we show the variation of the relic density ($\Omega_{\Phi_0}h^2$) with its mass 
$m_{\Phi_0}$ for different values of $\Delta M$. It is seen that with the increase of $\Delta M$,  its contribution 
towards the total relic of the dark matter decreases. More precisely, we note that with $\Delta M$ = 
20 GeV (30 GeV), it can contribute maximum of $6\%$ ($3\%$) towards the total relic density of the dark matter 
at $m_{\Phi_0}= 300$ GeV. Hence combining our understanding related to vacuum stability issue and to minimize 
the symmetric contribution from IHD to DM relic, we choose to work with $\Delta M$ range: 10-30 GeV. Effect of 
$\Delta M$ in having the lepton asymmetry and asymmetric DM generation are part of our study in the following 
section.

\noindent It is also interesting to point out that in an IHD scenario DM mass below 400 GeV is ruled out by 
Fermi-LAT constraints \cite{Borah:2017dfn}. However, IHD being the less abundant DM candidate as in this work,
effective IHD annihilation cross-section  for indirect detection is rescaled by a factor 
$\left(\frac{\Omega_{\Phi_0}h^2}{\Omega_{\rm DM}h^2}\right)^2$ \cite{Borah:2019aeq,Bhattacharya:2019fgs} and as a result limits
from Fermi-LAT becomes insignificant. 

\section{Results and discussion}
\label{result}
In this section, we present the results for the final yields of the lepton and the dark asymmetries obtained by 
solving the Boltzmann equations and investigate whether it can provide the required baryon asymmetry and 
dark matter abundance in order to satisfy the observed bounds. Simultaneously,  we study how things change 
with different neutrino mass hierarchies. For this purpose, we use the neutrino Yukawa couplings obtained using
CI parametrization with (i) given values of heavy right handed neutrino masses, and (ii) inert doublet parameters: 
$m_{{\Phi}_0}$ and $\Delta M$. The other coupling required to solve the Boltzmann equations are the dark sector 
Yukawa couplings $\lambda_{Di},~i=1,2,3$. As already stated, we assume $\lambda_{D_i}$ to be 
real and same, denoted by $\lambda_D$. It is interesting to mention that although we consider $\lambda_D$ 
to be real, a finite asymmetry in the dark sector $\epsilon_{\Psi}$ follows from the involvement of complex neutrino 
Yukawa couplings $Y$. As a result of it, there will not be any contribution from Fig.~\ref{figb}(b) and ~\ref{figb}(d). 
Under this circumstances, the Eq. (\ref{eq:epsilonDM1a}) reduces to 
\bea
\epsilon_{\Psi} = \frac{M_1}{8\pi} \frac{{\rm Im}[Y^*Y^T 
M^{-1} \lambda_{D}\lambda_{D}^\dagger]_{11}}{[2YY^\dagger + \lambda_{D}\lambda_{D}^{\dagger}]_{11}}\, .
\label{red-epsilonDM1}
\eea
 
As stated in Sec.{\ref{strategy}}, for a given set of neutrino parameters given in Table~\ref{nuparameters} (normal or inverted 
hierarchy), one can obtain Yukawa couplings $Y_{i \alpha}$ for different choices of model parameters 
$\Delta M,~m_{\Phi_0}$, RHN mass and CI parameter $z$ using Eqs.~(\ref{mnu1}-\ref{eq:R-matrix}). 
Varying the other parameter $\lambda_D$ (dark sector coupling),  asymmetries 
in lepton and dark sectors $\epsilon_{L,\Psi}$ (using Eqs. (\ref{eq:epsilonSM1a}) and (\ref{eq:epsilonDM1a})), 
total decay width of lightest RHN $\Gamma_1$ and its branching ratios to visible and dark sectors, $Br_{L,\Psi}$
are obtained. 
We then use $\epsilon_{L,\Psi}$, $\Gamma_1$ and $Br_{L,\Psi}$ to solve the Boltzmann equations 
(Eqs.~(\ref{eq:BE}-\ref{eq:BE1})) and obtain 
the final comoving density $Y^0_{\Delta B,\Delta\Psi}$ at present temperature $T_0$ for each value of $\lambda_D$. 

Our aim is to find out the relevant parameter space of the model which satisfies the correct baryon number density 
and dark matter relic abundance. This helps us to identify the corresponding allowed ranges of dark matter mass. While 
investigating the yields of lepton and dark asymmetries against $\lambda_D$ variation, we keep on changing 
other parameters as well, however one at a time, $i.e.$ changing a) mass splitting $\Delta M$, 
and b) different RHNs mass ratio.

\subsection{Case of NH} 

Using the prescription stated above, we first evaluate Yukawa couplings for NH case for a specific 
set of parameters 
\bea
z_R=0.030,~z_I=-0.086,~m_{\Phi_0}=300~{\rm GeV}, M_1=10^{12}~{\rm GeV}. 
\label{benchmark-1}
\eea
Note that, as stated before, this benchmark value of $m_{\Phi_0}$ is motivated by the fact that we 
are interested to keep the IHD within the intermediate mass range. We also consider heavy RHNs 
and $10^{12}$ GeV is only a representative value. However, choosing such high value, we can safely 
ignore the flavor effects \cite{Abada:2006fw,Nardi:2006fx,Abada:2006ea,Blanchet:2006be,Blanchet:2006ch,Dev:2017trv}. Then with $\Delta M=20$ GeV and 
a fixed ratio of RHN masses, $M_1:M_2:M_3=1:10:100$, using Eq. (\ref{yuk}) we obtain, 
\begin{align}
\label{eq:Y-matrix}
Y=\begin{pmatrix}
    0.0388759 - 0.00119665 i &  0.0545874 - 0.0042563 i & -0.0314376 - 0.00486761 i\\
    -0.0423476 - 0.0692572 i &  0.343607 + 0.00173362 i & 0.405694 - 4.97426\times10^{-6} i 
\end{pmatrix}. 
\hspace{0.5 cm}
\end{align}
We will also choose different sets of $\Delta M$ and mass ratio and correspondingly different $Y$ matrix 
would follow.  Throughout the work, solution to Boltzmann equations are obtained with $\lambda_D$ range 
$10^{-4}-10^{-1}$.  We note that condition for narrow width approximation, $\frac{\Gamma_1^2}{M_1H_1} 
\ll 1$,  mentioned in Sec.~\ref{Boltzmann} is valid for such chosen range of $\lambda_D$.

\begin{figure}[H]
\centering
\subfigure[Lepton Sector]{
\includegraphics[height=6.5 cm, width=8 cm,angle=0]{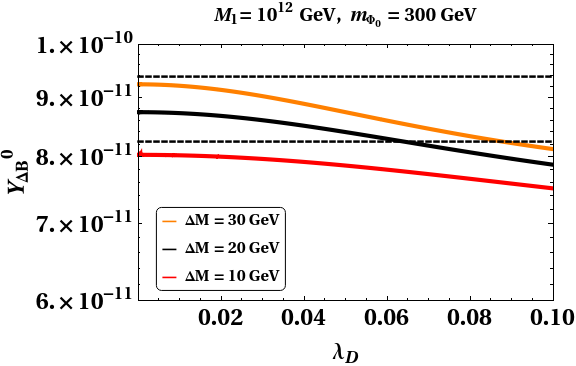}}
\subfigure [Dark Sector]{
\includegraphics[height=6.5 cm, width=8 cm,angle=0]{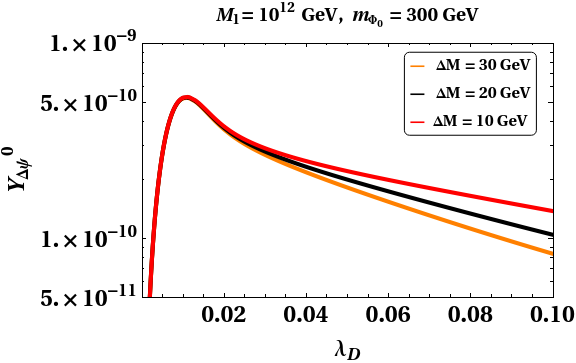}}
\caption{Variation of asymptotic yield $Y_{\Delta B}^0$ and $Y_{\Delta \Psi}^0$ with the dark 
sector Yukawa coupling $\lambda_D$ for $\Delta M= 10~ {\rm GeV},~20~ {\rm GeV},~30~ {\rm GeV}$.}
\label{fig2}
\end{figure}

In Fig.~\ref{fig2}(a) and (\ref{fig2}b), we plot the variation $Y_{\Delta B}^0$ ($Y_{\Delta \Psi}^0$) against $\lambda_D$ 
(after solving the Boltzmann equations) for different set of values of $\Delta M$ = 10, 20 and 30 GeV associated 
with the above benchmark set of parameters in Eq. (\ref{benchmark-1}). Here ratio of the RHN masses is taken as $M_1:M_2:M_3=1:10:100$. The horizontal black dashed lines in Fig. {\ref{fig2}}(a) represents the correct abundance 
yield of $Y_{\Delta B}^0$ today followed from the observed baryon asymmetry in the Universe\cite{Ade:2015xua}. 
From Fig.~\ref{fig2}(a), we notice that for $\Delta M=20$ GeV ($\Delta M=30$ GeV) baryon abundance in Universe 
is satisfied with $\lambda_D\le0.065$ ($\lambda_D\le0.09$). However, for $\Delta M=10$ GeV, $Y_{\Delta B}^0$ 
turns out to be inadequate for the entire range of $\lambda_D$ considered. Similarly, in the right panel we provide 
plots of variation of $Y_{ \Delta \Psi}^0$ against $\lambda_D$ with different values of $\Delta M$.

\begin{figure}[h]
\centering
\subfigure []{
\includegraphics[height=6.5 cm, width=8 cm,angle=0]{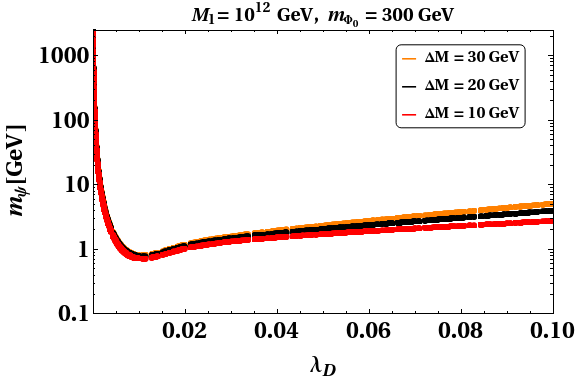}}
\subfigure []{
\includegraphics[height=6.5 cm, width=8 cm,angle=0]{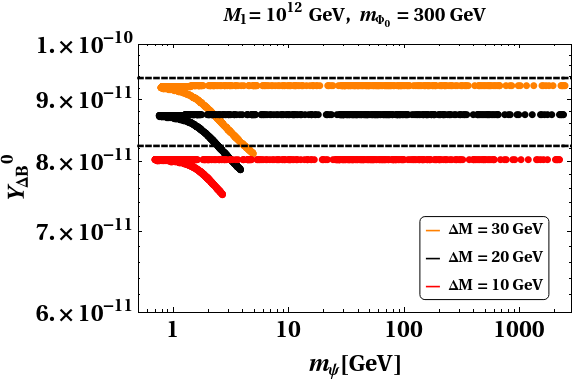}}
\caption{Left panel: Variation the asymmetric dark matter mass $m_{\Psi}$ with the dark sector Yukawa coupling $\l_D$, Right panel: Variation of asymptotic yield $Y_{\Delta B}^0$ with the dark 
matter mass $m_{\Psi}$. Both the plots are shown for $\Delta M= 10~ {\rm GeV},~20~ {\rm GeV},~30~ {\rm GeV}$.}
\label{fig4}
\end{figure}

In order to understand the patterns observed in Fig. \ref{fig2}, we first note that r.h.s of the Boltzmann equation 
for $Y_{\Delta L}$ (see Eq. (\ref{eq:BE1})) contains two terms, one is $\epsilon_L$ dependent and the other one 
is washout related, proportional to $Br_{\Psi}$. Since $Y$ is fixed, the nature of $Y_{\Delta B}^0$ against 
$\lambda_D$ is mostly governed by the $\epsilon_L$-related (first) term. Now from the expression of $\epsilon_{L}$ 
as in Eq. (\ref{eq:epsilonDM1}), we note its variation with $\lambda_D$ is insignificant due to the presence of $\lambda_D$ 
in both numerator and denominator. Hence a near to flatness is observed in left panel of Fig. \ref{fig2}(a). On the 
other hand, in case of dark sector, although the first term in the r.h.s. of corresponding Boltzmann equation for 
$Y_{\Delta L}$ dominates over the second term for small $\lambda_D$, the contribution form second term 
(proportional to $Br_{\Psi}$) increases with increasing $\lambda_D$ leading to large washout effects. As a result, 
when washout effect is negligible in dark sector, $Y_{\Delta \Psi}^0$ increases rapidly with increasing 
$\lambda_D$ as expected from Eq. (\ref{red-epsilonDM1}). This situation alters with higher $\lambda_D$ 
(beyond $\lambda_{D} \sim 0.01$) where a significant washout happens with further increase of $\lambda_D$ 
(associated to larger $Br_{\Psi}$). This produces an overall mild falling nature of $Y_{\Delta \Psi}^0$ against $\lambda_D$ 
 as observed in Fig.~\ref{fig2}(b) (for $\lambda_{D} > 0.01$) which now becomes distinguishable for different 
 $\Delta M$ values.

In Fig.~\ref{fig4}(a), we plot  the correlation between $m_{\Psi}$ and $\lambda_D$ for different choices 
of $\Delta M$. This correlation is simply obtained from the right panel of Fig.~\ref{fig2} by taking into account: 
(i) the required asymmetric contribution to DM relic from $\Psi$ ($\Omega_{\Psi} h^2$) corresponding to any 
value of $\lambda_D$ through Eq.~(\ref{planck}) and (\ref{relictot}) as the presence of IHD also provides a small 
but nonzero symmetric contribution to relic associated to specific choice of $\Delta M$; (ii) then from 
Fig.~\ref{fig2}(b), we find the respective $Y_{\D \Psi}^0$ and using Eq.~(\ref{relic}) the corresponding value of $m_{\Psi}$ 
is obtained. The typical nature (parabolic) of this correlation plot observed here is inherited from the right panel of 
Fig.~\ref{fig2} considering the reciprocity relation between $m_{\Psi}$ and $Y_{\D \Psi}^0$ as in Eq. (\ref{relic}). 

Then in Fig.~\ref{fig4}(b), we plot the correlation between dark matter mass $m_{\Psi}$ and $Y_{\Delta B}^0$ 
obtained directly from Fig.~\ref{fig2}(a) and Fig.~\ref{fig4}(a) for a given $\lambda_D$. From Fig.~\ref{fig2}(a) and \ref{fig2}(b), 
it is observed that for $\lambda_D \gtrsim 0.05$ order of magnitudes of dark matter abundance and baryon asymmetry are 
almost similar, $Y_{\D \Psi}^0 \sim Y_{\D B}^0$ and they do not change significantly with the increase of $\lambda_D$ 
beyond 0.05 resulting $m_{\Psi} \sim \mathcal{O}$ (GeV) as seen from Fig.~\ref{fig4}(a). On the other hand, for $\lambda_D 
< 0.05$, we have noticed a sharp variation in $Y_{\D \Psi}^0$ against $\lambda_D$ whereas $Y_{\Delta B}^0$ continues to 
exhibit the same moderate variation as observed by comparing Fig.~\ref{fig2}(a) and \ref{fig2}(b) resulting in a wider range 
of DM masses in this case: from GeV to TeV. This is also visible from Fig.~\ref{fig4}(a). Hence for this regime of DM mass (above GeV), 
$Y_{\D B}^0$ is effectively insensitive to the increase in $m_{\Psi}$ as seen in Fig.~\ref{fig4}(b). This finding that two regions 
of $\lambda_D$ (below and above 0.05) shows different dependency on $m_{\Psi}$ is due to the interplay between the generation 
and wash-out of asymmetries (as discussed before) both being functions of $\lambda_D$ and in line with observation in \cite{Falkowski:2011xh}, 
a characteristic of two-sector leptogenesis.

\begin{figure}[h]
\centering
\subfigure[]{
\includegraphics[height=7 cm, width=7 cm,angle=0]{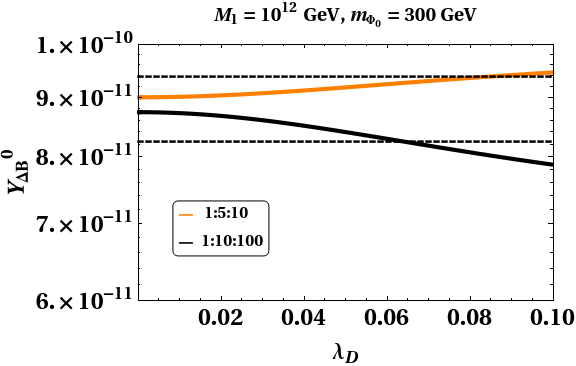}}
\subfigure []{
\includegraphics[height=7 cm, width=7 cm,angle=0]{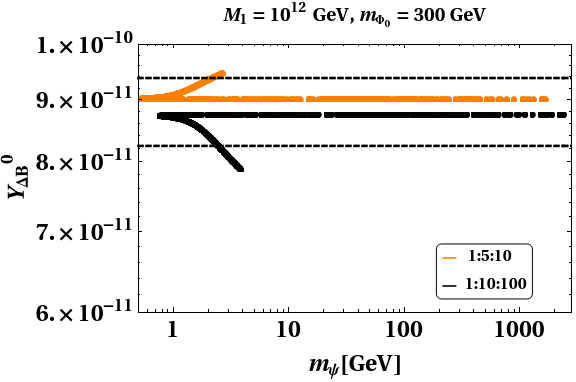}}
\caption{(a)Variation of asymptotic yield $Y_{\Delta B}^0$ with dark sector Yukawa coupling 
$\lambda_D$ for $\Delta M= 20$ GeV (b) $Y_{\Delta B}^0$ vs $m_{\Psi}$  plot satisfying DM relic 
abundance. In both the plots the variation is shown with two different RHN mass ratios $i.e.$ 1:5:10 and 1:10:100}
\label{fig5.1}
\end{figure}

We now investigate how $Y_{\Delta L, \Delta \Psi}^0$ change with RHN mass ratio  
$i.e. ~M_1:M_2:M_3$. We first find the $Y$ matrix for a fixed choice of $\Delta M=20$ GeV while 
considering $M_1:M_2:M_3=1:5:10$ keeping other parameters fixed as in Eq.~(\ref{benchmark-1}).
Then solve the Boltzmann equations (Eqs.~(\ref{eq:BE}-\ref{eq:BE1}) and results are shown in 
Fig.~\ref{fig5.1}(a) (Fig.~\ref{fig5.1}(b)) which are similar to Fig.~\ref{fig2}(a) (Fig.~\ref{fig4}(b)) where we 
also include the results for earlier mass ratio ($M_1:M_2:M_3=1:10:100$) for comparison purpose. 
Study of Boltzmann equations indicate that similar to the earlier observations made in Fig.~\ref{fig2}(a), 
here also the first term of Eq.~(\ref{eq:BE1}) corresponding to $\epsilon_L$ dominates over the washout 
term for visible sector. Whereas, variation of  $m_{\Psi}$ with $Y_{\Delta B}^0$ in Fig.~\ref{fig5.1}(b) directly 
reflects from Fig.~\ref{fig5.1}(a). Since $Y_{\Delta B}^0$ increases (decreases) with $\lambda_D$ for 
$M_1:M_2:M_3=1:5:10$ ($M_1:M_2:M_3=1:10:100$), behaviour of $m_{\Psi}$ versus $Y_{\Delta B}^0$ 
plots in Fig.~\ref{fig5.1}(b) are different. The parabolic nature of plots in Fig.~\ref{fig5.1}(b) clearly refers to 
presence strong washout in dark sector as emphasized in earlier discussions. In both the cases of RHN 
mass ratio considered, observed ADM mass ranges from few GeV to TeV, similar to the ones shown in 
Fig.~\ref{fig4}(b).

\subsection{Study with Inverted Hierarchy}

\begin{figure}[h]
\centering
\subfigure[]{
\includegraphics[height=7 cm, width=7 cm,angle=0]{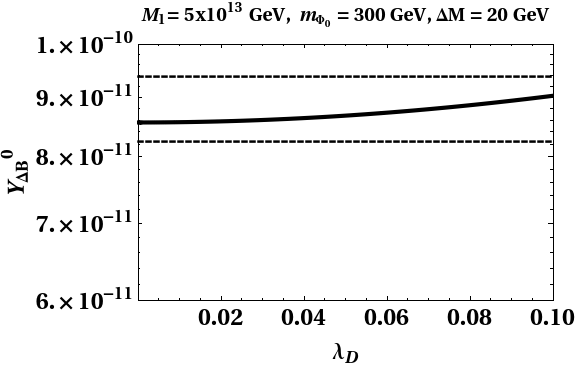}}
\subfigure []{
\includegraphics[height=7 cm, width=7 cm,angle=0]{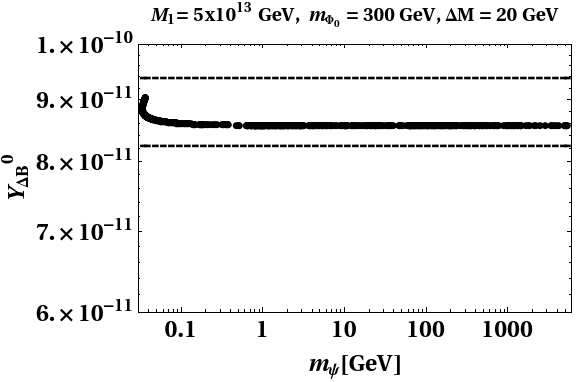}}
\caption{(a)Variation of asymptotic yield $Y_{\Delta B}^0$ with dark sector Yukawa coupling $\lambda_D$ for 
$\Delta M= 20$ GeV (b)  $Y_{\Delta B}^0$ vs $m_{\Psi}$  plot satisfying DM relic 
abundance for the inverted hierarchy.}
\label{fig5.2}
\end{figure}

So far, in Figs.~\ref{fig2} - \ref{fig5.1}, we have performed our calculations of lepton and 
dark sector asymmetry assuming normal hierarchy of neutrino mass. In this section, we repeat 
the study for IH case. Considering the value of $z_R=0.78$,  $z_I=-0.13$, $\Delta M=20$ GeV$, 
M_1=10^{13}$ GeV with rest of the parameters kept fixed as stated while discussing the NH case, 
$\lambda_D$ versus $Y_{\Delta B}^0$ ($m_{\Psi}$ versus $Y_{\Delta B}^0$) plots are generated in 
Fig.~\ref{fig5.2}(a) (Fig.\ref{fig5.2}(b)) by solving Boltzmann equations. We do not notice any significant 
change in the plots of Fig.~\ref{fig5.2}(a)-(b) when compared with NH case and find that for 
the specific range of $\lambda_D$, ADM mass $m_{\Psi}$ goes beyond GeV (to TeV as 
shown in the present figure). In this regime of $m_{\Psi}$ (from GeV to TeV), the one to one correspondence 
between $m_{\Psi}$ and $Y_{\Delta B}^0$ via $\lambda_D$ is lost as observed in case of normal hierarchy 
too. Such a feature is reminiscent of the combined effect of production and washout of asymmetry 
being a function of $\lambda_D$.

\subsection{Limits from experimental searches}

\begin{figure}[h]
\centering
\includegraphics[scale=0.5]{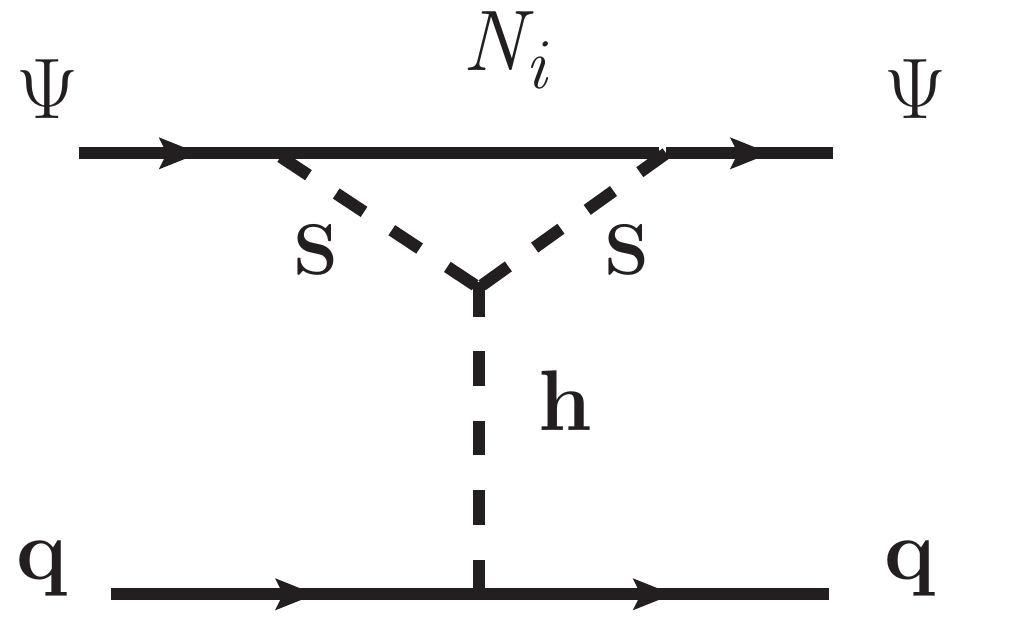}~~~~~~
\includegraphics[scale=0.5]{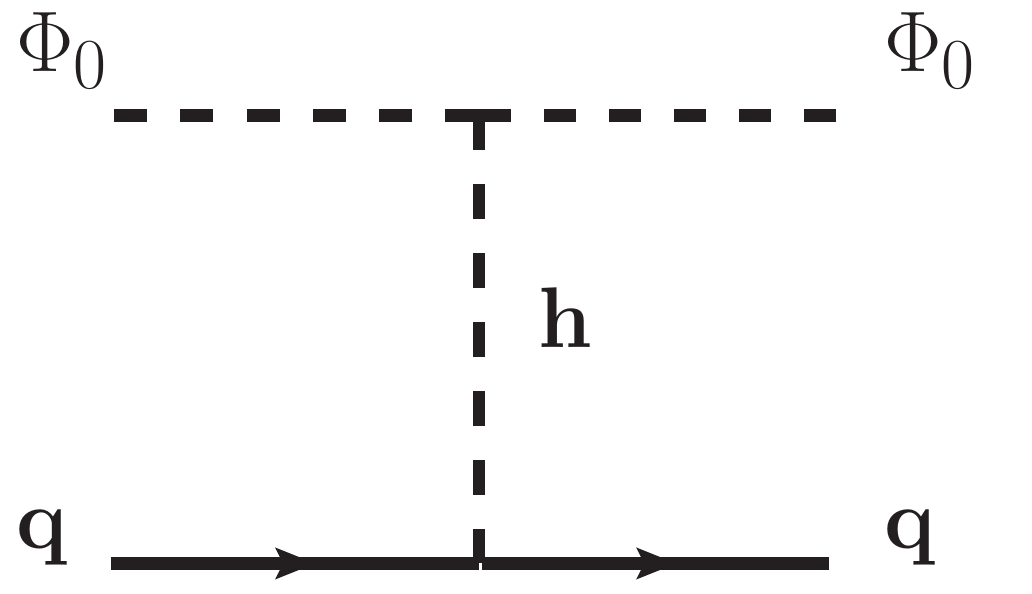}
\caption{Left panel: Loop suppressed interactions of the dark matter ($\Psi$) with quarks (nucleons) for direct search experiments.
Right panel: Direct detection of IHD candidate $\Phi_0$.} 
\label{dd}
\end{figure} 
 
In this section, we briefly discuss the direct detection limits for the dark matter candidate as well as 
experimental  bounds from LFV and Higgs signal strength. Since we consider 
$m_{\Phi^+}=m_{A_0}$, oblique parameter do not impose any constraint in the model and $\Delta 
T=0$ (see Eq.~(\ref{STU})). Direct detection diagram of asymmetric dark matter $\Psi$ and sub-dominant 
component $\Phi_0$ are shown in Fig.~\ref{dd}.
In the present model of asymmetric dark matter,  although the dark matter candidate 
$\Psi$ does not couple directly to the Standard Model Higgs boson or the 
gauge boson ($Z$), but it acquires an effective vertex $h\Psi\Psi$ at one loop. 
Hence, the primary scattering interaction for direct interaction is highly suppressed. Therefore,
in the present framework, the asymmetric dark matter evades all the direct detection bounds.
In a similar manner, one can calculate the direct detection of symmetric dark matter component $\Phi_0$. 
The expression for spin independent direct detection cross-section for $\Phi_0$ is expressed as
\begin{gather}
  \sigma_{\rm {\Phi_0}}^{\rm {SI}}= \frac{\lambda_L^2}{16\pi}\frac{1}{m_h^4} f^2
  \frac{m_N^4}{(m_{\Phi_0}+m_N)^2}\, ,
  \label{scalardd}
\end{gather}
where $m_N$ denotes the mass of the nucleon and $f\sim0.32$  \cite{Giedt:2009mr} is
Higg-nucleon coupling. It is to be noted that in presence of multi-particle nature
(recall that $\Phi_0$ contributes  3-6 $\%$ to the relic only), the effective cross-section
of $\Phi_0$ becomes: $[\frac{\Omega_{\Phi_0}h^2}{\Omega_{\rm DM}h^2}] \sigma_{\rm {\Phi_0}}^{\rm {SI}}$. 
Due to the suppression factor arising out of the ratio ($\frac{\Omega_{\Phi_0}h^2}{\Omega_{\rm DM}h^2}$), 
it satisfies direct detection bounds \cite{Borah:2019aeq,Bhattacharya:2019fgs,Bhattacharya:2019tqq,DuttaBanik:2020jrj} 
from XENON1T, LUX. 

\begin{table}[h]
\centering
\begin{tabular}{|c|c| c| c | c|c| c| c|c|c|c|c|}
\hline
 $M_1 ~\rm{[GeV]}$  & ~$m_{\Phi_0} ~\rm{[GeV]}$ & ~$\Delta M~\rm{[GeV]}$&  ~$\sigma_{\Phi_0}^{eff} ($pb$)$ & $Br(\m\rightarrow e\g)$ &$\m_{\g\g}$\\  \hline
$10^{12}$ & $300$  &  $20$ & $2.3\times10^{-12}$& $4.1\times10^{-54}$ &$0.0987$  \\
 \hline
\end{tabular}
\caption{Benchmark values allowed by the direct detection experiments (for $\Phi_0$) , $Br(\m\rightarrow e\g)$, 
Higgs signal strength ($\m_{\g\g}$).
 }
\label{BP}
\end{table}
In order to discuss the predictions for certain observables in the present set-up, like LFV decays and Higgs 
signal strength ($\m_{\g\g}$), we select a benchmark set of values of parameters for $M_1,m_{\Phi_0}$ and $\Delta M$ as shown 
in Table \ref{BP}. Then corresponding to this set of parameters, we provide values for effective direct detection 
cross-section of $\Phi_0$, $\m_{\g\g}$ and $Br(\mu\rightarrow e\gamma)$ in the same table which satisfy all the constraints. We calculate the values of $\m_{\g\g}$ and ($Br(\mu\rightarrow e\gamma)$) using Eq. (\ref{hgg4}) (Eq.~(\ref{lfv})). We found that the quantity $Br(\mu\rightarrow e\gamma)$ comes out 
to be many orders of magnitude smaller ($\mathcal{O}(10^{-54})$) (see Table \ref{BP}) than that of the present experimental 
bound ($\mathcal{O}(4.1\times10^{-13})$) as the ratio $\frac{M_{N_k}^2}{m_{\Phi^{\pm}}^2}$ is very large ($\mathcal{O}(10^{20})$ GeV)
\cite{Toma:2013zsa}. From Table~\ref{BP}, we conclude that the present asymmetric dark matter model is in 
agreement with the observed experimental bounds mentioned in Sec.~\ref{constraints}.

\section{Conclusion}
\label{summary}

In this framework, we explore a possibility of having a common origin of the dark matter, leptogenesis and 
neutrino mass by incorporating an inert Higgs doublet ($\Phi$), three right handed neutrinos ($N_i$) and a
dark sector consisting of a singlet scalar ($S$) and a singlet fermion ($\Psi$). While the interaction of the 
RHNs with the IHD is responsible for generating the neutrino mass at one loop, its simultaneous decay to 
the visible (SM leptons and the IHD) and the dark sectors (singlet fermion and scalar) generates asymmetries 
in both the sectors. A fraction of lepton asymmetry is converted to baryon asymmetry via Sphaleron transition
while the asymmetric component of $\Psi$ survives and accounts for the dark matter relic. We particularly 
focus on the intermediate mass range of the IHD, $i.e.~ 80 - 500$ GeV as in this regime, its contribution to 
the relic density of the DM remains sub-dominant and can further be reduced by increasing the mass 
splitting ($\Delta M$) among its components. In such a scenario where the IHD contributes negligibly to the 
relic density of the DM, we show that the asymmetric dark matter component of $\Psi$ can provide an 
explanation for the present day DM abundance of the Universe. 

The fate of the present framework at high scale is also tested by doing the RG evolution of all the couplings involved. 
One finds that the mass splitting, $\Delta M$ of IHD, plays a non trivial role in constraining the allowed parameter 
space form the high scale validity of the model and at the same time it also restricts the parameter space which 
explains the present day baryon asymmetry and the DM abundance of the Universe. We show that the $\Delta 
M\sim \mathcal{O}$(20~GeV) can make the EW vacuum stable while keeping the quartic couplings perturbative 
till the Planck scale. It turns out that with $\Delta M\sim \mathcal{O}$(20~GeV) there exists a small, but non-zero 
contribution to the relic density of DM from IHD, making the current framework effectively a two-component dark 
matter scenario.

We show in the present model that the asymmetries generated in the visible sector can provide an explanation 
for observed baryon asymmetry of the Universe via leptogenesis whereas the asymmetry generated in the dark 
sector is responsible for the present DM abundance of the Universe. We perform our analysis for both the normal 
and the inverted hierarchy of the light neutrino masses. 
For both NH and IH scenario, the present setup provide a
large range of the asymmetric dark matter mass, $m_{\Psi}$ from few GeV to TeV which remains consistent with 
the correct dark matter abundance of the Universe. We also discuss possible constraints on the model from  
charged lepton flavor violating decay like $\mu \rightarrow e\g$ arising from Yukawa interactions of the RHNs 
with IHD responsible for neutrino mass generation at one loop. We show that with the chosen set of model 
parameter $Br(\mu \rightarrow e\g)$ comes out to be many orders of magnitude smaller than that of the present 
experimental bound. We also find that the present model is consistent with various experimental observables 
such as Higgs signal strength ($\m_{\g\g}$) and oblique parameters along with direct and indirect search constraints on IHD.
In the present context the RHNs being heavy $\sim \mathcal{O}(10^{12})$ GeV, the effects of flavour in Boltzmann 
equations have been neglected. However, for the detailed study of leptogenesis with smaller values of 
RHN mass for which washout effects are significant and hence flavour effects can be relevant. This is expected to 
be pursued in a different work.

\vskip 1cm 

\noindent {\bf Acknowledgments} : Work of ADB and AS was initially supported by 
Department of Science and Technology, Government of India, under PDF/2016/002148. 
Work of ADB is also supported in part by National Science Foundation of China (11422545, 11947235). RR would like to
thank Arghyajit Datta and Devabrat Mahanta for various useful discussions during the course of this work.
\appendix

\section{1-loop $\beta-$functions}
\label{RGE}
Below we provide the 1-loop $\beta$-functions for all the couplings involved in the present setup. 
While generating the $\beta-$functions we have considered one IHD, one scalar singlet, 3 RHNs 
and a singlet Dirac fermion together with the SM particle spectrum. Since the new particles do not 
carry any colour charges and the Yukawa interactions of these particles with the SM Higgs are 
forbidden due to the symmetry assignment of the setup, no modification is observed in the 
$\beta-$function of the strong coupling $g_3$ and the top Yukawa coupling $y_t$. The hypercharge 
for all the BSM fields apart from the IHD is zero , whereas IHD being doublet, also carries a $SU(2)$ 
charge, this increase in the number of particles carrying a hyper charge and the $SU(2)$ charge 
leads to modification in the  $\beta-$function of $g_1$ and $g_2$ in comparison to that of the 
$\beta^{\rm{SM}}_{g_2}$ and $\beta^{\rm{SM}}_{g_2}$. 

\subsubsection{SM Couplings}
{\allowdisplaybreaks  \begin{align} 
 \beta_{g_1}&= \b_{g_1}^{\rm{SM}} +\b_{g_1}^{\rm{IHD}}~ = ~ \b_{g_1}^{\rm{SM}}+ \frac{g_1^3}{10}  \\  
\beta_{g_2}&= \b_{g_2}^{\rm{SM}} +\b_{g_2}^{\rm{IHD}}~ = ~\b_{g_2}^{\rm{SM}}+ \frac{g_2^3}{6} \\
\beta_{g_3} & = \b_{g_3}^{\rm{SM}}   \\ 
\beta_{y_t} &= \b_{y_t}^{\rm{SM}} \\  
\beta_{\l_H} & = \b_{\l_H}^{\rm{SM}}+\b_{\l_H}^{\rm{IHD}} +\b_{\l_H}^{\rm{S}}~ = ~\b_{\l_H}^{\rm{SM}}+2 \lambda_{1}^{2} +2 \lambda_{1} \lambda_{2} +\lambda_{2}^{2}+\lambda_{3}^{2}+\frac{1}{2}\l^2_{HS}        
\end{align}} 
\subsubsection{BSM couplings}
{\allowdisplaybreaks  \begin{align} 
\beta_{\l_S}&  = 3 \Big(4 \lambda_{HS}^{2}  + 4 \lambda_{\Phi S}^{2}  + \lambda_{S}^{2}\Big) -48 \lambda_D^4  + 8 \lambda_S \lambda_D^2 \\ 
\beta_{\l_{HS}}&  =-\frac{9}{10} g_{1}^{2} \lambda_{HS} -\frac{9}{2} g_{2}^{2} \lambda_{HS} +4 \lambda_{HS}^{2} +4 \lambda_{1} \lambda_{\Phi S} +2 \lambda_{2} \lambda_{\Phi S} +\lambda_{HS} \lambda_S +12 \lambda_{HS} \lambda +4 \lambda_{HS}\lambda_D^2 \nonumber \\ 
 & +6 \lambda_{HS}~y_t^2 \\
\beta_{\lambda_{\Phi S}}& =  
+4 \lambda_{1} \lambda_{HS} +2 \lambda_{2} \lambda_{HS} -\frac{9}{10} g_{1}^{2} \lambda_{\Phi S} -\frac{9}{2} g_{2}^{2} \lambda_{\Phi S} +12 \lambda_{\Phi} \lambda_{\Phi S} +4 \lambda_{\Phi S}^{2} +\lambda_{\Phi S} \lambda_S +4 \lambda_{\Phi S}  \lambda_D^2\nonumber \\ 
 & -8 \lambda_D^2  Y^{\dagger}  Y  +2 \lambda_{\Phi S} \mbox{Tr}\Big({Y  Y^{\dagger}}\Big) \\  
\beta_{\lambda_{\Phi}} & =  
+\frac{27}{200} g_{1}^{4} +\frac{9}{20} g_{1}^{2} g_{2}^{2} +\frac{9}{8} g_{2}^{4} -\frac{9}{5} g_{1}^{2} \lambda_{\Phi} -9 g_{2}^{2} \lambda_{\Phi} +24 \lambda_{\Phi}^{2} +2 \lambda_{1}^{2} +2 \lambda_{1} \lambda_{2} +\lambda_{2}^{2}+\lambda_{3}^{2}+\frac{1}{2} \lambda_{\Phi S}^{2} \nonumber \\ 
 &+4 \lambda_2 \mbox{Tr}\Big({Y  Y^{\dagger}}\Big) -2 \mbox{Tr}\Big({Y  Y^{\dagger}  Y  Y^{\dagger}}\Big) \\
\beta_{\l_1} & =  
\frac{27}{100} g_{1}^{4} -\frac{9}{10} g_{1}^{2} g_{2}^{2} +\frac{9}{4} g_{2}^{4} -\frac{9}{5} g_{1}^{2} \lambda_{1} -9 g_{2}^{2} \lambda_{1} +12 \lambda_{\Phi} \lambda_{1} +4 \lambda_{1}^{2} +4 \lambda_{\Phi} \lambda_{2} +2 \lambda_{2}^{2} +2 \lambda_{3}^{2}\nonumber \\ 
 & +\lambda_{HS} \lambda_{\phi S} +12 \lambda_{1} \lambda_H +4 \lambda_{2} \lambda_H  +6 \lambda_{1} y_t^2 +2 \lambda_{1} \mbox{Tr}\Big({Y Y^{\dagger}}\Big) \\ 
 \beta_{\l_2} & =  
\frac{9}{5} g_{1}^{2} g_{2}^{2} -\frac{9}{5} g_{1}^{2} \lambda_{2} -9 g_{2}^{2} \lambda_{2} +4 \lambda_{\Phi} \lambda_{2} +8 \lambda_{1} \lambda_{2} +4 \lambda_{2}^{2} +8 \lambda_{3}^{2} +4 \lambda_{2} \lambda_H  +6 \lambda_{2} y_t^2  +2 \lambda_{2} \mbox{Tr}\Big({Y  Y^{\dagger}}\Big) \\ 
 \beta_{\l_3} & =  
-\frac{9}{5} g_{1}^{2} \lambda_{3} -9 g_{2}^{2} \lambda_{3} +4 \lambda_{\Phi} \lambda_{3} +8 \lambda_{1} \lambda_{3} +12 \lambda_{2} \lambda_{3} +4 \lambda_{3} \lambda_H +6 \lambda_{3}y_t^2 +2 \lambda_{3} \mbox{Tr}\Big({Y  Y^{\dagger}}\Big) \\
 \beta_{Y} & =  
\frac{1}{20} \Big(10 \Big(3 {Y  Y^{\dagger}  Y}  + \lambda_D^2 Y + Y \Big(20 \mbox{Tr}\Big({Y  Y^{\dagger}}\Big)  -9 \Big(5 g_{2}^{2}  + g_{1}^{2}\Big)\Big)\Big)\\
 \beta_{\l_D} & =  
5 \lambda_D^3  + Y^{T}  Y  \lambda_D
\end{align}}

\bibliographystyle{apsrev}
\bibliography{ref.bib}

\begin{thebibliography}{129}
\expandafter\ifx\csname natexlab\endcsname\relax\def\natexlab#1{#1}\fi
\expandafter\ifx\csname bibnamefont\endcsname\relax
  \def\bibnamefont#1{#1}\fi
\expandafter\ifx\csname bibfnamefont\endcsname\relax
  \def\bibfnamefont#1{#1}\fi
\expandafter\ifx\csname citenamefont\endcsname\relax
  \def\citenamefont#1{#1}\fi
\expandafter\ifx\csname url\endcsname\relax
  \def\url#1{\texttt{#1}}\fi
\expandafter\ifx\csname urlprefix\endcsname\relax\def\urlprefix{URL }\fi
\providecommand{\bibinfo}[2]{#2}
\providecommand{\eprint}[2][]{\url{#2}}

\bibitem[{\citenamefont{Fukuda et~al.}(1998)}]{Fukuda:1998mi}
\bibinfo{author}{\bibfnamefont{Y.}~\bibnamefont{Fukuda}} \bibnamefont{et~al.}
  (\bibinfo{collaboration}{Super-Kamiokande}), \bibinfo{journal}{Phys. Rev.
  Lett.} \textbf{\bibinfo{volume}{81}}, \bibinfo{pages}{1562}
  (\bibinfo{year}{1998}), \eprint{hep-ex/9807003}.

\bibitem[{\citenamefont{Ahmad et~al.}(2002)}]{Ahmad:2002jz}
\bibinfo{author}{\bibfnamefont{Q.}~\bibnamefont{Ahmad}} \bibnamefont{et~al.}
  (\bibinfo{collaboration}{SNO}), \bibinfo{journal}{Phys. Rev. Lett.}
  \textbf{\bibinfo{volume}{89}}, \bibinfo{pages}{011301}
  (\bibinfo{year}{2002}), \eprint{nucl-ex/0204008}.

\bibitem[{\citenamefont{Hsu}(2006)}]{Hsu:2006gt}
\bibinfo{author}{\bibfnamefont{L.}~\bibnamefont{Hsu}}, \bibinfo{journal}{Nucl.
  Phys. B Proc. Suppl.} \textbf{\bibinfo{volume}{155}}, \bibinfo{pages}{158}
  (\bibinfo{year}{2006}).

\bibitem[{\citenamefont{Ahn et~al.}(2003)}]{Ahn:2002up}
\bibinfo{author}{\bibfnamefont{M.}~\bibnamefont{Ahn}} \bibnamefont{et~al.}
  (\bibinfo{collaboration}{K2K}), \bibinfo{journal}{Phys. Rev. Lett.}
  \textbf{\bibinfo{volume}{90}}, \bibinfo{pages}{041801}
  (\bibinfo{year}{2003}), \eprint{hep-ex/0212007}.

\bibitem[{\citenamefont{Julian}(1967)}]{Julian:1967zz}
\bibinfo{author}{\bibfnamefont{W.~H.} \bibnamefont{Julian}},
  \bibinfo{journal}{Astrophys. J.} \textbf{\bibinfo{volume}{148}},
  \bibinfo{pages}{175} (\bibinfo{year}{1967}).

\bibitem[{\citenamefont{Tegmark et~al.}(2004)}]{Tegmark:2003ud}
\bibinfo{author}{\bibfnamefont{M.}~\bibnamefont{Tegmark}} \bibnamefont{et~al.}
  (\bibinfo{collaboration}{SDSS}), \bibinfo{journal}{Phys. Rev. D}
  \textbf{\bibinfo{volume}{69}}, \bibinfo{pages}{103501}
  (\bibinfo{year}{2004}), \eprint{astro-ph/0310723}.

\bibitem[{\citenamefont{Riotto and Trodden}(1999)}]{Riotto:1999yt}
\bibinfo{author}{\bibfnamefont{A.}~\bibnamefont{Riotto}} \bibnamefont{and}
  \bibinfo{author}{\bibfnamefont{M.}~\bibnamefont{Trodden}},
  \bibinfo{journal}{Ann. Rev. Nucl. Part. Sci.} \textbf{\bibinfo{volume}{49}},
  \bibinfo{pages}{35} (\bibinfo{year}{1999}), \eprint{hep-ph/9901362}.

\bibitem[{\citenamefont{Dine and Kusenko}(2003)}]{Dine:2003ax}
\bibinfo{author}{\bibfnamefont{M.}~\bibnamefont{Dine}} \bibnamefont{and}
  \bibinfo{author}{\bibfnamefont{A.}~\bibnamefont{Kusenko}},
  \bibinfo{journal}{Rev. Mod. Phys.} \textbf{\bibinfo{volume}{76}},
  \bibinfo{pages}{1} (\bibinfo{year}{2003}), \eprint{hep-ph/0303065}.

\bibitem[{\citenamefont{Asaka et~al.}(2005)\citenamefont{Asaka, Blanchet, and
  Shaposhnikov}}]{Asaka:2005an}
\bibinfo{author}{\bibfnamefont{T.}~\bibnamefont{Asaka}},
  \bibinfo{author}{\bibfnamefont{S.}~\bibnamefont{Blanchet}}, \bibnamefont{and}
  \bibinfo{author}{\bibfnamefont{M.}~\bibnamefont{Shaposhnikov}},
  \bibinfo{journal}{Phys. Lett. B} \textbf{\bibinfo{volume}{631}},
  \bibinfo{pages}{151} (\bibinfo{year}{2005}), \eprint{hep-ph/0503065}.

\bibitem[{\citenamefont{Asaka and Shaposhnikov}(2005)}]{Asaka:2005pn}
\bibinfo{author}{\bibfnamefont{T.}~\bibnamefont{Asaka}} \bibnamefont{and}
  \bibinfo{author}{\bibfnamefont{M.}~\bibnamefont{Shaposhnikov}},
  \bibinfo{journal}{Phys. Lett. B} \textbf{\bibinfo{volume}{620}},
  \bibinfo{pages}{17} (\bibinfo{year}{2005}), \eprint{hep-ph/0505013}.

\bibitem[{\citenamefont{Ma}(2006)}]{Ma:2006km}
\bibinfo{author}{\bibfnamefont{E.}~\bibnamefont{Ma}}, \bibinfo{journal}{Phys.
  Rev. D} \textbf{\bibinfo{volume}{73}}, \bibinfo{pages}{077301}
  (\bibinfo{year}{2006}), \eprint{hep-ph/0601225}.

\bibitem[{\citenamefont{Fukugita and Yanagida}(1986)}]{Fukugita:1986hr}
\bibinfo{author}{\bibfnamefont{M.}~\bibnamefont{Fukugita}} \bibnamefont{and}
  \bibinfo{author}{\bibfnamefont{T.}~\bibnamefont{Yanagida}},
  \bibinfo{journal}{Phys. Lett. B} \textbf{\bibinfo{volume}{174}},
  \bibinfo{pages}{45} (\bibinfo{year}{1986}).

\bibitem[{\citenamefont{Buchmuller
  et~al.}(2005{\natexlab{a}})\citenamefont{Buchmuller, Di~Bari, and
  Plumacher}}]{Buchmuller:2004nz}
\bibinfo{author}{\bibfnamefont{W.}~\bibnamefont{Buchmuller}},
  \bibinfo{author}{\bibfnamefont{P.}~\bibnamefont{Di~Bari}}, \bibnamefont{and}
  \bibinfo{author}{\bibfnamefont{M.}~\bibnamefont{Plumacher}},
  \bibinfo{journal}{Annals Phys.} \textbf{\bibinfo{volume}{315}},
  \bibinfo{pages}{305} (\bibinfo{year}{2005}{\natexlab{a}}),
  \eprint{hep-ph/0401240}.

\bibitem[{\citenamefont{Anisimov et~al.}(2008)\citenamefont{Anisimov, Blanchet,
  and Di~Bari}}]{Anisimov:2007mw}
\bibinfo{author}{\bibfnamefont{A.}~\bibnamefont{Anisimov}},
  \bibinfo{author}{\bibfnamefont{S.}~\bibnamefont{Blanchet}}, \bibnamefont{and}
  \bibinfo{author}{\bibfnamefont{P.}~\bibnamefont{Di~Bari}},
  \bibinfo{journal}{JCAP} \textbf{\bibinfo{volume}{04}}, \bibinfo{pages}{033}
  (\bibinfo{year}{2008}), \eprint{0707.3024}.

\bibitem[{\citenamefont{Davidson et~al.}(2008)\citenamefont{Davidson, Nardi,
  and Nir}}]{Davidson:2008bu}
\bibinfo{author}{\bibfnamefont{S.}~\bibnamefont{Davidson}},
  \bibinfo{author}{\bibfnamefont{E.}~\bibnamefont{Nardi}}, \bibnamefont{and}
  \bibinfo{author}{\bibfnamefont{Y.}~\bibnamefont{Nir}},
  \bibinfo{journal}{Phys. Rept.} \textbf{\bibinfo{volume}{466}},
  \bibinfo{pages}{105} (\bibinfo{year}{2008}), \eprint{0802.2962}.

\bibitem[{\citenamefont{Baek et~al.}(2013)\citenamefont{Baek, Ko, and
  Park}}]{Baek:2013qwa}
\bibinfo{author}{\bibfnamefont{S.}~\bibnamefont{Baek}},
  \bibinfo{author}{\bibfnamefont{P.}~\bibnamefont{Ko}}, \bibnamefont{and}
  \bibinfo{author}{\bibfnamefont{W.-I.} \bibnamefont{Park}},
  \bibinfo{journal}{JHEP} \textbf{\bibinfo{volume}{07}}, \bibinfo{pages}{013}
  (\bibinfo{year}{2013}), \eprint{1303.4280}.

\bibitem[{\citenamefont{Buchmuller
  et~al.}(2005{\natexlab{b}})\citenamefont{Buchmuller, Peccei, and
  Yanagida}}]{Buchmuller:2005eh}
\bibinfo{author}{\bibfnamefont{W.}~\bibnamefont{Buchmuller}},
  \bibinfo{author}{\bibfnamefont{R.}~\bibnamefont{Peccei}}, \bibnamefont{and}
  \bibinfo{author}{\bibfnamefont{T.}~\bibnamefont{Yanagida}},
  \bibinfo{journal}{Ann. Rev. Nucl. Part. Sci.} \textbf{\bibinfo{volume}{55}},
  \bibinfo{pages}{311} (\bibinfo{year}{2005}{\natexlab{b}}),
  \eprint{hep-ph/0502169}.

\bibitem[{\citenamefont{Davoudiasl and Zhang}(2015)}]{Davoudiasl:2015jja}
\bibinfo{author}{\bibfnamefont{H.}~\bibnamefont{Davoudiasl}} \bibnamefont{and}
  \bibinfo{author}{\bibfnamefont{Y.}~\bibnamefont{Zhang}},
  \bibinfo{journal}{Phys. Rev. D} \textbf{\bibinfo{volume}{92}},
  \bibinfo{pages}{016005} (\bibinfo{year}{2015}), \eprint{1504.07244}.

\bibitem[{\citenamefont{Guo et~al.}(2017)\citenamefont{Guo, Li, Liu,
  Ramsey-Musolf, and Shu}}]{Guo:2016ixx}
\bibinfo{author}{\bibfnamefont{H.-K.} \bibnamefont{Guo}},
  \bibinfo{author}{\bibfnamefont{Y.-Y.} \bibnamefont{Li}},
  \bibinfo{author}{\bibfnamefont{T.}~\bibnamefont{Liu}},
  \bibinfo{author}{\bibfnamefont{M.}~\bibnamefont{Ramsey-Musolf}},
  \bibnamefont{and} \bibinfo{author}{\bibfnamefont{J.}~\bibnamefont{Shu}},
  \bibinfo{journal}{Phys. Rev. D} \textbf{\bibinfo{volume}{96}},
  \bibinfo{pages}{115034} (\bibinfo{year}{2017}), \eprint{1609.09849}.

\bibitem[{\citenamefont{Hern\'andez et~al.}(2016)\citenamefont{Hern\'andez,
  Kekic, L\'opez-Pav\'on, Racker, and Salvado}}]{Hernandez:2016kel}
\bibinfo{author}{\bibfnamefont{P.}~\bibnamefont{Hern\'andez}},
  \bibinfo{author}{\bibfnamefont{M.}~\bibnamefont{Kekic}},
  \bibinfo{author}{\bibfnamefont{J.}~\bibnamefont{L\'opez-Pav\'on}},
  \bibinfo{author}{\bibfnamefont{J.}~\bibnamefont{Racker}}, \bibnamefont{and}
  \bibinfo{author}{\bibfnamefont{J.}~\bibnamefont{Salvado}},
  \bibinfo{journal}{JHEP} \textbf{\bibinfo{volume}{08}}, \bibinfo{pages}{157}
  (\bibinfo{year}{2016}), \eprint{1606.06719}.

\bibitem[{\citenamefont{Narendra
  et~al.}(2018{\natexlab{a}})\citenamefont{Narendra, Sahoo, and
  Sahu}}]{Narendra:2017uxl}
\bibinfo{author}{\bibfnamefont{N.}~\bibnamefont{Narendra}},
  \bibinfo{author}{\bibfnamefont{N.}~\bibnamefont{Sahoo}}, \bibnamefont{and}
  \bibinfo{author}{\bibfnamefont{N.}~\bibnamefont{Sahu}},
  \bibinfo{journal}{Nucl. Phys. B} \textbf{\bibinfo{volume}{936}},
  \bibinfo{pages}{76} (\bibinfo{year}{2018}{\natexlab{a}}),
  \eprint{1712.02960}.

\bibitem[{\citenamefont{Dolan et~al.}(2018)\citenamefont{Dolan, Dutka, and
  Volkas}}]{Dolan:2018qpy}
\bibinfo{author}{\bibfnamefont{M.~J.} \bibnamefont{Dolan}},
  \bibinfo{author}{\bibfnamefont{T.~P.} \bibnamefont{Dutka}}, \bibnamefont{and}
  \bibinfo{author}{\bibfnamefont{R.~R.} \bibnamefont{Volkas}},
  \bibinfo{journal}{JCAP} \textbf{\bibinfo{volume}{06}}, \bibinfo{pages}{012}
  (\bibinfo{year}{2018}), \eprint{1802.08373}.

\bibitem[{\citenamefont{Ipek et~al.}(2018)\citenamefont{Ipek, Plascencia, and
  Turner}}]{Ipek:2018sai}
\bibinfo{author}{\bibfnamefont{S.}~\bibnamefont{Ipek}},
  \bibinfo{author}{\bibfnamefont{A.~D.} \bibnamefont{Plascencia}},
  \bibnamefont{and} \bibinfo{author}{\bibfnamefont{J.}~\bibnamefont{Turner}},
  \bibinfo{journal}{JHEP} \textbf{\bibinfo{volume}{12}}, \bibinfo{pages}{111}
  (\bibinfo{year}{2018}), \eprint{1806.00460}.

\bibitem[{\citenamefont{Das et~al.}(2020{\natexlab{a}})\citenamefont{Das, Das,
  and Khan}}]{Das:2019ntw}
\bibinfo{author}{\bibfnamefont{P.}~\bibnamefont{Das}},
  \bibinfo{author}{\bibfnamefont{M.~K.} \bibnamefont{Das}}, \bibnamefont{and}
  \bibinfo{author}{\bibfnamefont{N.}~\bibnamefont{Khan}},
  \bibinfo{journal}{JHEP} \textbf{\bibinfo{volume}{03}}, \bibinfo{pages}{018}
  (\bibinfo{year}{2020}{\natexlab{a}}), \eprint{1911.07243}.

\bibitem[{\citenamefont{Domcke et~al.}(2020)\citenamefont{Domcke, Drewes,
  Hufnagel, and Lucente}}]{Domcke:2020ety}
\bibinfo{author}{\bibfnamefont{V.}~\bibnamefont{Domcke}},
  \bibinfo{author}{\bibfnamefont{M.}~\bibnamefont{Drewes}},
  \bibinfo{author}{\bibfnamefont{M.}~\bibnamefont{Hufnagel}}, \bibnamefont{and}
  \bibinfo{author}{\bibfnamefont{M.}~\bibnamefont{Lucente}}
  (\bibinfo{year}{2020}), \eprint{2009.11678}.

\bibitem[{\citenamefont{Das et~al.}(2020{\natexlab{b}})\citenamefont{Das, Das,
  and Khan}}]{Das:2020vca}
\bibinfo{author}{\bibfnamefont{P.}~\bibnamefont{Das}},
  \bibinfo{author}{\bibfnamefont{M.~K.} \bibnamefont{Das}}, \bibnamefont{and}
  \bibinfo{author}{\bibfnamefont{N.}~\bibnamefont{Khan}}
  (\bibinfo{year}{2020}{\natexlab{b}}), \eprint{2010.13084}.

\bibitem[{\citenamefont{Chen et~al.}(2020)\citenamefont{Chen, Dutta~Banik, and
  Liu}}]{Chen:2019etb}
\bibinfo{author}{\bibfnamefont{S.-L.} \bibnamefont{Chen}},
  \bibinfo{author}{\bibfnamefont{A.}~\bibnamefont{Dutta~Banik}},
  \bibnamefont{and} \bibinfo{author}{\bibfnamefont{Z.-K.} \bibnamefont{Liu}},
  \bibinfo{journal}{JCAP} \textbf{\bibinfo{volume}{03}}, \bibinfo{pages}{009}
  (\bibinfo{year}{2020}), \eprint{1912.07185}.

\bibitem[{\citenamefont{Kashiwase and Suematsu}(2012)}]{Kashiwase:2012xd}
\bibinfo{author}{\bibfnamefont{S.}~\bibnamefont{Kashiwase}} \bibnamefont{and}
  \bibinfo{author}{\bibfnamefont{D.}~\bibnamefont{Suematsu}},
  \bibinfo{journal}{Phys. Rev. D} \textbf{\bibinfo{volume}{86}},
  \bibinfo{pages}{053001} (\bibinfo{year}{2012}), \eprint{1207.2594}.

\bibitem[{\citenamefont{Pilaftsis}(1997)}]{Pilaftsis:1997jf}
\bibinfo{author}{\bibfnamefont{A.}~\bibnamefont{Pilaftsis}},
  \bibinfo{journal}{Phys. Rev. D} \textbf{\bibinfo{volume}{56}},
  \bibinfo{pages}{5431} (\bibinfo{year}{1997}), \eprint{hep-ph/9707235}.

\bibitem[{\citenamefont{Pilaftsis and Underwood}(2004)}]{Pilaftsis:2003gt}
\bibinfo{author}{\bibfnamefont{A.}~\bibnamefont{Pilaftsis}} \bibnamefont{and}
  \bibinfo{author}{\bibfnamefont{T.~E.} \bibnamefont{Underwood}},
  \bibinfo{journal}{Nucl. Phys. B} \textbf{\bibinfo{volume}{692}},
  \bibinfo{pages}{303} (\bibinfo{year}{2004}), \eprint{hep-ph/0309342}.

\bibitem[{\citenamefont{Dev et~al.}(2018{\natexlab{a}})\citenamefont{Dev,
  Garny, Klaric, Millington, and Teresi}}]{Dev:2017wwc}
\bibinfo{author}{\bibfnamefont{B.}~\bibnamefont{Dev}},
  \bibinfo{author}{\bibfnamefont{M.}~\bibnamefont{Garny}},
  \bibinfo{author}{\bibfnamefont{J.}~\bibnamefont{Klaric}},
  \bibinfo{author}{\bibfnamefont{P.}~\bibnamefont{Millington}},
  \bibnamefont{and} \bibinfo{author}{\bibfnamefont{D.}~\bibnamefont{Teresi}},
  \bibinfo{journal}{Int. J. Mod. Phys. A} \textbf{\bibinfo{volume}{33}},
  \bibinfo{pages}{1842003} (\bibinfo{year}{2018}{\natexlab{a}}),
  \eprint{1711.02863}.

\bibitem[{\citenamefont{Barbieri et~al.}(2006)\citenamefont{Barbieri, Hall, and
  Rychkov}}]{Barbieri:2006dq}
\bibinfo{author}{\bibfnamefont{R.}~\bibnamefont{Barbieri}},
  \bibinfo{author}{\bibfnamefont{L.~J.} \bibnamefont{Hall}}, \bibnamefont{and}
  \bibinfo{author}{\bibfnamefont{V.~S.} \bibnamefont{Rychkov}},
  \bibinfo{journal}{Phys. Rev. D} \textbf{\bibinfo{volume}{74}},
  \bibinfo{pages}{015007} (\bibinfo{year}{2006}), \eprint{hep-ph/0603188}.

\bibitem[{\citenamefont{Cirelli et~al.}(2006)\citenamefont{Cirelli, Fornengo,
  and Strumia}}]{Cirelli:2005uq}
\bibinfo{author}{\bibfnamefont{M.}~\bibnamefont{Cirelli}},
  \bibinfo{author}{\bibfnamefont{N.}~\bibnamefont{Fornengo}}, \bibnamefont{and}
  \bibinfo{author}{\bibfnamefont{A.}~\bibnamefont{Strumia}},
  \bibinfo{journal}{Nucl. Phys. B} \textbf{\bibinfo{volume}{753}},
  \bibinfo{pages}{178} (\bibinfo{year}{2006}), \eprint{hep-ph/0512090}.

\bibitem[{\citenamefont{Lopez~Honorez et~al.}(2007)\citenamefont{Lopez~Honorez,
  Nezri, Oliver, and Tytgat}}]{LopezHonorez:2006gr}
\bibinfo{author}{\bibfnamefont{L.}~\bibnamefont{Lopez~Honorez}},
  \bibinfo{author}{\bibfnamefont{E.}~\bibnamefont{Nezri}},
  \bibinfo{author}{\bibfnamefont{J.~F.} \bibnamefont{Oliver}},
  \bibnamefont{and} \bibinfo{author}{\bibfnamefont{M.~H.}
  \bibnamefont{Tytgat}}, \bibinfo{journal}{JCAP} \textbf{\bibinfo{volume}{02}},
  \bibinfo{pages}{028} (\bibinfo{year}{2007}), \eprint{hep-ph/0612275}.

\bibitem[{\citenamefont{Cao et~al.}(2007)\citenamefont{Cao, Ma, and
  Rajasekaran}}]{Cao:2007rm}
\bibinfo{author}{\bibfnamefont{Q.-H.} \bibnamefont{Cao}},
  \bibinfo{author}{\bibfnamefont{E.}~\bibnamefont{Ma}}, \bibnamefont{and}
  \bibinfo{author}{\bibfnamefont{G.}~\bibnamefont{Rajasekaran}},
  \bibinfo{journal}{Phys. Rev. D} \textbf{\bibinfo{volume}{76}},
  \bibinfo{pages}{095011} (\bibinfo{year}{2007}), \eprint{0708.2939}.

\bibitem[{\citenamefont{Majumdar and Ghosal}(2008)}]{Majumdar:2006nt}
\bibinfo{author}{\bibfnamefont{D.}~\bibnamefont{Majumdar}} \bibnamefont{and}
  \bibinfo{author}{\bibfnamefont{A.}~\bibnamefont{Ghosal}},
  \bibinfo{journal}{Mod. Phys. Lett. A} \textbf{\bibinfo{volume}{23}},
  \bibinfo{pages}{2011} (\bibinfo{year}{2008}), \eprint{hep-ph/0607067}.

\bibitem[{\citenamefont{Lundstrom et~al.}(2009)\citenamefont{Lundstrom,
  Gustafsson, and Edsjo}}]{Lundstrom:2008ai}
\bibinfo{author}{\bibfnamefont{E.}~\bibnamefont{Lundstrom}},
  \bibinfo{author}{\bibfnamefont{M.}~\bibnamefont{Gustafsson}},
  \bibnamefont{and} \bibinfo{author}{\bibfnamefont{J.}~\bibnamefont{Edsjo}},
  \bibinfo{journal}{Phys. Rev. D} \textbf{\bibinfo{volume}{79}},
  \bibinfo{pages}{035013} (\bibinfo{year}{2009}), \eprint{0810.3924}.

\bibitem[{\citenamefont{Dolle and Su}(2009)}]{Dolle:2009fn}
\bibinfo{author}{\bibfnamefont{E.~M.} \bibnamefont{Dolle}} \bibnamefont{and}
  \bibinfo{author}{\bibfnamefont{S.}~\bibnamefont{Su}}, \bibinfo{journal}{Phys.
  Rev. D} \textbf{\bibinfo{volume}{80}}, \bibinfo{pages}{055012}
  (\bibinfo{year}{2009}), \eprint{0906.1609}.

\bibitem[{\citenamefont{Lopez~Honorez and Yaguna}(2010)}]{Honorez:2010re}
\bibinfo{author}{\bibfnamefont{L.}~\bibnamefont{Lopez~Honorez}}
  \bibnamefont{and} \bibinfo{author}{\bibfnamefont{C.~E.}
  \bibnamefont{Yaguna}}, \bibinfo{journal}{JHEP} \textbf{\bibinfo{volume}{09}},
  \bibinfo{pages}{046} (\bibinfo{year}{2010}), \eprint{1003.3125}.

\bibitem[{\citenamefont{Lopez~Honorez and Yaguna}(2011)}]{LopezHonorez:2010tb}
\bibinfo{author}{\bibfnamefont{L.}~\bibnamefont{Lopez~Honorez}}
  \bibnamefont{and} \bibinfo{author}{\bibfnamefont{C.~E.}
  \bibnamefont{Yaguna}}, \bibinfo{journal}{JCAP} \textbf{\bibinfo{volume}{01}},
  \bibinfo{pages}{002} (\bibinfo{year}{2011}), \eprint{1011.1411}.

\bibitem[{\citenamefont{Chowdhury et~al.}(2012)\citenamefont{Chowdhury,
  Nemevsek, Senjanovic, and Zhang}}]{Chowdhury:2011ga}
\bibinfo{author}{\bibfnamefont{T.~A.} \bibnamefont{Chowdhury}},
  \bibinfo{author}{\bibfnamefont{M.}~\bibnamefont{Nemevsek}},
  \bibinfo{author}{\bibfnamefont{G.}~\bibnamefont{Senjanovic}},
  \bibnamefont{and} \bibinfo{author}{\bibfnamefont{Y.}~\bibnamefont{Zhang}},
  \bibinfo{journal}{JCAP} \textbf{\bibinfo{volume}{02}}, \bibinfo{pages}{029}
  (\bibinfo{year}{2012}), \eprint{1110.5334}.

\bibitem[{\citenamefont{Arhrib et~al.}(2014)\citenamefont{Arhrib, Tsai, Yuan,
  and Yuan}}]{Arhrib:2013ela}
\bibinfo{author}{\bibfnamefont{A.}~\bibnamefont{Arhrib}},
  \bibinfo{author}{\bibfnamefont{Y.-L.~S.} \bibnamefont{Tsai}},
  \bibinfo{author}{\bibfnamefont{Q.}~\bibnamefont{Yuan}}, \bibnamefont{and}
  \bibinfo{author}{\bibfnamefont{T.-C.} \bibnamefont{Yuan}},
  \bibinfo{journal}{JCAP} \textbf{\bibinfo{volume}{06}}, \bibinfo{pages}{030}
  (\bibinfo{year}{2014}), \eprint{1310.0358}.

\bibitem[{\citenamefont{Plascencia}(2015)}]{Plascencia:2015xwa}
\bibinfo{author}{\bibfnamefont{A.~D.} \bibnamefont{Plascencia}},
  \bibinfo{journal}{JHEP} \textbf{\bibinfo{volume}{09}}, \bibinfo{pages}{026}
  (\bibinfo{year}{2015}), \eprint{1507.04996}.

\bibitem[{\citenamefont{Borah and Gupta}(2017)}]{Borah:2017dfn}
\bibinfo{author}{\bibfnamefont{D.}~\bibnamefont{Borah}} \bibnamefont{and}
  \bibinfo{author}{\bibfnamefont{A.}~\bibnamefont{Gupta}},
  \bibinfo{journal}{Phys. Rev. D} \textbf{\bibinfo{volume}{96}},
  \bibinfo{pages}{115012} (\bibinfo{year}{2017}), \eprint{1706.05034}.

\bibitem[{\citenamefont{Biswas and Shaw}(2017)}]{Biswas:2017dxt}
\bibinfo{author}{\bibfnamefont{A.}~\bibnamefont{Biswas}} \bibnamefont{and}
  \bibinfo{author}{\bibfnamefont{A.}~\bibnamefont{Shaw}}
  (\bibinfo{year}{2017}), \bibinfo{note}{[Erratum: JCAP 07, E01 (2019)]},
  \eprint{1709.01099}.

\bibitem[{\citenamefont{Falkowski et~al.}(2011)\citenamefont{Falkowski,
  Ruderman, and Volansky}}]{Falkowski:2011xh}
\bibinfo{author}{\bibfnamefont{A.}~\bibnamefont{Falkowski}},
  \bibinfo{author}{\bibfnamefont{J.~T.} \bibnamefont{Ruderman}},
  \bibnamefont{and} \bibinfo{author}{\bibfnamefont{T.}~\bibnamefont{Volansky}},
  \bibinfo{journal}{JHEP} \textbf{\bibinfo{volume}{05}}, \bibinfo{pages}{106}
  (\bibinfo{year}{2011}), \eprint{1101.4936}.

\bibitem[{\citenamefont{Kaplan et~al.}(2009)\citenamefont{Kaplan, Luty, and
  Zurek}}]{Kaplan:2009ag}
\bibinfo{author}{\bibfnamefont{D.~E.} \bibnamefont{Kaplan}},
  \bibinfo{author}{\bibfnamefont{M.~A.} \bibnamefont{Luty}}, \bibnamefont{and}
  \bibinfo{author}{\bibfnamefont{K.~M.} \bibnamefont{Zurek}},
  \bibinfo{journal}{Phys. Rev. D} \textbf{\bibinfo{volume}{79}},
  \bibinfo{pages}{115016} (\bibinfo{year}{2009}), \eprint{0901.4117}.

\bibitem[{\citenamefont{Zurek}(2014)}]{Zurek:2013wia}
\bibinfo{author}{\bibfnamefont{K.~M.} \bibnamefont{Zurek}},
  \bibinfo{journal}{Phys. Rept.} \textbf{\bibinfo{volume}{537}},
  \bibinfo{pages}{91} (\bibinfo{year}{2014}), \eprint{1308.0338}.

\bibitem[{\citenamefont{Hamze et~al.}(2015)\citenamefont{Hamze, Kilic, Koeller,
  Trendafilova, and Yu}}]{Hamze:2014wca}
\bibinfo{author}{\bibfnamefont{A.}~\bibnamefont{Hamze}},
  \bibinfo{author}{\bibfnamefont{C.}~\bibnamefont{Kilic}},
  \bibinfo{author}{\bibfnamefont{J.}~\bibnamefont{Koeller}},
  \bibinfo{author}{\bibfnamefont{C.}~\bibnamefont{Trendafilova}},
  \bibnamefont{and} \bibinfo{author}{\bibfnamefont{J.-H.} \bibnamefont{Yu}},
  \bibinfo{journal}{Phys. Rev. D} \textbf{\bibinfo{volume}{91}},
  \bibinfo{pages}{035009} (\bibinfo{year}{2015}), \eprint{1410.3030}.

\bibitem[{\citenamefont{Kitabayashi and Kurosawa}(2016)}]{Kitabayashi:2015oda}
\bibinfo{author}{\bibfnamefont{T.}~\bibnamefont{Kitabayashi}} \bibnamefont{and}
  \bibinfo{author}{\bibfnamefont{Y.}~\bibnamefont{Kurosawa}},
  \bibinfo{journal}{Phys. Rev. D} \textbf{\bibinfo{volume}{93}},
  \bibinfo{pages}{033002} (\bibinfo{year}{2016}), \eprint{1509.05564}.

\bibitem[{\citenamefont{Frandsen and Shoemaker}(2016)}]{Frandsen:2016bke}
\bibinfo{author}{\bibfnamefont{M.~T.} \bibnamefont{Frandsen}} \bibnamefont{and}
  \bibinfo{author}{\bibfnamefont{I.~M.} \bibnamefont{Shoemaker}},
  \bibinfo{journal}{JCAP} \textbf{\bibinfo{volume}{05}}, \bibinfo{pages}{064}
  (\bibinfo{year}{2016}), \eprint{1603.09354}.

\bibitem[{\citenamefont{Murase and Shoemaker}(2016)}]{Murase:2016nwx}
\bibinfo{author}{\bibfnamefont{K.}~\bibnamefont{Murase}} \bibnamefont{and}
  \bibinfo{author}{\bibfnamefont{I.~M.} \bibnamefont{Shoemaker}},
  \bibinfo{journal}{Phys. Rev. D} \textbf{\bibinfo{volume}{94}},
  \bibinfo{pages}{063512} (\bibinfo{year}{2016}), \eprint{1606.03087}.

\bibitem[{\citenamefont{Agrawal et~al.}(2017)\citenamefont{Agrawal, Kilic,
  Swaminathan, and Trendafilova}}]{Agrawal:2016uwf}
\bibinfo{author}{\bibfnamefont{P.}~\bibnamefont{Agrawal}},
  \bibinfo{author}{\bibfnamefont{C.}~\bibnamefont{Kilic}},
  \bibinfo{author}{\bibfnamefont{S.}~\bibnamefont{Swaminathan}},
  \bibnamefont{and}
  \bibinfo{author}{\bibfnamefont{C.}~\bibnamefont{Trendafilova}},
  \bibinfo{journal}{Phys. Rev. D} \textbf{\bibinfo{volume}{95}},
  \bibinfo{pages}{015031} (\bibinfo{year}{2017}), \eprint{1608.04745}.

\bibitem[{\citenamefont{Nagata et~al.}(2017)\citenamefont{Nagata, Olive, and
  Zheng}}]{Nagata:2016knk}
\bibinfo{author}{\bibfnamefont{N.}~\bibnamefont{Nagata}},
  \bibinfo{author}{\bibfnamefont{K.~A.} \bibnamefont{Olive}}, \bibnamefont{and}
  \bibinfo{author}{\bibfnamefont{J.}~\bibnamefont{Zheng}},
  \bibinfo{journal}{JCAP} \textbf{\bibinfo{volume}{02}}, \bibinfo{pages}{016}
  (\bibinfo{year}{2017}), \eprint{1611.04693}.

\bibitem[{\citenamefont{Baldes and Petraki}(2017)}]{Baldes:2017gzw}
\bibinfo{author}{\bibfnamefont{I.}~\bibnamefont{Baldes}} \bibnamefont{and}
  \bibinfo{author}{\bibfnamefont{K.}~\bibnamefont{Petraki}},
  \bibinfo{journal}{JCAP} \textbf{\bibinfo{volume}{09}}, \bibinfo{pages}{028}
  (\bibinfo{year}{2017}), \eprint{1703.00478}.

\bibitem[{\citenamefont{Gresham
  et~al.}(2018{\natexlab{a}})\citenamefont{Gresham, Lou, and
  Zurek}}]{Gresham:2017cvl}
\bibinfo{author}{\bibfnamefont{M.~I.} \bibnamefont{Gresham}},
  \bibinfo{author}{\bibfnamefont{H.~K.} \bibnamefont{Lou}}, \bibnamefont{and}
  \bibinfo{author}{\bibfnamefont{K.~M.} \bibnamefont{Zurek}},
  \bibinfo{journal}{Phys. Rev. D} \textbf{\bibinfo{volume}{97}},
  \bibinfo{pages}{036003} (\bibinfo{year}{2018}{\natexlab{a}}),
  \eprint{1707.02316}.

\bibitem[{\citenamefont{HajiSadeghi et~al.}(2019)\citenamefont{HajiSadeghi,
  Smolenski, and Wudka}}]{HajiSadeghi:2017zrl}
\bibinfo{author}{\bibfnamefont{S.}~\bibnamefont{HajiSadeghi}},
  \bibinfo{author}{\bibfnamefont{S.}~\bibnamefont{Smolenski}},
  \bibnamefont{and} \bibinfo{author}{\bibfnamefont{J.}~\bibnamefont{Wudka}},
  \bibinfo{journal}{Phys. Rev. D} \textbf{\bibinfo{volume}{99}},
  \bibinfo{pages}{023514} (\bibinfo{year}{2019}), \eprint{1709.00436}.

\bibitem[{\citenamefont{Tsao}(2018)}]{Tsao:2017vtn}
\bibinfo{author}{\bibfnamefont{K.-H.} \bibnamefont{Tsao}}, \bibinfo{journal}{J.
  Phys. G} \textbf{\bibinfo{volume}{45}}, \bibinfo{pages}{075001}
  (\bibinfo{year}{2018}), \eprint{1710.06572}.

\bibitem[{\citenamefont{Gresham
  et~al.}(2018{\natexlab{b}})\citenamefont{Gresham, Lou, and
  Zurek}}]{Gresham:2018anj}
\bibinfo{author}{\bibfnamefont{M.~I.} \bibnamefont{Gresham}},
  \bibinfo{author}{\bibfnamefont{H.~K.} \bibnamefont{Lou}}, \bibnamefont{and}
  \bibinfo{author}{\bibfnamefont{K.~M.} \bibnamefont{Zurek}},
  \bibinfo{journal}{Phys. Rev. D} \textbf{\bibinfo{volume}{98}},
  \bibinfo{pages}{096001} (\bibinfo{year}{2018}{\natexlab{b}}),
  \eprint{1805.04512}.

\bibitem[{\citenamefont{Narendra
  et~al.}(2018{\natexlab{b}})\citenamefont{Narendra, Patra, Sahu, and
  Shil}}]{Narendra:2018vfw}
\bibinfo{author}{\bibfnamefont{N.}~\bibnamefont{Narendra}},
  \bibinfo{author}{\bibfnamefont{S.}~\bibnamefont{Patra}},
  \bibinfo{author}{\bibfnamefont{N.}~\bibnamefont{Sahu}}, \bibnamefont{and}
  \bibinfo{author}{\bibfnamefont{S.}~\bibnamefont{Shil}},
  \bibinfo{journal}{Phys. Rev. D} \textbf{\bibinfo{volume}{98}},
  \bibinfo{pages}{095016} (\bibinfo{year}{2018}{\natexlab{b}}),
  \eprint{1805.04860}.

\bibitem[{\citenamefont{Ibe et~al.}(2018)\citenamefont{Ibe, Kamada, Kobayashi,
  and Nakano}}]{Ibe:2018juk}
\bibinfo{author}{\bibfnamefont{M.}~\bibnamefont{Ibe}},
  \bibinfo{author}{\bibfnamefont{A.}~\bibnamefont{Kamada}},
  \bibinfo{author}{\bibfnamefont{S.}~\bibnamefont{Kobayashi}},
  \bibnamefont{and} \bibinfo{author}{\bibfnamefont{W.}~\bibnamefont{Nakano}},
  \bibinfo{journal}{JHEP} \textbf{\bibinfo{volume}{11}}, \bibinfo{pages}{203}
  (\bibinfo{year}{2018}), \eprint{1805.06876}.

\bibitem[{\citenamefont{Van~Dong et~al.}(2019)\citenamefont{Van~Dong, Huong,
  Camargo, Queiroz, and Valle}}]{Dong:2018aak}
\bibinfo{author}{\bibfnamefont{P.}~\bibnamefont{Van~Dong}},
  \bibinfo{author}{\bibfnamefont{D.}~\bibnamefont{Huong}},
  \bibinfo{author}{\bibfnamefont{D.~A.} \bibnamefont{Camargo}},
  \bibinfo{author}{\bibfnamefont{F.~S.} \bibnamefont{Queiroz}},
  \bibnamefont{and} \bibinfo{author}{\bibfnamefont{J.~W.} \bibnamefont{Valle}},
  \bibinfo{journal}{Phys. Rev. D} \textbf{\bibinfo{volume}{99}},
  \bibinfo{pages}{055040} (\bibinfo{year}{2019}), \eprint{1805.08251}.

\bibitem[{\citenamefont{Narendra
  et~al.}(2019{\natexlab{a}})\citenamefont{Narendra, Sahu, and
  Shil}}]{Narendra:2019cyt}
\bibinfo{author}{\bibfnamefont{N.}~\bibnamefont{Narendra}},
  \bibinfo{author}{\bibfnamefont{N.}~\bibnamefont{Sahu}}, \bibnamefont{and}
  \bibinfo{author}{\bibfnamefont{S.}~\bibnamefont{Shil}}
  (\bibinfo{year}{2019}{\natexlab{a}}), \eprint{1910.12762}.

\bibitem[{\citenamefont{An et~al.}(2010)\citenamefont{An, Chen, Mohapatra, and
  Zhang}}]{An:2009vq}
\bibinfo{author}{\bibfnamefont{H.}~\bibnamefont{An}},
  \bibinfo{author}{\bibfnamefont{S.-L.} \bibnamefont{Chen}},
  \bibinfo{author}{\bibfnamefont{R.~N.} \bibnamefont{Mohapatra}},
  \bibnamefont{and} \bibinfo{author}{\bibfnamefont{Y.}~\bibnamefont{Zhang}},
  \bibinfo{journal}{JHEP} \textbf{\bibinfo{volume}{03}}, \bibinfo{pages}{124}
  (\bibinfo{year}{2010}), \eprint{0911.4463}.

\bibitem[{\citenamefont{Arina and Sahu}(2012)}]{Arina:2011cu}
\bibinfo{author}{\bibfnamefont{C.}~\bibnamefont{Arina}} \bibnamefont{and}
  \bibinfo{author}{\bibfnamefont{N.}~\bibnamefont{Sahu}},
  \bibinfo{journal}{Nucl. Phys. B} \textbf{\bibinfo{volume}{854}},
  \bibinfo{pages}{666} (\bibinfo{year}{2012}), \eprint{1108.3967}.

\bibitem[{\citenamefont{Josse-Michaux and
  Molinaro}(2011)}]{JosseMichaux:2011ba}
\bibinfo{author}{\bibfnamefont{F.-X.} \bibnamefont{Josse-Michaux}}
  \bibnamefont{and} \bibinfo{author}{\bibfnamefont{E.}~\bibnamefont{Molinaro}},
  \bibinfo{journal}{Phys. Rev. D} \textbf{\bibinfo{volume}{84}},
  \bibinfo{pages}{125021} (\bibinfo{year}{2011}), \eprint{1108.0482}.

\bibitem[{\citenamefont{Arina}(2014)}]{Arina:2012jp}
\bibinfo{author}{\bibfnamefont{C.}~\bibnamefont{Arina}}, \bibinfo{journal}{J.
  Phys. Conf. Ser.} \textbf{\bibinfo{volume}{485}}, \bibinfo{pages}{012039}
  (\bibinfo{year}{2014}), \eprint{1209.1288}.

\bibitem[{\citenamefont{Gu}(2017)}]{Gu:2016xno}
\bibinfo{author}{\bibfnamefont{P.-H.} \bibnamefont{Gu}},
  \bibinfo{journal}{JHEP} \textbf{\bibinfo{volume}{04}}, \bibinfo{pages}{159}
  (\bibinfo{year}{2017}), \eprint{1611.03256}.

\bibitem[{\citenamefont{Fornal et~al.}(2017)\citenamefont{Fornal, Shirman,
  Tait, and West}}]{Fornal:2017owa}
\bibinfo{author}{\bibfnamefont{B.}~\bibnamefont{Fornal}},
  \bibinfo{author}{\bibfnamefont{Y.}~\bibnamefont{Shirman}},
  \bibinfo{author}{\bibfnamefont{T.~M.~P.} \bibnamefont{Tait}},
  \bibnamefont{and} \bibinfo{author}{\bibfnamefont{J.~R.} \bibnamefont{West}},
  \bibinfo{journal}{Phys. Rev. D} \textbf{\bibinfo{volume}{96}},
  \bibinfo{pages}{035001} (\bibinfo{year}{2017}), \eprint{1703.00199}.

\bibitem[{\citenamefont{Yang}(2019)}]{Yang:2018zrj}
\bibinfo{author}{\bibfnamefont{W.-M.} \bibnamefont{Yang}},
  \bibinfo{journal}{Nucl. Phys.} \textbf{\bibinfo{volume}{B}},
  \bibinfo{pages}{114643} (\bibinfo{year}{2019}), \eprint{1807.03036}.

\bibitem[{\citenamefont{Biswas et~al.}(2019)\citenamefont{Biswas, Choubey,
  Covi, and Khan}}]{Biswas:2018sib}
\bibinfo{author}{\bibfnamefont{A.}~\bibnamefont{Biswas}},
  \bibinfo{author}{\bibfnamefont{S.}~\bibnamefont{Choubey}},
  \bibinfo{author}{\bibfnamefont{L.}~\bibnamefont{Covi}}, \bibnamefont{and}
  \bibinfo{author}{\bibfnamefont{S.}~\bibnamefont{Khan}},
  \bibinfo{journal}{JHEP} \textbf{\bibinfo{volume}{05}}, \bibinfo{pages}{193}
  (\bibinfo{year}{2019}), \eprint{1812.06122}.

\bibitem[{\citenamefont{Narendra
  et~al.}(2019{\natexlab{b}})\citenamefont{Narendra, Patra, Sahu, and
  Shil}}]{Narendra:2019pag}
\bibinfo{author}{\bibfnamefont{N.}~\bibnamefont{Narendra}},
  \bibinfo{author}{\bibfnamefont{S.}~\bibnamefont{Patra}},
  \bibinfo{author}{\bibfnamefont{N.}~\bibnamefont{Sahu}}, \bibnamefont{and}
  \bibinfo{author}{\bibfnamefont{S.}~\bibnamefont{Shil}},
  \bibinfo{journal}{Springer Proc. Phys.} \textbf{\bibinfo{volume}{234}},
  \bibinfo{pages}{335} (\bibinfo{year}{2019}{\natexlab{b}}).

\bibitem[{\citenamefont{Ghosh et~al.}(2018)\citenamefont{Ghosh, Saha, and
  Sil}}]{Ghosh:2017fmr}
\bibinfo{author}{\bibfnamefont{P.}~\bibnamefont{Ghosh}},
  \bibinfo{author}{\bibfnamefont{A.~K.} \bibnamefont{Saha}}, \bibnamefont{and}
  \bibinfo{author}{\bibfnamefont{A.}~\bibnamefont{Sil}},
  \bibinfo{journal}{Phys. Rev. D} \textbf{\bibinfo{volume}{97}},
  \bibinfo{pages}{075034} (\bibinfo{year}{2018}), \eprint{1706.04931}.

\bibitem[{\citenamefont{Bhattacharya
  et~al.}(2020{\natexlab{a}})\citenamefont{Bhattacharya, Ghosh, Saha, and
  Sil}}]{Bhattacharya:2019fgs}
\bibinfo{author}{\bibfnamefont{S.}~\bibnamefont{Bhattacharya}},
  \bibinfo{author}{\bibfnamefont{P.}~\bibnamefont{Ghosh}},
  \bibinfo{author}{\bibfnamefont{A.~K.} \bibnamefont{Saha}}, \bibnamefont{and}
  \bibinfo{author}{\bibfnamefont{A.}~\bibnamefont{Sil}},
  \bibinfo{journal}{JHEP} \textbf{\bibinfo{volume}{03}}, \bibinfo{pages}{090}
  (\bibinfo{year}{2020}{\natexlab{a}}), \eprint{1905.12583}.

\bibitem[{\citenamefont{Jangid et~al.}(2020)\citenamefont{Jangid,
  Bandyopadhyay, Bhupal~Dev, and Kumar}}]{Bandyopadhyay:2020vfc}
\bibinfo{author}{\bibfnamefont{S.}~\bibnamefont{Jangid}},
  \bibinfo{author}{\bibfnamefont{P.}~\bibnamefont{Bandyopadhyay}},
  \bibinfo{author}{\bibfnamefont{P.}~\bibnamefont{Bhupal~Dev}},
  \bibnamefont{and} \bibinfo{author}{\bibfnamefont{A.}~\bibnamefont{Kumar}},
  \bibinfo{journal}{JHEP} \textbf{\bibinfo{volume}{08}}, \bibinfo{pages}{154}
  (\bibinfo{year}{2020}), \eprint{2001.01764}.

\bibitem[{\citenamefont{Isidori et~al.}(2001)\citenamefont{Isidori, Ridolfi,
  and Strumia}}]{Isidori:2001bm}
\bibinfo{author}{\bibfnamefont{G.}~\bibnamefont{Isidori}},
  \bibinfo{author}{\bibfnamefont{G.}~\bibnamefont{Ridolfi}}, \bibnamefont{and}
  \bibinfo{author}{\bibfnamefont{A.}~\bibnamefont{Strumia}},
  \bibinfo{journal}{Nucl. Phys.} \textbf{\bibinfo{volume}{B609}},
  \bibinfo{pages}{387} (\bibinfo{year}{2001}), \eprint{hep-ph/0104016}.

\bibitem[{\citenamefont{Greenwood et~al.}(2009)\citenamefont{Greenwood,
  Halstead, Poltis, and Stojkovic}}]{Greenwood:2008qp}
\bibinfo{author}{\bibfnamefont{E.}~\bibnamefont{Greenwood}},
  \bibinfo{author}{\bibfnamefont{E.}~\bibnamefont{Halstead}},
  \bibinfo{author}{\bibfnamefont{R.}~\bibnamefont{Poltis}}, \bibnamefont{and}
  \bibinfo{author}{\bibfnamefont{D.}~\bibnamefont{Stojkovic}},
  \bibinfo{journal}{Phys. Rev. D} \textbf{\bibinfo{volume}{79}},
  \bibinfo{pages}{103003} (\bibinfo{year}{2009}), \eprint{0810.5343}.

\bibitem[{\citenamefont{Ellis et~al.}(2009)\citenamefont{Ellis, Espinosa,
  Giudice, Hoecker, and Riotto}}]{Ellis:2009tp}
\bibinfo{author}{\bibfnamefont{J.}~\bibnamefont{Ellis}},
  \bibinfo{author}{\bibfnamefont{J.~R.} \bibnamefont{Espinosa}},
  \bibinfo{author}{\bibfnamefont{G.~F.} \bibnamefont{Giudice}},
  \bibinfo{author}{\bibfnamefont{A.}~\bibnamefont{Hoecker}}, \bibnamefont{and}
  \bibinfo{author}{\bibfnamefont{A.}~\bibnamefont{Riotto}},
  \bibinfo{journal}{Phys. Lett.} \textbf{\bibinfo{volume}{B679}},
  \bibinfo{pages}{369} (\bibinfo{year}{2009}), \eprint{0906.0954}.

\bibitem[{\citenamefont{Elias-Miro et~al.}(2012)\citenamefont{Elias-Miro,
  Espinosa, Giudice, Isidori, Riotto, and Strumia}}]{EliasMiro:2011aa}
\bibinfo{author}{\bibfnamefont{J.}~\bibnamefont{Elias-Miro}},
  \bibinfo{author}{\bibfnamefont{J.~R.} \bibnamefont{Espinosa}},
  \bibinfo{author}{\bibfnamefont{G.~F.} \bibnamefont{Giudice}},
  \bibinfo{author}{\bibfnamefont{G.}~\bibnamefont{Isidori}},
  \bibinfo{author}{\bibfnamefont{A.}~\bibnamefont{Riotto}}, \bibnamefont{and}
  \bibinfo{author}{\bibfnamefont{A.}~\bibnamefont{Strumia}},
  \bibinfo{journal}{Phys. Lett.} \textbf{\bibinfo{volume}{B709}},
  \bibinfo{pages}{222} (\bibinfo{year}{2012}), \eprint{1112.3022}.

\bibitem[{\citenamefont{Alekhin et~al.}(2012)\citenamefont{Alekhin, Djouadi,
  and Moch}}]{Alekhin:2012py}
\bibinfo{author}{\bibfnamefont{S.}~\bibnamefont{Alekhin}},
  \bibinfo{author}{\bibfnamefont{A.}~\bibnamefont{Djouadi}}, \bibnamefont{and}
  \bibinfo{author}{\bibfnamefont{S.}~\bibnamefont{Moch}},
  \bibinfo{journal}{Phys. Lett.} \textbf{\bibinfo{volume}{B716}},
  \bibinfo{pages}{214} (\bibinfo{year}{2012}), \eprint{1207.0980}.

\bibitem[{\citenamefont{Degrassi et~al.}(2012)\citenamefont{Degrassi, Di~Vita,
  Elias-Miro, Espinosa, Giudice, Isidori, and Strumia}}]{Degrassi:2012ry}
\bibinfo{author}{\bibfnamefont{G.}~\bibnamefont{Degrassi}},
  \bibinfo{author}{\bibfnamefont{S.}~\bibnamefont{Di~Vita}},
  \bibinfo{author}{\bibfnamefont{J.}~\bibnamefont{Elias-Miro}},
  \bibinfo{author}{\bibfnamefont{J.~R.} \bibnamefont{Espinosa}},
  \bibinfo{author}{\bibfnamefont{G.~F.} \bibnamefont{Giudice}},
  \bibinfo{author}{\bibfnamefont{G.}~\bibnamefont{Isidori}}, \bibnamefont{and}
  \bibinfo{author}{\bibfnamefont{A.}~\bibnamefont{Strumia}},
  \bibinfo{journal}{JHEP} \textbf{\bibinfo{volume}{08}}, \bibinfo{pages}{098}
  (\bibinfo{year}{2012}), \eprint{1205.6497}.

\bibitem[{\citenamefont{Buttazzo et~al.}(2013)\citenamefont{Buttazzo, Degrassi,
  Giardino, Giudice, Sala, Salvio, and Strumia}}]{Buttazzo:2013uya}
\bibinfo{author}{\bibfnamefont{D.}~\bibnamefont{Buttazzo}},
  \bibinfo{author}{\bibfnamefont{G.}~\bibnamefont{Degrassi}},
  \bibinfo{author}{\bibfnamefont{P.~P.} \bibnamefont{Giardino}},
  \bibinfo{author}{\bibfnamefont{G.~F.} \bibnamefont{Giudice}},
  \bibinfo{author}{\bibfnamefont{F.}~\bibnamefont{Sala}},
  \bibinfo{author}{\bibfnamefont{A.}~\bibnamefont{Salvio}}, \bibnamefont{and}
  \bibinfo{author}{\bibfnamefont{A.}~\bibnamefont{Strumia}},
  \bibinfo{journal}{JHEP} \textbf{\bibinfo{volume}{12}}, \bibinfo{pages}{089}
  (\bibinfo{year}{2013}), \eprint{1307.3536}.

\bibitem[{\citenamefont{Anchordoqui et~al.}(2013)\citenamefont{Anchordoqui,
  Antoniadis, Goldberg, Huang, Lust, Taylor, and Vlcek}}]{Anchordoqui:2012fq}
\bibinfo{author}{\bibfnamefont{L.~A.} \bibnamefont{Anchordoqui}},
  \bibinfo{author}{\bibfnamefont{I.}~\bibnamefont{Antoniadis}},
  \bibinfo{author}{\bibfnamefont{H.}~\bibnamefont{Goldberg}},
  \bibinfo{author}{\bibfnamefont{X.}~\bibnamefont{Huang}},
  \bibinfo{author}{\bibfnamefont{D.}~\bibnamefont{Lust}},
  \bibinfo{author}{\bibfnamefont{T.~R.} \bibnamefont{Taylor}},
  \bibnamefont{and} \bibinfo{author}{\bibfnamefont{B.}~\bibnamefont{Vlcek}},
  \bibinfo{journal}{JHEP} \textbf{\bibinfo{volume}{02}}, \bibinfo{pages}{074}
  (\bibinfo{year}{2013}), \eprint{1208.2821}.

\bibitem[{\citenamefont{Tang}(2013)}]{Tang:2013bz}
\bibinfo{author}{\bibfnamefont{Y.}~\bibnamefont{Tang}}, \bibinfo{journal}{Mod.
  Phys. Lett.} \textbf{\bibinfo{volume}{A28}}, \bibinfo{pages}{1330002}
  (\bibinfo{year}{2013}), \eprint{1301.5812}.

\bibitem[{\citenamefont{Salvio}(2015)}]{Salvio:2015cja}
\bibinfo{author}{\bibfnamefont{A.}~\bibnamefont{Salvio}},
  \bibinfo{journal}{Phys. Lett. B} \textbf{\bibinfo{volume}{743}},
  \bibinfo{pages}{428} (\bibinfo{year}{2015}), \eprint{1501.03781}.

\bibitem[{\citenamefont{Salvio}(2019)}]{Salvio:2018rv}
\bibinfo{author}{\bibfnamefont{A.}~\bibnamefont{Salvio}},
  \bibinfo{journal}{Phys. Rev. D} \textbf{\bibinfo{volume}{99}},
  \bibinfo{pages}{015037} (\bibinfo{year}{2019}), \eprint{1810.00792}.

\bibitem[{\citenamefont{Dutta~Banik et~al.}(2018)\citenamefont{Dutta~Banik,
  Saha, and Sil}}]{DuttaBanik:2018emv}
\bibinfo{author}{\bibfnamefont{A.}~\bibnamefont{Dutta~Banik}},
  \bibinfo{author}{\bibfnamefont{A.~K.} \bibnamefont{Saha}}, \bibnamefont{and}
  \bibinfo{author}{\bibfnamefont{A.}~\bibnamefont{Sil}},
  \bibinfo{journal}{Phys. Rev. D} \textbf{\bibinfo{volume}{98}},
  \bibinfo{pages}{075013} (\bibinfo{year}{2018}), \eprint{1806.08080}.

\bibitem[{\citenamefont{Bhattacharya
  et~al.}(2020{\natexlab{b}})\citenamefont{Bhattacharya, Chakrabarty, Roshan,
  and Sil}}]{Bhattacharya:2019tqq}
\bibinfo{author}{\bibfnamefont{S.}~\bibnamefont{Bhattacharya}},
  \bibinfo{author}{\bibfnamefont{N.}~\bibnamefont{Chakrabarty}},
  \bibinfo{author}{\bibfnamefont{R.}~\bibnamefont{Roshan}}, \bibnamefont{and}
  \bibinfo{author}{\bibfnamefont{A.}~\bibnamefont{Sil}},
  \bibinfo{journal}{JCAP} \textbf{\bibinfo{volume}{04}}, \bibinfo{pages}{013}
  (\bibinfo{year}{2020}{\natexlab{b}}), \eprint{1910.00612}.

\bibitem[{\citenamefont{Borah et~al.}(2020)\citenamefont{Borah, Roshan, and
  Sil}}]{Borah:2020nsz}
\bibinfo{author}{\bibfnamefont{D.}~\bibnamefont{Borah}},
  \bibinfo{author}{\bibfnamefont{R.}~\bibnamefont{Roshan}}, \bibnamefont{and}
  \bibinfo{author}{\bibfnamefont{A.}~\bibnamefont{Sil}},
  \bibinfo{journal}{Phys. Rev. D} \textbf{\bibinfo{volume}{102}},
  \bibinfo{pages}{075034} (\bibinfo{year}{2020}), \eprint{2007.14904}.

\bibitem[{\citenamefont{Jangid and Bandyopadhyay}(2020)}]{Jangid:2020qgo}
\bibinfo{author}{\bibfnamefont{S.}~\bibnamefont{Jangid}} \bibnamefont{and}
  \bibinfo{author}{\bibfnamefont{P.}~\bibnamefont{Bandyopadhyay}},
  \bibinfo{journal}{Eur. Phys. J. C} \textbf{\bibinfo{volume}{80}},
  \bibinfo{pages}{715} (\bibinfo{year}{2020}), \eprint{2003.11821}.

\bibitem[{\citenamefont{Bandyopadhyay et~al.}(2020)\citenamefont{Bandyopadhyay,
  Jangid, and Mitra}}]{Bandyopadhyay:2020djh}
\bibinfo{author}{\bibfnamefont{P.}~\bibnamefont{Bandyopadhyay}},
  \bibinfo{author}{\bibfnamefont{S.}~\bibnamefont{Jangid}}, \bibnamefont{and}
  \bibinfo{author}{\bibfnamefont{M.}~\bibnamefont{Mitra}}
  (\bibinfo{year}{2020}), \eprint{2008.11956}.

\bibitem[{\citenamefont{Borah et~al.}(2019)\citenamefont{Borah, Roshan, and
  Sil}}]{Borah:2019aeq}
\bibinfo{author}{\bibfnamefont{D.}~\bibnamefont{Borah}},
  \bibinfo{author}{\bibfnamefont{R.}~\bibnamefont{Roshan}}, \bibnamefont{and}
  \bibinfo{author}{\bibfnamefont{A.}~\bibnamefont{Sil}},
  \bibinfo{journal}{Phys. Rev. D} \textbf{\bibinfo{volume}{100}},
  \bibinfo{pages}{055027} (\bibinfo{year}{2019}), \eprint{1904.04837}.

\bibitem[{\citenamefont{Dutta~Banik et~al.}(2020)\citenamefont{Dutta~Banik,
  Roshan, and Sil}}]{DuttaBanik:2020jrj}
\bibinfo{author}{\bibfnamefont{A.}~\bibnamefont{Dutta~Banik}},
  \bibinfo{author}{\bibfnamefont{R.}~\bibnamefont{Roshan}}, \bibnamefont{and}
  \bibinfo{author}{\bibfnamefont{A.}~\bibnamefont{Sil}} (\bibinfo{year}{2020}),
  \eprint{2009.01262}.

\bibitem[{\citenamefont{Kannike}(2012)}]{Kannike:2012pe}
\bibinfo{author}{\bibfnamefont{K.}~\bibnamefont{Kannike}},
  \bibinfo{journal}{Eur. Phys. J. C} \textbf{\bibinfo{volume}{72}},
  \bibinfo{pages}{2093} (\bibinfo{year}{2012}), \eprint{1205.3781}.

\bibitem[{\citenamefont{Chakrabortty et~al.}(2014)\citenamefont{Chakrabortty,
  Konar, and Mondal}}]{Chakrabortty:2013mha}
\bibinfo{author}{\bibfnamefont{J.}~\bibnamefont{Chakrabortty}},
  \bibinfo{author}{\bibfnamefont{P.}~\bibnamefont{Konar}}, \bibnamefont{and}
  \bibinfo{author}{\bibfnamefont{T.}~\bibnamefont{Mondal}},
  \bibinfo{journal}{Phys. Rev.} \textbf{\bibinfo{volume}{D89}},
  \bibinfo{pages}{095008} (\bibinfo{year}{2014}), \eprint{1311.5666}.

\bibitem[{\citenamefont{Peskin and Takeuchi}(1992)}]{Peskin:1991sw}
\bibinfo{author}{\bibfnamefont{M.~E.} \bibnamefont{Peskin}} \bibnamefont{and}
  \bibinfo{author}{\bibfnamefont{T.}~\bibnamefont{Takeuchi}},
  \bibinfo{journal}{Phys. Rev.} \textbf{\bibinfo{volume}{D46}},
  \bibinfo{pages}{381} (\bibinfo{year}{1992}).

\bibitem[{\citenamefont{Grimus et~al.}(2008)\citenamefont{Grimus, Lavoura,
  Ogreid, and Osland}}]{Grimus:2008nb}
\bibinfo{author}{\bibfnamefont{W.}~\bibnamefont{Grimus}},
  \bibinfo{author}{\bibfnamefont{L.}~\bibnamefont{Lavoura}},
  \bibinfo{author}{\bibfnamefont{O.~M.} \bibnamefont{Ogreid}},
  \bibnamefont{and} \bibinfo{author}{\bibfnamefont{P.}~\bibnamefont{Osland}},
  \bibinfo{journal}{Nucl. Phys.} \textbf{\bibinfo{volume}{B801}},
  \bibinfo{pages}{81} (\bibinfo{year}{2008}), \eprint{0802.4353}.

\bibitem[{\citenamefont{Arhrib et~al.}(2012)\citenamefont{Arhrib, Benbrik, and
  Gaur}}]{Arhrib:2012ia}
\bibinfo{author}{\bibfnamefont{A.}~\bibnamefont{Arhrib}},
  \bibinfo{author}{\bibfnamefont{R.}~\bibnamefont{Benbrik}}, \bibnamefont{and}
  \bibinfo{author}{\bibfnamefont{N.}~\bibnamefont{Gaur}},
  \bibinfo{journal}{Phys. Rev.} \textbf{\bibinfo{volume}{D85}},
  \bibinfo{pages}{095021} (\bibinfo{year}{2012}), \eprint{1201.2644}.

\bibitem[{\citenamefont{Tanabashi et~al.}(2018)}]{Tanabashi:2018oca}
\bibinfo{author}{\bibfnamefont{M.}~\bibnamefont{Tanabashi}}
  \bibnamefont{et~al.} (\bibinfo{collaboration}{Particle Data Group}),
  \bibinfo{journal}{Phys. Rev.} \textbf{\bibinfo{volume}{D98}},
  \bibinfo{pages}{030001} (\bibinfo{year}{2018}).

\bibitem[{\citenamefont{Swiezewska and Krawczyk}(2013)}]{Swiezewska:2012eh}
\bibinfo{author}{\bibfnamefont{B.}~\bibnamefont{Swiezewska}} \bibnamefont{and}
  \bibinfo{author}{\bibfnamefont{M.}~\bibnamefont{Krawczyk}},
  \bibinfo{journal}{Phys. Rev.} \textbf{\bibinfo{volume}{D88}},
  \bibinfo{pages}{035019} (\bibinfo{year}{2013}), \eprint{1212.4100}.

\bibitem[{\citenamefont{Aaboud et~al.}(2018)}]{Aaboud:2018xdt}
\bibinfo{author}{\bibfnamefont{M.}~\bibnamefont{Aaboud}} \bibnamefont{et~al.}
  (\bibinfo{collaboration}{ATLAS}), \bibinfo{journal}{Phys. Rev.}
  \textbf{\bibinfo{volume}{D98}}, \bibinfo{pages}{052005}
  (\bibinfo{year}{2018}), \eprint{1802.04146}.

\bibitem[{\citenamefont{Sirunyan et~al.}(2018)}]{Sirunyan:2018ouh}
\bibinfo{author}{\bibfnamefont{A.~M.} \bibnamefont{Sirunyan}}
  \bibnamefont{et~al.} (\bibinfo{collaboration}{CMS}), \bibinfo{journal}{JHEP}
  \textbf{\bibinfo{volume}{11}}, \bibinfo{pages}{185} (\bibinfo{year}{2018}),
  \eprint{1804.02716}.

\bibitem[{\citenamefont{Aghanim et~al.}(2018)}]{Aghanim:2018eyx}
\bibinfo{author}{\bibfnamefont{N.}~\bibnamefont{Aghanim}} \bibnamefont{et~al.}
  (\bibinfo{collaboration}{Planck}) (\bibinfo{year}{2018}),
  \eprint{1807.06209}.

\bibitem[{\citenamefont{Akerib et~al.}(2017)}]{Akerib:2016vxi}
\bibinfo{author}{\bibfnamefont{D.}~\bibnamefont{Akerib}} \bibnamefont{et~al.}
  (\bibinfo{collaboration}{LUX}), \bibinfo{journal}{Phys. Rev. Lett.}
  \textbf{\bibinfo{volume}{118}}, \bibinfo{pages}{021303}
  (\bibinfo{year}{2017}), \eprint{1608.07648}.

\bibitem[{\citenamefont{Aprile et~al.}(2018)}]{Aprile:2018dbl}
\bibinfo{author}{\bibfnamefont{E.}~\bibnamefont{Aprile}} \bibnamefont{et~al.}
  (\bibinfo{collaboration}{XENON}), \bibinfo{journal}{Phys. Rev. Lett.}
  \textbf{\bibinfo{volume}{121}}, \bibinfo{pages}{111302}
  (\bibinfo{year}{2018}), \eprint{1805.12562}.

\bibitem[{\citenamefont{Tan et~al.}(2016)}]{Tan:2016zwf}
\bibinfo{author}{\bibfnamefont{A.}~\bibnamefont{Tan}} \bibnamefont{et~al.}
  (\bibinfo{collaboration}{PandaX-II}), \bibinfo{journal}{Phys. Rev. Lett.}
  \textbf{\bibinfo{volume}{117}}, \bibinfo{pages}{121303}
  (\bibinfo{year}{2016}), \eprint{1607.07400}.

\bibitem[{\citenamefont{Cui et~al.}(2017)}]{Cui:2017nnn}
\bibinfo{author}{\bibfnamefont{X.}~\bibnamefont{Cui}} \bibnamefont{et~al.}
  (\bibinfo{collaboration}{PandaX-II}), \bibinfo{journal}{Phys. Rev. Lett.}
  \textbf{\bibinfo{volume}{119}}, \bibinfo{pages}{181302}
  (\bibinfo{year}{2017}), \eprint{1708.06917}.

\bibitem[{\citenamefont{Ma and Raidal}(2001)}]{Ma:2001mr}
\bibinfo{author}{\bibfnamefont{E.}~\bibnamefont{Ma}} \bibnamefont{and}
  \bibinfo{author}{\bibfnamefont{M.}~\bibnamefont{Raidal}},
  \bibinfo{journal}{Phys. Rev. Lett.} \textbf{\bibinfo{volume}{87}},
  \bibinfo{pages}{011802} (\bibinfo{year}{2001}), \bibinfo{note}{[Erratum:
  Phys.Rev.Lett. 87, 159901 (2001)]}, \eprint{hep-ph/0102255}.

\bibitem[{\citenamefont{Toma and Vicente}(2014)}]{Toma:2013zsa}
\bibinfo{author}{\bibfnamefont{T.}~\bibnamefont{Toma}} \bibnamefont{and}
  \bibinfo{author}{\bibfnamefont{A.}~\bibnamefont{Vicente}},
  \bibinfo{journal}{JHEP} \textbf{\bibinfo{volume}{01}}, \bibinfo{pages}{160}
  (\bibinfo{year}{2014}), \eprint{1312.2840}.

\bibitem[{\citenamefont{Baek et~al.}(2014)\citenamefont{Baek, Okada, and
  Toma}}]{Baek:2014awa}
\bibinfo{author}{\bibfnamefont{S.}~\bibnamefont{Baek}},
  \bibinfo{author}{\bibfnamefont{H.}~\bibnamefont{Okada}}, \bibnamefont{and}
  \bibinfo{author}{\bibfnamefont{T.}~\bibnamefont{Toma}},
  \bibinfo{journal}{Phys. Lett. B} \textbf{\bibinfo{volume}{732}},
  \bibinfo{pages}{85} (\bibinfo{year}{2014}), \eprint{1401.6921}.

\bibitem[{\citenamefont{Das et~al.}(2017)\citenamefont{Das, Nomura, Okada, and
  Roy}}]{Das:2017ski}
\bibinfo{author}{\bibfnamefont{A.}~\bibnamefont{Das}},
  \bibinfo{author}{\bibfnamefont{T.}~\bibnamefont{Nomura}},
  \bibinfo{author}{\bibfnamefont{H.}~\bibnamefont{Okada}}, \bibnamefont{and}
  \bibinfo{author}{\bibfnamefont{S.}~\bibnamefont{Roy}},
  \bibinfo{journal}{Phys. Rev. D} \textbf{\bibinfo{volume}{96}},
  \bibinfo{pages}{075001} (\bibinfo{year}{2017}), \eprint{1704.02078}.

\bibitem[{\citenamefont{Baldini et~al.}(2016)}]{TheMEG:2016wtm}
\bibinfo{author}{\bibfnamefont{A.}~\bibnamefont{Baldini}} \bibnamefont{et~al.}
  (\bibinfo{collaboration}{MEG}), \bibinfo{journal}{Eur. Phys. J. C}
  \textbf{\bibinfo{volume}{76}}, \bibinfo{pages}{434} (\bibinfo{year}{2016}),
  \eprint{1605.05081}.

\bibitem[{\citenamefont{Ahriche et~al.}(2018)\citenamefont{Ahriche, Jueid, and
  Nasri}}]{Ahriche:2017iar}
\bibinfo{author}{\bibfnamefont{A.}~\bibnamefont{Ahriche}},
  \bibinfo{author}{\bibfnamefont{A.}~\bibnamefont{Jueid}}, \bibnamefont{and}
  \bibinfo{author}{\bibfnamefont{S.}~\bibnamefont{Nasri}},
  \bibinfo{journal}{Phys. Rev. D} \textbf{\bibinfo{volume}{97}},
  \bibinfo{pages}{095012} (\bibinfo{year}{2018}), \eprint{1710.03824}.

\bibitem[{\citenamefont{Maki et~al.}(1962)\citenamefont{Maki, Nakagawa, and
  Sakata}}]{Maki:1962mu}
\bibinfo{author}{\bibfnamefont{Z.}~\bibnamefont{Maki}},
  \bibinfo{author}{\bibfnamefont{M.}~\bibnamefont{Nakagawa}}, \bibnamefont{and}
  \bibinfo{author}{\bibfnamefont{S.}~\bibnamefont{Sakata}},
  \bibinfo{journal}{Prog. Theor. Phys.} \textbf{\bibinfo{volume}{28}},
  \bibinfo{pages}{870} (\bibinfo{year}{1962}).

\bibitem[{\citenamefont{Davidson and Ibarra}(2002)}]{Davidson:2002qv}
\bibinfo{author}{\bibfnamefont{S.}~\bibnamefont{Davidson}} \bibnamefont{and}
  \bibinfo{author}{\bibfnamefont{A.}~\bibnamefont{Ibarra}},
  \bibinfo{journal}{Phys. Lett. B} \textbf{\bibinfo{volume}{535}},
  \bibinfo{pages}{25} (\bibinfo{year}{2002}), \eprint{hep-ph/0202239}.

\bibitem[{\citenamefont{Edsjo and Gondolo}(1997)}]{Edsjo:1997bg}
\bibinfo{author}{\bibfnamefont{J.}~\bibnamefont{Edsjo}} \bibnamefont{and}
  \bibinfo{author}{\bibfnamefont{P.}~\bibnamefont{Gondolo}},
  \bibinfo{journal}{Phys. Rev. D} \textbf{\bibinfo{volume}{56}},
  \bibinfo{pages}{1879} (\bibinfo{year}{1997}), \eprint{hep-ph/9704361}.

\bibitem[{\citenamefont{Hugle et~al.}(2018)\citenamefont{Hugle, Platscher, and
  Schmitz}}]{Hugle:2018qbw}
\bibinfo{author}{\bibfnamefont{T.}~\bibnamefont{Hugle}},
  \bibinfo{author}{\bibfnamefont{M.}~\bibnamefont{Platscher}},
  \bibnamefont{and} \bibinfo{author}{\bibfnamefont{K.}~\bibnamefont{Schmitz}},
  \bibinfo{journal}{Phys. Rev. D} \textbf{\bibinfo{volume}{98}},
  \bibinfo{pages}{023020} (\bibinfo{year}{2018}), \eprint{1804.09660}.

\bibitem[{\citenamefont{Mahanta and Borah}(2019)}]{Mahanta:2019gfe}
\bibinfo{author}{\bibfnamefont{D.}~\bibnamefont{Mahanta}} \bibnamefont{and}
  \bibinfo{author}{\bibfnamefont{D.}~\bibnamefont{Borah}},
  \bibinfo{journal}{JCAP} \textbf{\bibinfo{volume}{11}}, \bibinfo{pages}{021}
  (\bibinfo{year}{2019}), \eprint{1906.03577}.

\bibitem[{\citenamefont{Mahanta and Borah}(2020)}]{Mahanta:2019sfo}
\bibinfo{author}{\bibfnamefont{D.}~\bibnamefont{Mahanta}} \bibnamefont{and}
  \bibinfo{author}{\bibfnamefont{D.}~\bibnamefont{Borah}},
  \bibinfo{journal}{JCAP} \textbf{\bibinfo{volume}{04}}, \bibinfo{pages}{032}
  (\bibinfo{year}{2020}), \eprint{1912.09726}.

\bibitem[{\citenamefont{Casas and Ibarra}(2001)}]{Casas:2001sr}
\bibinfo{author}{\bibfnamefont{J.}~\bibnamefont{Casas}} \bibnamefont{and}
  \bibinfo{author}{\bibfnamefont{A.}~\bibnamefont{Ibarra}},
  \bibinfo{journal}{Nucl. Phys. B} \textbf{\bibinfo{volume}{618}},
  \bibinfo{pages}{171} (\bibinfo{year}{2001}), \eprint{hep-ph/0103065}.

\bibitem[{\citenamefont{Staub}(2014)}]{Staub:2013tta}
\bibinfo{author}{\bibfnamefont{F.}~\bibnamefont{Staub}},
  \bibinfo{journal}{Comput. Phys. Commun.} \textbf{\bibinfo{volume}{185}},
  \bibinfo{pages}{1773} (\bibinfo{year}{2014}), \eprint{1309.7223}.

\bibitem[{\citenamefont{Abada et~al.}(2006{\natexlab{a}})\citenamefont{Abada,
  Davidson, Josse-Michaux, Losada, and Riotto}}]{Abada:2006fw}
\bibinfo{author}{\bibfnamefont{A.}~\bibnamefont{Abada}},
  \bibinfo{author}{\bibfnamefont{S.}~\bibnamefont{Davidson}},
  \bibinfo{author}{\bibfnamefont{F.-X.} \bibnamefont{Josse-Michaux}},
  \bibinfo{author}{\bibfnamefont{M.}~\bibnamefont{Losada}}, \bibnamefont{and}
  \bibinfo{author}{\bibfnamefont{A.}~\bibnamefont{Riotto}},
  \bibinfo{journal}{JCAP} \textbf{\bibinfo{volume}{04}}, \bibinfo{pages}{004}
  (\bibinfo{year}{2006}{\natexlab{a}}), \eprint{hep-ph/0601083}.

\bibitem[{\citenamefont{Nardi et~al.}(2006)\citenamefont{Nardi, Nir, Roulet,
  and Racker}}]{Nardi:2006fx}
\bibinfo{author}{\bibfnamefont{E.}~\bibnamefont{Nardi}},
  \bibinfo{author}{\bibfnamefont{Y.}~\bibnamefont{Nir}},
  \bibinfo{author}{\bibfnamefont{E.}~\bibnamefont{Roulet}}, \bibnamefont{and}
  \bibinfo{author}{\bibfnamefont{J.}~\bibnamefont{Racker}},
  \bibinfo{journal}{JHEP} \textbf{\bibinfo{volume}{01}}, \bibinfo{pages}{164}
  (\bibinfo{year}{2006}), \eprint{hep-ph/0601084}.

\bibitem[{\citenamefont{Abada et~al.}(2006{\natexlab{b}})\citenamefont{Abada,
  Davidson, Ibarra, Josse-Michaux, Losada, and Riotto}}]{Abada:2006ea}
\bibinfo{author}{\bibfnamefont{A.}~\bibnamefont{Abada}},
  \bibinfo{author}{\bibfnamefont{S.}~\bibnamefont{Davidson}},
  \bibinfo{author}{\bibfnamefont{A.}~\bibnamefont{Ibarra}},
  \bibinfo{author}{\bibfnamefont{F.-X.} \bibnamefont{Josse-Michaux}},
  \bibinfo{author}{\bibfnamefont{M.}~\bibnamefont{Losada}}, \bibnamefont{and}
  \bibinfo{author}{\bibfnamefont{A.}~\bibnamefont{Riotto}},
  \bibinfo{journal}{JHEP} \textbf{\bibinfo{volume}{09}}, \bibinfo{pages}{010}
  (\bibinfo{year}{2006}{\natexlab{b}}), \eprint{hep-ph/0605281}.

\bibitem[{\citenamefont{Blanchet and Di~Bari}(2007)}]{Blanchet:2006be}
\bibinfo{author}{\bibfnamefont{S.}~\bibnamefont{Blanchet}} \bibnamefont{and}
  \bibinfo{author}{\bibfnamefont{P.}~\bibnamefont{Di~Bari}},
  \bibinfo{journal}{JCAP} \textbf{\bibinfo{volume}{03}}, \bibinfo{pages}{018}
  (\bibinfo{year}{2007}), \eprint{hep-ph/0607330}.

\bibitem[{\citenamefont{Blanchet et~al.}(2007)\citenamefont{Blanchet, Di~Bari,
  and Raffelt}}]{Blanchet:2006ch}
\bibinfo{author}{\bibfnamefont{S.}~\bibnamefont{Blanchet}},
  \bibinfo{author}{\bibfnamefont{P.}~\bibnamefont{Di~Bari}}, \bibnamefont{and}
  \bibinfo{author}{\bibfnamefont{G.}~\bibnamefont{Raffelt}},
  \bibinfo{journal}{JCAP} \textbf{\bibinfo{volume}{03}}, \bibinfo{pages}{012}
  (\bibinfo{year}{2007}), \eprint{hep-ph/0611337}.

\bibitem[{\citenamefont{Dev et~al.}(2018{\natexlab{b}})\citenamefont{Dev,
  Di~Bari, Garbrecht, Lavignac, Millington, and Teresi}}]{Dev:2017trv}
\bibinfo{author}{\bibfnamefont{P.~S.~B.} \bibnamefont{Dev}},
  \bibinfo{author}{\bibfnamefont{P.}~\bibnamefont{Di~Bari}},
  \bibinfo{author}{\bibfnamefont{B.}~\bibnamefont{Garbrecht}},
  \bibinfo{author}{\bibfnamefont{S.}~\bibnamefont{Lavignac}},
  \bibinfo{author}{\bibfnamefont{P.}~\bibnamefont{Millington}},
  \bibnamefont{and} \bibinfo{author}{\bibfnamefont{D.}~\bibnamefont{Teresi}},
  \bibinfo{journal}{Int. J. Mod. Phys. A} \textbf{\bibinfo{volume}{33}},
  \bibinfo{pages}{1842001} (\bibinfo{year}{2018}{\natexlab{b}}),
  \eprint{1711.02861}.

\bibitem[{\citenamefont{Ade et~al.}(2016)}]{Ade:2015xua}
\bibinfo{author}{\bibfnamefont{P.}~\bibnamefont{Ade}} \bibnamefont{et~al.}
  (\bibinfo{collaboration}{Planck}), \bibinfo{journal}{Astron. Astrophys.}
  \textbf{\bibinfo{volume}{594}}, \bibinfo{pages}{A13} (\bibinfo{year}{2016}),
  \eprint{1502.01589}.

\bibitem[{\citenamefont{Giedt et~al.}(2009)\citenamefont{Giedt, Thomas, and
  Young}}]{Giedt:2009mr}
\bibinfo{author}{\bibfnamefont{J.}~\bibnamefont{Giedt}},
  \bibinfo{author}{\bibfnamefont{A.~W.} \bibnamefont{Thomas}},
  \bibnamefont{and} \bibinfo{author}{\bibfnamefont{R.~D.} \bibnamefont{Young}},
  \bibinfo{journal}{Phys. Rev. Lett.} \textbf{\bibinfo{volume}{103}},
  \bibinfo{pages}{201802} (\bibinfo{year}{2009}), \eprint{0907.4177}.

\end{thebibliography}

\end{document}